\documentclass[twocolumn, twocolappendix]{aastex7}

\usepackage{newtxtext,newtxmath}
\usepackage{anyfontsize}
\usepackage{xcolor}
\usepackage{newtxmath}
\usepackage{subcaption}

\shorttitle{GMRT Survey of Magnetic Massive Stars -- I: Emission from Single Stars}
\shortauthors{Biswas et al.}

\begin{document}


\title{GMRT Survey of Radio Emission from Magnetic Massive Stars -- I: Emission from Single Stars at sub-GHz Frequencies}


\author[0000-0002-1741-6286]{Ayan Biswas}
\affiliation{National Centre for Radio Astrophysics, Tata Institute of Fundamental Research, Ganeshkhind, Pune-411007, India}
\affiliation{Department of Physics, Engineering Physics \& Astronomy, Queen’s University, Kingston, Ontario K7L 3N6, Canada}
\affiliation{Department of Physics \& Space Science, Royal Military College of Canada, PO Box 17000, Station Forces, Kingston, ON K7K 7B4, Canada}
\email[show]{abiswas@ncra.tifr.res.in, ayan.scc2@gmail.com}

\author[0000-0002-1854-0131]{Gregg A. Wade}
\affiliation{Department of Physics \& Space Science, Royal Military College of Canada, PO Box 17000, Station Forces, Kingston, ON K7K 7B4, Canada}
\affiliation{Department of Physics, Engineering Physics \& Astronomy, Queen’s University, Kingston, Ontario K7L 3N6, Canada}
\email{Gregg.Wade@rmc.ca} 

\author[0000-0001-8704-1822]{Barnali Das}
\affiliation{National Centre for Radio Astrophysics, Tata Institute of Fundamental Research, Ganeshkhind, Pune-411007, India}
\email{barnali@ncra.tifr.res.in} 

\author[0000-0002-5633-7548]{Veronique Petit}
\affiliation{Department of Physics and Astronomy, Bartol Research Institute, University of Delaware, Newark, DE 19716, USA}
\email{VPetit@udel.edu} 

\author[0000-0003-1387-5044]{Matthew E. Shultz}
\affiliation{Department of Physics and Astronomy, Bartol Research Institute, University of Delaware, Newark, DE 19716, USA}
\email{matt.shultz@gmail.com} 

\correspondingauthor{Ayan Biswas}

 
\begin{abstract}

With the growing subset of magnetic massive stars, it is now possible to conduct a systematic survey of radio emission from magnetic hot stars to better understand the underlying emission mechanisms. Previous surveys of radio emission from hot star magnetospheres have focused on high frequencies ($>$2 GHz). At lower frequencies, additional emission and absorption mechanisms are expected, increasing the complexity of the observed emission. In this work, we survey towards lower frequencies while also increasing the sample size. We report the study of 28 magnetic hot stars with the Giant Metrewave Radio Telescope (GMRT) during cycles 27 and 28 of its operation. Among these, we found 11 detections and 17 non-detections. We also include 16 additional targets observed with GMRT from the literature. We investigated the dependence of low-frequency radio luminosity on different stellar parameters and searched for a scaling relationship at low frequencies. We further test the centrifugal breakout model for gyrosynchrotron emission. The observed low-frequency radio luminosities show a clear dependence of radio emission on magnetic field strength and rotation period, consistent with high-frequency studies. We observe a trend in scaling relationships with frequency and comment on the statistical behavior of gyrosynchrotron spectra. The observed low-frequency behavior likely reflects a combination of free-free absorption and the location of the low-frequency turnover in the gyrosynchrotron spectrum, which may vary among stars depending on their magnetospheric properties and can suppress detectable sub-GHz emission. One of the detected stars, HD 37742, is the first magnetic O-type star detected at sub-GHz frequencies.
\end{abstract}




\section{Introduction} \label{sec:intro}

Massive stars serve as the energetic and nucleosynthetic powerhouses of galaxies. Prior to their explosive demise as supernovae, they enrich galaxies with heavy elements and stimulate fresh waves of star formation. Their high luminosity illuminates and ionizes the surrounding interstellar medium, while their vigorous stellar winds drive powerful, high-velocity outflows. Despite their relatively small numbers, they significantly impact the energetics, structure, and chemical enrichment of their host galaxy \citep{Crowther2010}.

\begin{figure*}
    \centering
    \includegraphics[width=\linewidth]{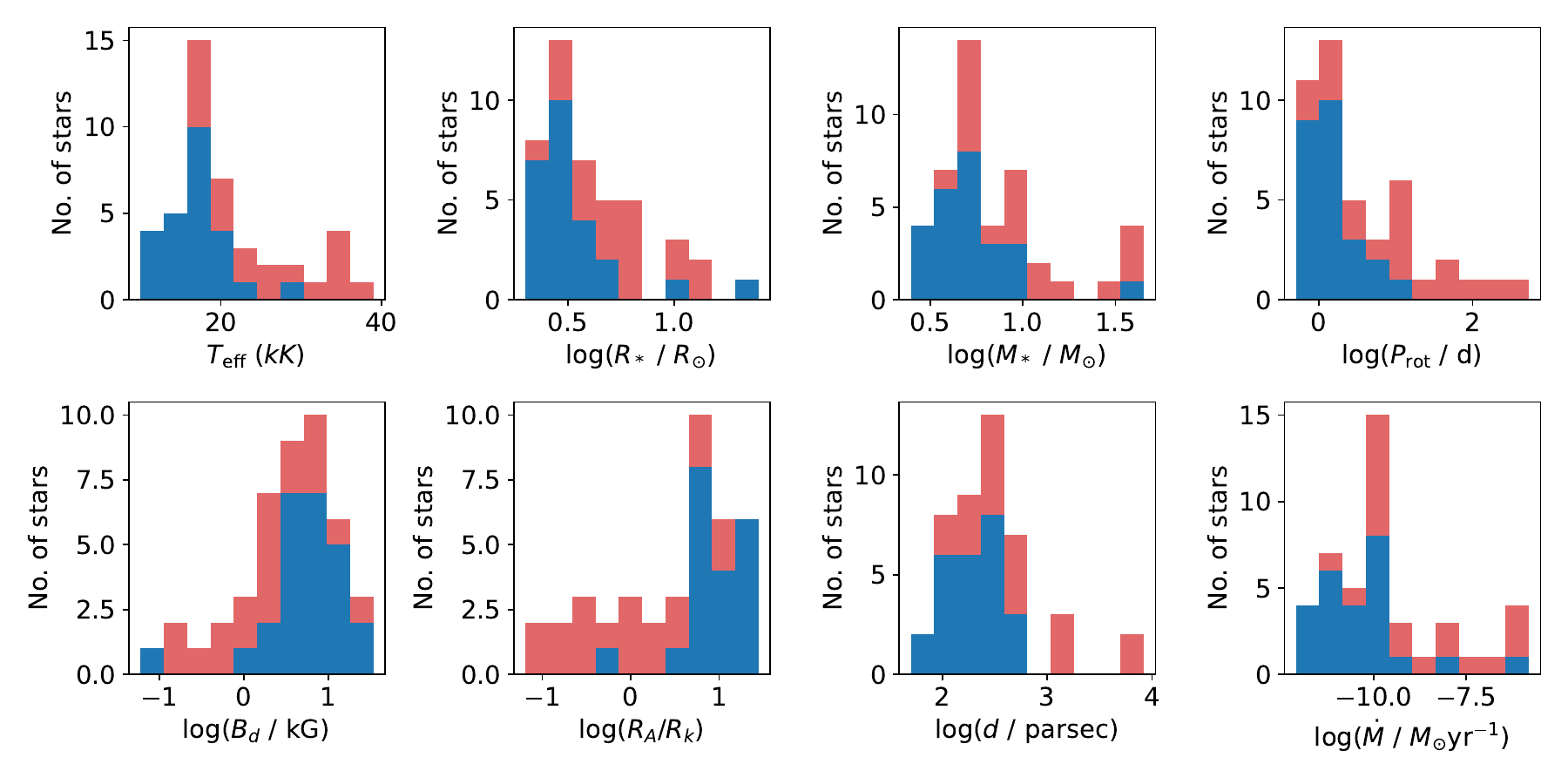}
    \caption{Histogram showing the distribution of all the magnetic hot stars observed with GMRT/uGMRT with respect to some selected parameters. The parameters shown are (from top left, clockwise): effective temperature $(T_{\rm eff} \ / \ {\rm kK})$, stellar radius $(R_* \ / \ R_{\odot}$), stellar mass $(M_* \ / \ M_{\odot}$), rotation period $(P_{\rm rot} \ / \ {\rm d}$), dipolar field strength $(B_{\rm d} \ / \ {\rm kG}$), ratio of Alfv{\'{e}}n radius and the Kepler co-rotation radius ($R_A/R_K$), distance $(d \ / \ {\rm pc}$), and mass-loss rate $(\dot{M} \ / \ M_{\odot} {\rm yr}^{-1}$). The background histogram in red represents the complete sample, while the blue coloured area represent the distribution of detected stars. Details on the sample can be found in Sec. \ref{sec:target_selection}.     }
    \label{fig:hist_plot}
\end{figure*}

Recent systematic investigations, such as the Magnetism in Massive Stars (MiMeS) survey \citep{Wade2016, Grunhut2017} or the B Fields in OB Stars (BOB) survey \citep{Morel2015}, have revealed strong ($\sim$kG), organized (typically dipolar) magnetic fields in a growing subset of massive stars \citep{Petit2013, Alecian2015}. In parallel, analytic models and magnetohydrodynamical (MHD) simulations \citep{ud-Doula2002, ud-Doula2006, Townsend2005, ud-Doula2014, Owocki2016} have explored the dynamical interaction of such fields with rotation and mass-loss.  As the magnetic field strength decreases with distance from the star, the wind pressure eventually becomes equal to the magnetic pressure, defining the Alfv{\'{e}}n radius.  Magnetospheric emission, in general, is produced within the magnetically dominated region below Alfv{\'{e}}n surface, also known as the inner magnetosphere. There exist extensive studies of massive star magnetospheres at various wavelengths (e.g. H$\alpha$, \citealt{Petit2013,Shultz2020}; X-rays, \citealt{Naze2014}; infrared, \citealt{Oksala2015}). Radio emissions in massive stars are expected to be produced from different parts of the magnetosphere, giving a complementary study of the stellar magnetosphere. Despite the importance, systematic studies of radio magnetospheres in hot stars have only been possible recently due to the growing subset of newly detected magnetic stars and the availability of modern sensitive radio telescopes. 

Historically, around one-fourth of the magnetic B/A type stars have been found to exhibit nonthermal radio emission \citep{Drake1987, Andre1988, Linsky1992, Leone1994, Leone1996}. These studies were usually performed in high-frequency ($>1.5$ GHz) radio bands. \citealt{Kurapati2017} conducted a systematic survey of magnetic O stars in the frequency range 2--35 GHz. Recently, \cite{Leto2021} and \cite{Shultz2022} carried out a survey of radio emission from magnetic stars, also focusing at higher frequencies ($>1.5$ GHz). However, the study of magnetic massive stars in low-frequency ($<1.5$ GHz) radio bands is just gaining momentum in the recent years. At low frequencies, \cite{Hajduk2022} searched the LOFAR Two Metre Survey (LoTSS, \citealt{Shimwell2022}) to find radio emission in the frequency range of 120--168 MHz and detected two of 65 targets. The first pilot survey of OB stars at sub-GHz bands with the Giant Metrewave Radio Telescope (GMRT) was reported by \cite{Chandra2015} where they detected 5 of 8 OB stars. In a recent study, \cite{Keszthelyi2024} detected one magnetic hot star among 5 target B-type stars with the upgraded GMRT (uGMRT, \citealt{Gupta2017}), which was also independently observed in this study using GMRT. \cite{Das2022c} surveyed selected magnetic hot stars near the magnetic nulls in search of coherent radio emission at uGMRT frequencies. In lower frequencies, additional nonthermal emission can take place that can go undiagnosed at high frequencies. A short review of the possible low-frequency radio emission mechanisms is given in the following section.

\subsection{Radio Emission Mechanism}

In single stars with high mass-loss rates, radio emission usually arises from the thermal free-free emission from the ionized stellar wind plasma \citep{Wright1975}. {At lower frequencies, this emission can become partially or strongly absorbed because of free-free absorption, modifying the observed radio spectrum and potentially producing a low-frequency turnover. The location of this turnover is expected to depend on the local magnetospheric and wind properties, and therefore may vary from star to star, providing information about the mass-loss rate, wind density, and magnetospheric structure \citep{Leto2026}.}

In the presence of a magnetic field, the wind electrons can reach sub-relativistic energies. These accelerated electrons travel along field lines and emit radio by the gyrosynchrotron mechanism \citep{Drake1987, Linsky1992}.  Such emission exhibits periodic rotational modulation, which correlates with that of the stellar disk averaged line-of-sight magnetic field $\langle B_\mathrm{z}\rangle$ \citep{Leone1993, Lim1996}. The gyrosynchrotron emission varies periodically, with a period equal to the stellar rotational period. This emission is likely to be maximum when the dipole axis is closest to the line-of-sight, i.e., extrema of $\langle B_\mathrm{z}\rangle$.

\startlongtable
\begin{deluxetable*}{lllccccccccccc}
\tablecaption{Details (spectral type, effective temperature, luminosity, radius, mass, rotational period, polar strength of the magnetic dipole at the stellar surface, Kepler co-rotation radius, Alfv{\'{e}}n radius, mass-loss rate and type of magnetosphere) of all magnetic OBA stars observed with GMRT/uGMRT at sub-GHz frequencies according to the selection criteria described in Sec. \ref{sec:target_selection}. We estimated the mass-loss rate using the procedure described by \cite{Vink2021}. In the 13th column. CM represents Centrifugal Magnetosphere, and DM stands for Dynamical Magnetosphere (i.e. if $R_A<R_K$). The last column represents the reference from which the parameter values are adopted. The bold-faced stars represent the `core sample' of this survey, as discussed in Sec. \ref{sec:target_selection}.  For the stars with superscript `C', and `D' in the first column, radio results are taken from \cite{Chandra2015}, and \cite{Das2022c}, respectively, constituting the `supplementary sample' discussed in Sec. \ref{sec:target_selection}. Stars with superscript `S' are the common sample between \cite{Shultz2022} and this study. The stars with superscript `*' show no evidence of a magnetosphere in the optical or UV bands. HD 55522 is a sample from both this study, and \cite{Keszthelyi2024}, independently. Periods with superscript `t' represent new rotation period estimates from this study using TESS data. \label{tab:properties}}
\tablehead{
\colhead{ID} & \colhead{Star} & \colhead{Spectral} & \colhead{$T_{\rm eff}$} & \colhead{$\log L_*$} & \colhead{$R_*$} & \colhead{$M_*$} & \colhead{Period} & \colhead{$B_p$} & \colhead{$R_K$} & \colhead{$R_A$} & \colhead{$\log (\dot{M})$} & \colhead{Type} & \colhead{Ref.} \\
\colhead{} & \colhead{} & \colhead{Type} & \colhead{(kK)} & \colhead{($L_\odot$)} & \colhead{($R_\odot$)} & \colhead{($M_\odot$)} & \colhead{(days)} & \colhead{(kG)} & \colhead{($R_*$)} & \colhead{($R_*$)} & \colhead{($M_\odot$/yr)} & \colhead{(CM/DM)} & \colhead{} 
}
\startdata
1 & \textbf{CPD-28 2561} & O6.5 f?p & 35 $\pm$ 2 & 5.4 $\pm$ 0.2 & 13 & 35 & 73.41 & 2.6 & 19 & 4 & -6.1 & DM & 1 \\
2 & HD 37022$^{\rm C}$ & O7 Vp & 39 $\pm$ 1 & 5.3 $\pm$ 0.1 & 9.9 & 45 & 15.42 & 1.1 & 9.4 & 2.4 & -6.4 & DM & 2,3,4 \\
3 & \textbf{HD 191612}$^{\rm S}$ & O8fpe & 35 $\pm$ 1 & 5.4 $\pm$ 0.2 & 14.5 & 30 & 537.2 & 2.5 & 57 & 3.7 & -6.0 & DM & 5,6 \\
4 & \textbf{NGC 1624-2} & O7f?cp & 35 $\pm$ 2 & 5.1 $\pm$ 0.2 & 9.7 & 34 & 158.0 & $>$20 & 41 & $>$11 & -6.8 & DM & 4,7 \\
5 & HD 57682$^{\rm C}$ & O9 V & 34 $\pm$ 1 & 4.8 $\pm$ 0.2 & 7 & 17 & 63.57 & 1.7 & 24 & 3.7 & -7.1 & DM & 4,8,9 \\
6 & \textbf{HD 37742}$^{*}$$^{\rm , S}$ & O9.5 Ib & 29 $\pm$ 1 & 5.6 $\pm$ 0.1 & 25 & 40 & 7.0 & 0.14 & 2.1 & 1.1 & -6.0 & DM & 4,10-12 \\
7 & \textbf{HD 149438} & B0.2 V & 32 $\pm$ 1 & 4.5 $\pm$ 0.1 & 5.6 & 11 & 41.033 & 0.20 & 20 & 1.8 & -7.6 & DM & 4, 13, 14 \\
8 & \textbf{HD 66665} & B0.5 V & 28 $\pm$ 1 & 4.2 $\pm$ 0.5 & 5.5 & 9.0 & 21 & 0.67 & 12 & 4.0 & -8.4 & DM & 15,16 \\
9 & \textbf{HD 205021}$^{\rm S}$ & B1 IV & 26 $\pm$ 1 & 4.2 $\pm$ 0.1 & 6.5 & 12 & 12.00092 & 0.36 & 7.3 & 4.0 & -8.6 & DM & 15,17 \\
10 & \textbf{HD 163472} & B1/B2 V & 25 $\pm$ 1 & 3.8 $\pm$ 0.1 & 4.6 & 10.3 & 3.638833 & 1.1 & 4.6 & 7 & -9.5 & CM & 15,18  \\
11 & HD 37479$^{\rm C}$ & B2 Vp & 23 $\pm$ 2 & 3.6 $\pm$ 0.2 & 3.9 & 5 & 1.19 & 9.6 & 2.1 & 31 & -9.7 & CM & 15 \\
12 & \textbf{HD 184927} & B2 Vp & 22 $\pm$ 1 & 3.6 $\pm$ 0.2 & 4.3 & 5.5 & 9.530 & 3.9 & 7.7 & 11 & -8.5 & DM & 19 \\
13 & HD 37017$^{\rm C}$ & B2 Vp & 21 $\pm$ 2 & 3.4 $\pm$ 0.2 & 3.6 & 8.4 & 0.90 & 6.2 & 2.1 & 24 & -9.2 & CM & 15 \\
14 & \textbf{HD 37776} & B2 Vp & 22 $\pm$ 1 & 3.3 $\pm$ 0.1 & 3.5 & 8.3 & 1.538756 & 6.1 & 3.1 & 24 & -9.4 & CM & 15,20 \\
15 & HD 147932$^{\rm D}$ & B2 V & 21 $\pm$ 1 & 3.6 $\pm$ 0.2 & 4.5 & 8 & 0.747326 & 2.7 & 1.4 & 10 & -8.7  & CM & 21    \\
16 & \textbf{HD 3360}$^{\rm S}$ & B2 IV & 20.8 $\pm$ 0.2 & 3.8 $\pm$ 0.1 & 6.2 & 8.6 & 5.37045 & 0.15 & 4.2 & 4.6 & -8.3 & CM & 15,22 \\
17 & \textbf{HD 200775}$^{*}$$^{\rm , S}$ & B2 Ve & 19 $\pm$ 2 & 4.0 $\pm$ 0.3 & 11 & 10 & 4.328 & 1.0 & 2.3 & 7.9 & -8.0 & CM & 15,23,24 \\
18 & \textbf{HD 35912}$^{*}$ & B2 V & 18 $\pm$ 1 & 3.3 $\pm$ 0.3 & 4.4 & 7.2 & 1.1988$^t$ & $>$6.0 & 1.7 & $>$23 & -9.4 & CM & 4,25,26 \\
19 & \textbf{HD 182180} & B2 Vn & 17 $\pm$ 1 & 3.0 $\pm$ 0.1 & 3.2 & 6.5 & 0.5214404 & 9.5 & 1.5 & 24 & -10.0 & CM & 15 \\
20 & \textbf{HD 55522}$^{*}$$^{\rm S}$ & B2 IV/V & 17.4 $\pm$ 0.4 & 3.0 $\pm$ 0.2 & 4.2 & 5.9 & 2.729$^t$ & 3.1 & 4.4 & 16 & -9.9 & CM & 15,27 \\
21 & \textbf{HD 142184} & B2 V & 17 $\pm$ 1 & 2.8 $\pm$ 0.1 & 3.1 & 5.5 & 0.50828 & 10 & 1.6 & 45 & -10.4 & CM & 28,29 \\
22 & HD 36485$^{\rm C}$ & B3 Vp & 20 $\pm$ 2 & 3.5 $\pm$ 0.2 & 4.5 & 7.1 & 1.48 & 8.9 & 2.4 & 24 & -8.9 & CM & 4,30 \\
23 & \textbf{HD 208057}$^{*}$ & B3 V & 17 $\pm$ 1 & 3.0 $\pm$ 0.1 & 3.8 & 5.3 & 1.1787$^t$ & 0.6 & 2.2 & 8 & -9.9 & CM & 15,31,t \\
24 & \textbf{HD 35298}$^{*}$ & B3 Vw & 16 $\pm$ 1 & 2.4 $\pm$ 0.1 & 2.4 & 4.3 & 1.85336 & 11 & 4.2 & 34 & -11.1 & CM & 15,32 \\
25 & \textbf{HD 130807}$^{*}$ & B5 & 18 $\pm$ 2 & 3.1 $\pm$ 0.1 & 2.9 & 5.4 & 2.95333 & 9 & 5.3 & 32 & -9.6 & CM & 15,33 \\
26 & \textbf{HD 142990} & B5 V & 18 $\pm$ 1 & 2.9 $\pm$ 0.2 & 2.8 & 5.6 & 0.97907 & 4.7 & 2.6 & 21 & -10.1 & CM & 15,34 \\
27 & \textbf{HD 37058}$^{*}$ & B3 VpC & 19 $\pm$ 1 & 2.9 $\pm$ 0.1 & 2.8 & 5.8 & 14.612 & 2.5 & 16 & 15 & -10.1 & CM  & 15,35 \\
28 & \textbf{HD 35502} & B5 V & 18 $\pm$ 1 & 3.0 $\pm$ 0.1 & 3.0 & 5.8 & 0.853807 & 7.3 & 2.2 & 26 & -9.9 & CM & 15,36 \\
29 & \textbf{HD 176582} & B5 IV & 17 $\pm$ 1 & 2.9 $\pm$ 0.1 & 3.1 & 5.6 & 1.581984 & 5.4  & 3.1 & 24 & -10.1 & CM & 15,37 \\
30 & \textbf{HD 189775}$^{*}$$^{\rm , S}$ & B5 V & 18 $\pm$ 1 & 2.9 $\pm$ 0.1 & 3.2 & 5.6 & 2.6048 & 4.3 & 4.2 & 21 & -10.1 & CM & 15,38 \\
31 & \textbf{HD 61556}$^{*}$ & B5 V & 19 $\pm$ 1 & 3.1 $\pm$ 0.2 & 3.3 & 6.1 & 1.9093 & 2.8 & 3.2 & 15 & -9.7 & CM & 39 \\
32 & \textbf{HD 175362}$^{\rm S}$ & B5 V & 17.6 $\pm$ 0.4 & 2.6 $\pm$ 0.1 & 2.7 & 5.3 & 3.6738 & 17 & 6.5 & 46 & -10.7 & CM & 15,38 \\
33 & HD 19832$^{\rm D}$ & B6 IV/V & 12.4 $\pm$ 0.4 & 2.1 $\pm$ 0.2 & 2.3 & 3.4 & 0.72776 & 2.7 & 2.0 & 45 & -11.8 & CM & 13,34 \\ 
34 & \textbf{HD 125823} & B7 IIIp & 19 $\pm$ 2 & 3.2 $\pm$ 0.1 & 3.0 & 5.9 & 8.8169 & 1.8 & 11 & 12 & -9.7 & CM & 13,34 \\
35 & \textbf{HD 36526}$^{*}$ & B8 Vp & 16 $\pm$ 3 & 2.5 $\pm$ 0.3 & 2.4 & 4.3 & 1.5405 & 11.2 & 3.8 & 47 & -11.0 & CM & 15,34 \\
36 & HD 142301$^{\rm D}$ & B8 III & 15.9 $\pm$ 0.2 & 2.6 $\pm$ 0.1 & 2.4 & 4.5 & 1.459 & 12.5 & 12.5 & 59 & -10.9 & CM & 34  \\
37 & HD 215441$^{\rm C}$ & B8-9p & 15 $\pm$ 2 & 1.8 & 2.6 & 3.5 & 9.49 & 64 & ... & ... & -11.6 & CM & 40,t \\
38 & HD 133880$^{\rm C}$ & B8 IVp & 13 $\pm$ 1 & 2.0 & 2.0 & 3.2 & 0.88 & 19 & 2.2 & 60 & -11.0 & CM & 40  \\
39 & HD 79158$^{\rm D}$ &  B8III & 13.3 $\pm$ 0.1 & 2.6 $\pm$ 0.1 & 3.4 & 4.3 & 3.835 & 3.1 & ... & ... & -10.8 & CM & 13,41 \\
40 & HD 145501$^{\rm D}$ & B8/9 V & 14.5 $\pm$ 0.5 & 2.5 $\pm$ 0.2 & 2.26 & 4.0 & 1.026 & 5.8 & 2.92 & 48 & -11.2 & CM & 34 \\
41 & HD 170000$^{\rm D}$ & B8 V & 11.6 $\pm$ 0.0 & 2.4 $\pm$ 0.0 & 3.7 & 3.5 & 1.716 & 1.8 & ... & ... & -11.1 & CM & 13 \\
42 & HD 45583$^{\rm D}$ & ApSi & 13.3 $\pm$ 0.3 & 2.1 $\pm$ 0.1 & 2.1 & 3.2 & 1.177 & 9.1 & 3.2 & 82 & -11.8 & CM & 34 \\
43 & HD 124224$^{\rm D}$ & ApSi & 12.3 $\pm$ 0.2 & 1.9 $\pm$ 0.0 & 2.2 & 3.0 & 0.5207 & 4.0 & ... & ... & -12.1 & CM & 13  \\
44 & HD 12447$^{\rm D}$ & A0 & 10.0 $\pm$ 0.2 & 1.7 $\pm$ 0.1 & 2.7 & 2.49 & 1.4907 & 1.7 & ... & ... & -11.4 & CM & 13  \\
\enddata
\tablecomments{References: (1) \cite{Wade2015}; (2) \cite{Simon2006}; (3) \cite{Stahl2008}; (4) \cite{Petit2013}; (5) \cite{Howarth2007}; (6) \cite{Wade2011}; (7) \citep{Wade2012}; (8) \cite{Grunhut2009}; (9) \cite{Grunhut2012}; (10) \cite{Bouret2008}; (11) \cite{Hummel2013}; (12) \cite{Blazere2015}; (13) \cite{Shultz2022}; (14) \cite{Donati2006}; (15) \cite{Shultz2019}; (16) \cite{Petit2011}; (17) \cite{Donati2001}; (18) \cite{Neiner2003}; (19) \cite{Yakunin2015}; (20) \cite{Kochukhov2011}; (21) \cite{Leto2020a}; (22) \cite{Briquet2016}; (23) \cite{Alecian2008}; (24) \cite{Bisyarina2015}; (25) \cite{Bychkov2005}; (26) \cite{Simon2010}; (27) \cite{Briquet2007}; (28) \cite{Grunhut2012_hr5907}; (29) \cite{Leto2018}; (30) \cite{Leone2010}; (31) \cite{Henrichs2009}; (32) \cite{Yakunin2013}; (33) \cite{Buysschaert2019}; (34) \cite{Shultz2019b}; (35) \cite{Pedersen1979}; (36) \cite{Sikora2016}; (37) \cite{Bohlender2011}; (38) \cite{Shultz2018b}; (39) \cite{Shultz2015b}; (40) \cite{Chandra2015}; (41) \cite{Oksala2018}
}
\end{deluxetable*}

In the original gyrosynchrotron model, nonthermal radio emission comes from the magnetosphere beyond the Alfv{\'{e}}n surface \citep{Trigilio2004}. The outflowing plasma, along with the open field lines caused by wind ram pressure, create a current sheet where electrons are accelerated, fueling gyrosynchrotron emission \citep{Trigilio2004}. In this framework, the primary role of rotation is to modulate the disk-averaged line-of-sight magnetic field. Recently, a strong dependence of the average radio luminosity on rotation has recently challenged this theory \citep{Leto2021, Shultz2022}. It should be noted that \cite{Linsky1992} predicted that the radio luminosity may inversely depend on the rotation period. However, the authors could not find any strong correlation of radio luminosity with rotation period from their survey of chemically peculiar stars.

The discovery of a rotationally dependent scaling law challenged this model, leading to the development of the new centrifugal breakout (CBO) mechanism \citep{Shultz2022, Owocki2022}. In this scenario, gyrosynchrotron emission requires rapid rotation as well as a strong magnetic ﬁeld. In such a case, the centrifugal support of the plasma above the Kepler co-rotation radius ($R_K$) leads to the formation of a centrifugal magnetosphere (CM), in which plasma accumulates to a high density.  However, below $R_K$ the magnetosphere remains dynamical, similar to the non-rotating case. When the gas pressure overloads the ability of the magnetic field to confine the plasma, the plasma is ejected outward by a CBO event. Magnetic reconnection during CBO leads to flaring, which accelerates electrons to high energies, thereby providing the source electrons to populate the radio magnetosphere. Note also that the fraction of the wind plasma captured by the CM is much higher than that captured by the current sheet (CS) in the non-rotating case. The total energy available for CBO events follows the scaling law given by \cite{Owocki2022}:

\begin{equation}
    L_{\rm CBO} \approx \dot{M} \Omega^2 R_*^2 \left( \frac{B_{\rm eq}^2 R_*^2}{\dot{M} v_{\rm orb}}   \right)^{1/p},
\end{equation}

\noindent
where $\dot{M}$ is the mass-loss rate, $\Omega$ is the rotational frequency, $R_*$ is the stellar radius, $B_{\rm eq}$ is the equatorial field strength, $v_{\rm orb}$ is the surface orbital speed ($=\sqrt{G M_*/R_*}$), and $p$ is the effective multipole index. For a split-monopole case (i.e. $p=1$), we get:

\begin{equation}
    L_{{\rm CBO} (p = 1)} \propto \Omega R_*^3 B_{\rm eq}^2.
\end{equation}

This relationship is similar to the observed scaling relationship ($L_{\rm rad} \propto \Omega^2 R_*^4 B_{d}^2$) obtained by \cite{Leto2021} and \cite{Shultz2022} from high-frequency radio studies. In the case of a pure dipole (i.e. $p=2$), the relationship is modified as:

\begin{equation}
    L_{{\rm CBO} (p = 2)} \propto \Omega^{1.5} R_*^{2.5} B_{\rm eq}.
\end{equation}

In this case, we get a weaker linear scaling with $B_{\rm d}$, and also dependence on the mass-loss rate.

In addition to this periodically variable emission, a fraction of magnetic A/B stars have been found to produce highly directed coherent radio emission with a very high degree of circular polarization \citep{Trigilio2000}. This kind of emission appears as enhancements superimposed on the gyrosynchrotron ‘continuum’. The rotational phases of the arrival of the enhancements coincide approximately with those of the nulls of the longitudinal magnetic field $\langle B_\mathrm{z}\rangle$, which implies that the direction of emission is approximately perpendicular to the magnetic field axis. From these characteristics, the emission mechanism has been inferred to be the electron cyclotron maser emission (ECME; \citealt{Trigilio2000, Trigilio2004}), and has since been detected in numerous stars \citep{Das2020a, Das2022}.

In the case of a binary system, the presence of a companion star may modify the observed radio emission and the inferred mass-loss rate. The winds of both stars may contribute to the observed emission. Detailed observations and simulations have revealed the formation of shocks due to collision of winds from individual stars of the binary system \citep{Usov1992}. The wind-wind interaction can accelerate particles up to relativistic energies, which in the presence of magnetic fields can produce nonthermal synchrotron radio emission \citep{Becker2007}. {This scenario provides an opportunity to investigate shock-driven particle acceleration processes in massive star environments and the associated nonthermal emission from colliding-wind binaries.} This emission is expected to modulate according to the orbital period of the binary system. Additionally, in case both stars are magnetic and in close orbit, the magnetospheres of both stars may interact and give excess nonthermal emission in addition to individual components producing gyrosynchrotron emission \citep{Biswas2023}.

So, in general, radio emission is expected from 3 components: (i) a continuous thermal free-free emission, arising due to mass-loss flow, at low brightness temperature, (ii) nonthermal gyrosynchrotron and/or ECME emission due to the organized magnetic field, (iii) non-steady, and possibly variable synchrotron emission with a higher brightness temperature due to binarity. 

\begin{figure*}
\centering
\gridline{\fig{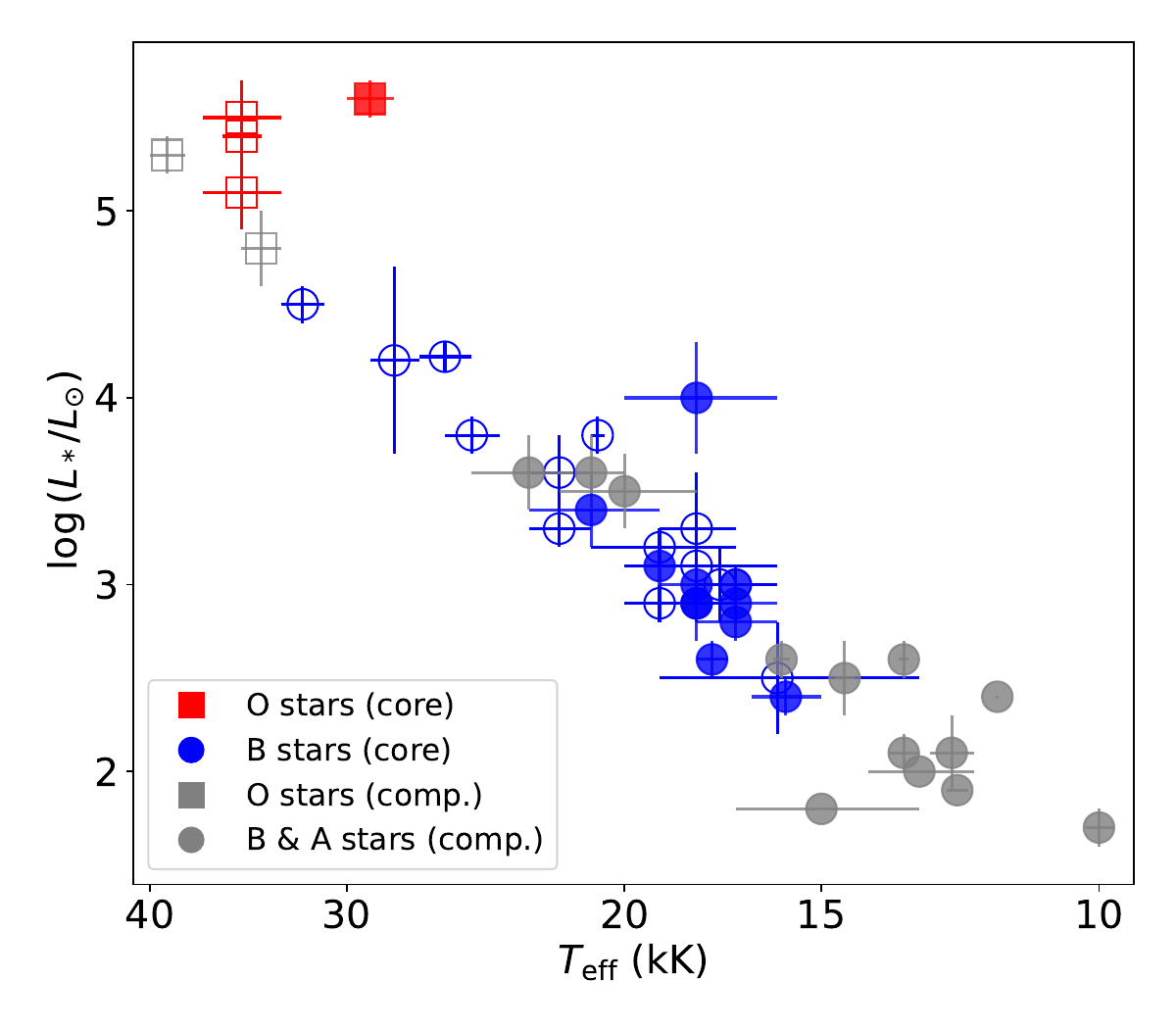}{0.495\linewidth}{(a)} \label{fig:Lbol}
          \fig{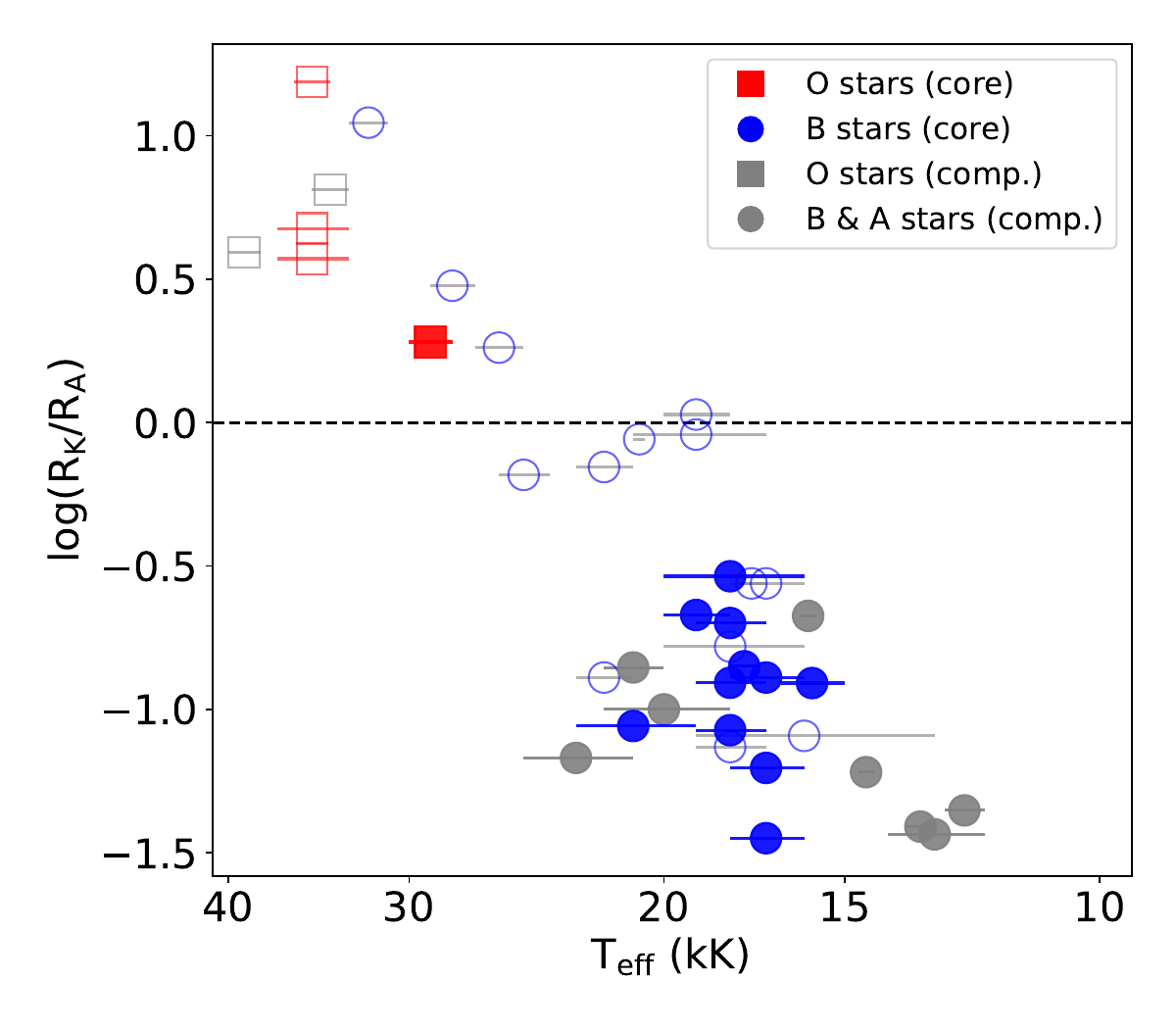}{0.495\linewidth}{(b)} \label{fig:Rk}}
    \caption{(a) HR diagram showing the survey targets observed with GMRT, and (b) plot of $R_{\rm K}/R_{\rm A}$ vs. $T_{\rm eff}$, divided into two sections: above the horizontal dashed line showing stars with dynamic magnetospheres ($R_{\rm A}< R_{\rm K}$), and the bottom part showing stars with a CM ($R_{\rm A}> R_{\rm K}$). For both figures, red/blue filled squares/circles represent stars detected with GMRT under projects 27\textunderscore 048 and 28\_075. Grey markers represent the additional 16 targets observed with GMRT (from \citealt{Chandra2015}), or uGMRT (from \citealt{Das2022c}). Hollow markers represent the undetected targets. The sole detected O-star near the top-left corner is HD 37742.}
    \label{fig:survey_targets}
\end{figure*}


\subsection{Motivation}

Low-frequency ($\nu < 1.5$\,GHz) radio studies of magnetic massive stars are still rare.  
Previous surveys (e.g.\ \citealt{Linsky1992, Leone1994, Kurapati2017, Leto2021, Leto2026}) have concentrated almost entirely on high-frequency ($2$--$35$\,GHz) observations, with only a handful of targeted sub-GHz measurements.  
To date, systematic investigations at these frequencies have been limited to the GMRT pilot survey \citep{Chandra2015}, the RAMBO project \citep{Keszthelyi2024}, and the VAST-MeMeS survey \citep{Das2025}.  
In particular, no dedicated low-frequency survey of {magnetic O-type stars} has yet been carried out, despite extensive optical, ultra-violet (UV), and X-ray studies of these objects under the MiMeS program. 

Extending radio observations into the sub-GHz regime is crucial for several reasons:  
{
\begin{enumerate}
    \item it probes emission processes and absorption effects not accessible at higher frequencies, since the free-free optical depth increases toward lower frequencies, making the circumstellar plasma progressively more opaque in the sub-GHz regime and thereby significantly modifying or suppressing the observed radio spectrum;
    \item it enables the detection and characterization of nonthermal processes such as gyrosynchrotron radiation and electron cyclotron maser emission (ECME). Unlike what is assumed in \cite{Shultz2022}, there is evidence that there could be a significant contribution from sub-GHz emission to the total incoherent radio luminosity as demonstrated in the VAST-MeMeS project, \citep{Das2025}. Thus, for robust estimation of incoherent radio luminosity, we must acquire observations at sub-GHz frequencies;
    \item even upper limits can provide important constraints on the magnetospheric environment and the shape of the radio spectrum, including possible low-frequency attenuation due to free-free absorption or a gyrosynchrotron turnover shifted above the observing band as a consequence of the individual stellar and magnetospheric properties.
\end{enumerate}
}

Here we extend earlier work \citep[e.g.][]{Linsky1992, Chandra2015, Kurapati2017, Leto2021, Shultz2022, Das2022c, Keszthelyi2024, Leto2026} by assembling the most comprehensive sub-GHz study of magnetic O- and B-type stars to date. This study is centered on two recent uGMRT observing programs (projects 27\textunderscore 048 and 28\textunderscore 075), which together targeted 28 stars. Seven of these were previously reported by \cite{Shultz2022}. We supplement this with additional measurements from \cite{Chandra2015} and \cite{Das2022c}, yielding a combined sample of 44 stars with observations having identical observing frequency and bandwidth. This mix of new detections, re-analyzed archival data, and literature results enables both the discovery of previously unreported sub-GHz emitters and the first uniform statistical characterization of magnetic massive stars in this frequency range.

\subsection{Strategy}

The study is designed to:  
\begin{enumerate}
    \item Survey a well-defined, unbiased set of magnetic massive stars for sub-GHz radio emission.
    \item Determine the dominant emission mechanism(s) and assess variability where possible.
    \item Measure the low-frequency spectral index, with multiband observations obtained at similar rotational phases when feasible.
    \item Examine the correlations between incoherent radio emission and stellar/magnetospheric parameters.
    \item Obtain a low-frequency scaling relationship, compare it with high-frequency results, and evaluate the CBO mechanism.
\end{enumerate}

{ For stars with dynamical magnetospheres (DM; $R_{\rm K}>R_{\rm A}$; \citealt{Petit2013}), where $R_{\rm K}$ is the Kepler co-rotation radius estimated from the stellar rotational period and $R_{\rm A}$ is the Alfv\'en radius determined from magnetic confinement of the stellar wind  following \cite{ud-Doula2002}}, we schedule observations near phases of the maximum longitudinal magnetic field $\langle B_z\rangle$ to optimize detectability and coordinate 1390\,MHz and 610\,MHz observations at the same phase when possible. CMs, with their short ($\sim1$\,d) rotational periods, are observed without strict phase constraints, enabling the assessment of rotational variability. { While this work focuses mainly on the magnetospheres of massive OB stars, the sample also includes a few magnetic A-type stars. Unlike OB stars, which possess strong radiatively driven winds, A-type stars are expected to host much weaker winds, potentially leading to different magnetospheric and radio-emission properties. The A-star sample considered here is limited to the hottest early A-type stars, which are closest in properties to late B-type stars.}

\begin{figure*}
\centering
\gridline{\fig{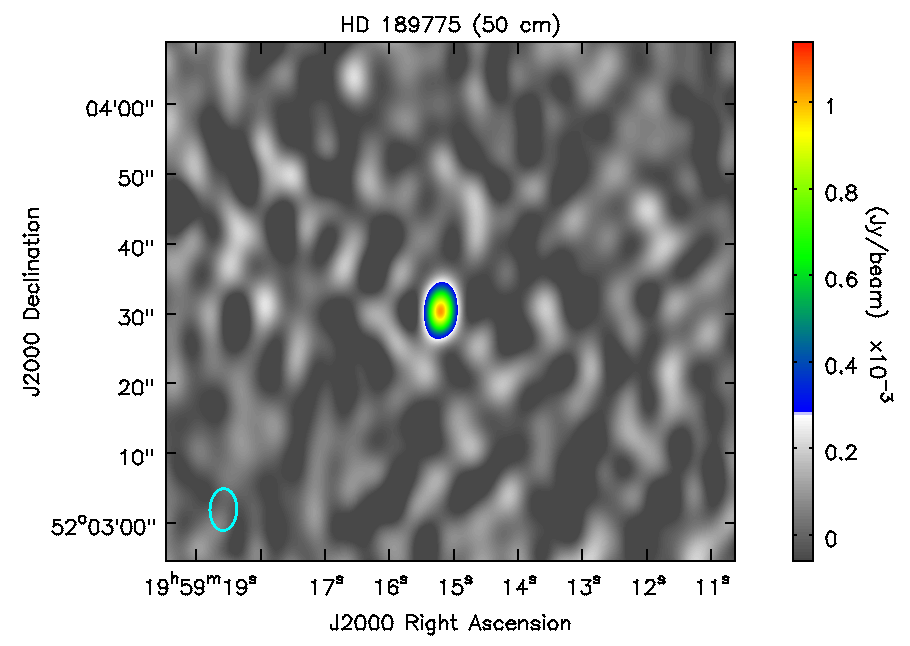}{0.495\linewidth}{(a)} \label{fig:rad1}
    \fig{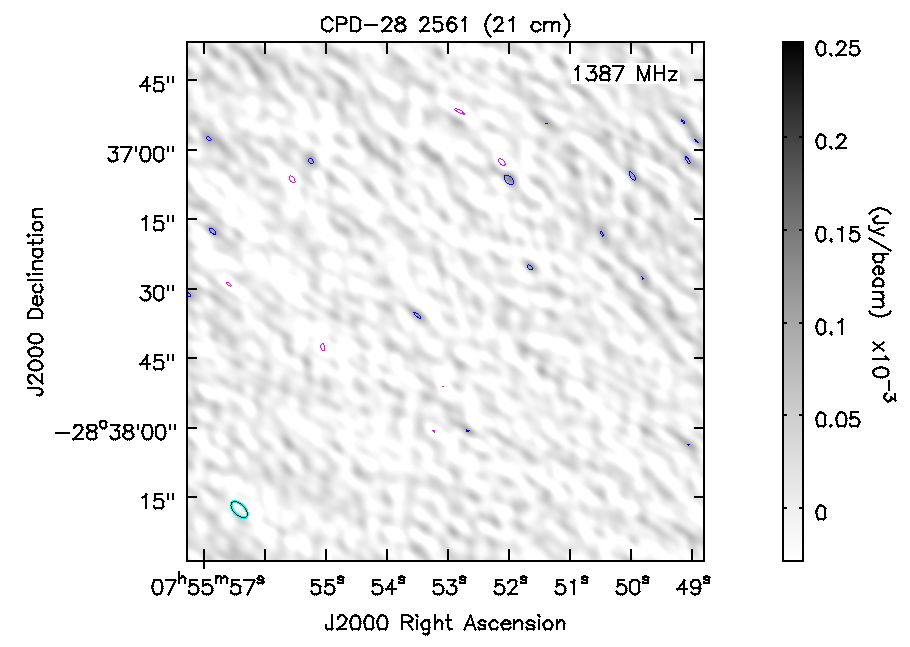}{0.495\linewidth}{(b)} \label{fig:rad2}}
\caption{Sample Stokes I images of two survey targets. (a) 610 MHz band image (zoomed) of HD 189775, one of the unique detections from this survey. The noise in the image is $\sim73 \ \mu$Jy. (b) Zoomed in 1390 MHz band image of the target CPD-28 2561, an example of a non-detection. The 3-$\sigma$  negative and positive levels are drawn with red and blue contours. However, the size of these contours are much smaller than the beam size (cyan ellipse near bottom-left corner of each image), and are not reliable detections.  \label{fig:gmrt_img}}
\end{figure*}

This paper is structured as follows. Section \ref{sec:target_selection} gives the target selection criteria for the survey; the observation details and data analysis procedure are described in Section \ref{Observations}; details on individual detections and statistical analysis are shown in Section \ref{Results}; in Section \ref{Discussion}, we discuss the emission mechanisms observed at low frequencies and obtain a scaling relation for the sub-GHz frequency band; finally in Section \ref{Conclusion}, we summarize the results and conclude. Additionally, we include details on the non-detections in Appendix \ref{sec:App_non_det}, and period analysis of selected targets in Appendix \ref{App:TESS}.

\begin{deluxetable*}{lllcccccccc}
\tablecaption{Observation details and imaging results of the GMRT observations under projects 27\textunderscore 048 and 28\_075. $F_{1390}$ and $F_{610}$ represent the flux densities in the 1390 MHz and 610 MHz, respectively. HJD stands for Heliocentric Julian Date. For stars that were not detected, we report the 3$\sigma$ upper limit. The superscripts beside each star indicate the works from which the flux density values are taken: s: \cite{Shultz2022}, **: this work. \label{tab:gmrtflux}}
\tablehead{
\colhead{ID} & \colhead{Star} & \multicolumn{4}{c}{1390 MHz} & \multicolumn{4}{c}{610 MHz} \\
\colhead{} & \colhead{} & \colhead{HJD$-$2450000} & \colhead{$\phi_{\rm mean}$} & \colhead{$F_{1390}$ (mJy)} & \colhead{rms ($\mu$Jy)} & \colhead{HJD$-$2450000} & \colhead{$\phi_{\rm mean}$} & \colhead{$F_{610}$ (mJy)} & \colhead{rms ($\mu$Jy)}
}
\startdata
 & O-stars \\
1 & CPD-28 2561$^{**}$ & 6973.613 $\pm$ 0.150 & 0.71 & $<$ 0.093 & 30.3 & ...   & ... & ... & ... \\
  &  &  6967.569 $\pm$ 0.027 & 0.63 & $<$ 0.124 & 41.2 &  ...  & ... & ... & ... \\
3 & HD 191612$^{**}$ &  6952.140 $\pm$ 0.059 & 0.58 & $<$ 0.219 & 73.1 &  7231.403 $\pm$ 0.031 & 0.10 & $<$0.345 & 115.2 \\
4 & NGC 1624-2$^{**}$ & 6598.458 $\pm$ 0.047 & 0.00 & $<$ 0.217 & 72.2 & ... &  ...  & ... & ... \\
6 & \textbf{HD 37742}$^{**}$ & 6984.390 $\pm$ 0.161 & 0.37 & 0.315 $\pm$ 0.051 & 53.7 & 7280.487 $\pm$ 0.049 &  0.73  & 0.905 $\pm$ 0.134  & 104.7 \\
 & B-stars \\
7 & HD 149438$^{**}$ & 7033.753 $\pm$ 0.053 & 0.56 & $<$ 0.123 & 40.9  & 7231.155 $\pm$ 0.030 &  0.37  & $<$0.389 & 129.5  \\
8 & HD 66665$^{**}$ & 7031.457 $\pm$ 0.078 & 0.72 & $<$ 0.150 & 50.1 & 7276.538 $\pm$ 0.047 & 0.72   & $<$0.308 &  102.5 \\
9 & HD 205021$^{\rm s}$ & 6982.218 $\pm$ 0.051 & 0.64 & $<$ 0.149 & 49.7 & 7231.571 $\pm$ 0.029 &   0.41 & $<$0.252 & 84.0  \\
10 & HD 163472$^{**}$ & 6970.003 $\pm$ 0.052 & 0.31 & $<$ 0.139 & 46.3 & 7228.242 $\pm$ 0.027 & 0.40  &  $<$ 0.193  & 64.4 \\
12 & HD 184927$^{\rm s}$ & 7032.963 $\pm$ 0.056 & 0.17 & $<$0.114 & 38.0 & 7277.098 $\pm$ 0.047 & 0.79  & $<$0.187 & 62.3  \\
14 & HD 37776$^{\rm **}$ & 6999.259 $\pm$ 0.126 &  0.47 & $<$ 0.094 & 31.4 &  7277.691 $\pm$ 0.026 & 0.42   & $<$ 0.292 & 97.3  \\
16 & HD 3360$^{\rm s}$ & 6982.346 $\pm$ 0.054 & 0.61 & $<$ 0.097 & 32.3 & ... &  ...   & ... &  ...  \\
17 & \textbf{HD 200775}$^{\rm s,**}$ & 6982.091 $\pm$ 0.054 &  0.56  & 0.371 $\pm$ 0.107 & 56.2 & 7229.531 $\pm$ 0.029 &   0.64   & $0.281 \pm 0.076$ &  87.3  \\
18 & HD 35912$^{**}$ & 6987.353 $\pm$ 0.168 & 0.54 & $<$0.156 & 51.9 & 7276.645 $\pm$ 0.029 & 0.86   & $<$ 0.149 & 49.7  \\
19 & \textbf{HD 182180}$^{\rm **}$ & 7033.878 $\pm$ 0.044 &  0.95 & ... & ...  & 7290.166 $\pm$ 0.048 &  0.44    & 3.570 $\pm$ 0.130 &  71.3  \\
20 & HD 55522$^{\rm s, **}$ & 7081.173 $\pm$ 0.129 &  0.95 & $<$ 0.118 & 39.2 & 7280.563 $\pm$ 0.025 &   0.00   & $<$ 0.219 &  72.9  \\
21 & \textbf{HD 142184}$^{**}$ & 7036.592 $\pm$ 0.049 &  0.44 & 12.99 $\pm$ 0.12 & 35.7 & 7231.077 $\pm$ 0.024 &   0.29   & 3.66 $\pm$ 0.19 & 86.5 \\
23 & HD 208057$^{\rm **}$ & 7014.093 $\pm$ 0.049 &  0.98 & $<$ 0.129 & 43.0 & 7231.497 $\pm$ 0.040 &   0.42   & $<$ 0.209 &  69.5  \\
24 & \textbf{HD 35298}$^{\rm **}$ & 6987.320 $\pm$ 0.168 &  0.24 & 0.324 $\pm$ 0.077 & 46.0 & 7276.440 $\pm$ 0.031 &   0.13   & 0.275 $\pm$ 0.092 &  91.7 \\
25 & HD 130807$^{**}$ & 7033.631 $\pm$ 0.044 & 0.72  & $<$ 0.167 & 55.6 & 7435.580 $\pm$ 0.035 &  0.82    & $<$ 0.290 & 96.5   \\
26 & \textbf{HD 142990}$^{\rm **}$ & 6969.880 $\pm$ 0.044 & 0.99  & 5.490 $\pm$ 0.280 & 69.8 & 7276.096 $\pm$ 0.043  &    0.57  & ... & ...   \\
27 & HD 37058$^{**}$ & 6984.365 $\pm$ 0.166 & -  & $<$ 0.160 & 53.4 & 7277.442 $\pm$ 0.039 &   ...   & $<$ 0.225 &  74.9  \\
28 & \textbf{HD 35502}$^{\rm **}$ & 6987.336 $\pm$ 0.168 &  0.93 & 2.088 $\pm$ 0.086 & 49.1 & 7280.398 $\pm$  0.039 &   0.17   & ... &  ... \\
29 & \textbf{HD 176582}$^{\rm **}$ & 6999.050 $\pm$ 0.053 &  0.78 & 0.344 $\pm$ 0.072 & 44.4 & 7231.237 $\pm$ 0.031 &   0.55   & 1.290 $\pm$ 0.120 &  51.2  \\
30 & \textbf{HD 189775}$^{\rm s}$ & 6951.218 $\pm$ 0.056 & 0.36  & 0.41 $\pm$ 0.07 & 43.3 & 7231.320 $\pm$ 0.031  &   0.80   & 1.094 $\pm$ 0.125 & 72.6  \\
31 & \textbf{HD 61556}$^{\rm **}$ & 7081.201 $\pm$ 0.129 & 0.62  & 0.643 $\pm$ 0.089 & 50.2 & ... &   ...   & ...  & ... \\
32 & \textbf{HD 175362}$^{\rm s}$ & 6951.072 $\pm$ 0.060 & 0.91 & 0.275 $\pm$ 0.070 & 47.4 & 7224.324 $\pm$ 0.022 &  0.29  & 0.540 $\pm$ 0.106 & 78.3  \\
34 & HD 125823$^{**}$ & 7031.596 $\pm$ 0.437 &  0.24 & $<$ 0.175 & 58.3 & 7435.503 $\pm$ 0.031 &   0.05   & $<$ 0.290 & 96.8   \\
35 & HD 36526$^{\rm **}$ &  ... &   ... &  ... &  ... & 7277.530 $\pm$ 0.028 &   0.26   & $<$ 0.281 &  93.8 \\
\enddata
\end{deluxetable*}

\section{Sample Selection} \label{sec:target_selection}

We applied the following criteria during target selection:
\begin{enumerate}
    \item Magnetic OB stars for which the basic stellar parameters, periods, and magnetic field strengths are known (e.g.\ from \citealt{Petit2013, Shultz2019}).
    \item Targets that are observable with the GMRT (i.e.\ DEC $>-50^{\circ}$), thus consisting the majority of known magnetic massive stars.
    \item Single stars or binaries in long-period orbits, to avoid major contamination from binary interactions (e.g.\ \citealt{Biswas2023}). A separate survey for close binary magnetic stars will be presented in the accompanying Paper~II. 
    \item No bias in selection based on stellar parameters (e.g.\ mass, radius, temperature, rotation period) or magnetospheric parameters (e.g.\ magnetic field strength, obliquity). Any known magnetic OB stars (as of 2014) that are observable with GMRT were included as targets for this study.
\end{enumerate}

Applying these criteria to the magnetic OB star catalog of \cite{Petit2013} and \cite{Shultz2019} yields the following:

\begin{itemize}
    \item \textbf{Core sample:} 28 stars observed in uGMRT projects 27\_048 (1390\,MHz) and 28\_075 (610\,MHz).  
          These include both entirely new targets and the 7 stars first reported by \cite{Shultz2022}. For these 7 stars, since \cite{Shultz2022} used the same calibration and imaging procedures followed in this study, we have taken the flux densities from \cite{Shultz2022} in case the noise levels were satisfactory. We have done further analysis of these data (e.g. search for polarized emission, spectral index study), and for some observations, they were carefully re-analyzed in case the noise levels were high (shown in Table \ref{tab:gmrtflux}). It should be noted that these 7 targets and corresponding 14 observations were finally not used by \cite{Shultz2022} in deriving their CBO scaling relationships.
    \item \textbf{Supplementary sample:} 16 stars that meet the same selection criteria but for which suitable GMRT/uGMRT measurements already exist in the literature \citep{Chandra2015, Das2022c}. We adopt the literature fluxes directly for this sample.
\end{itemize}

The final combined dataset thus comprises 44 stars (28 core + 16 supplementary), enabling both individual source analysis and a uniform statistical characterization of magnetic massive stars in the sub-GHz regime. This approach preserves the unbiased nature of the sample while maximizing completeness by incorporating compatible archival and literature results. All of the observations for these 44 stars leverage identical instrumental configurations of GMRT. It should be noted that GMRT went through an upgrade after these observations (i.e. uGMRT), and the latest surveys will have a superior sensitivity. 

Fig.~\ref{fig:hist_plot} shows the distribution of the stellar parameters for the whole sample, and for the stars with radio emission detected at sub-GHz frequencies. Their positions in the H-R diagram (Fig.~\ref{fig:survey_targets}a) demonstrate that the sample spans a broad range of main-sequence magnetic OB stars. The $R_{\rm K}/R_{\rm A}$ vs.\ $T_{\rm eff}$ diagram (Fig.~\ref{fig:survey_targets}b) highlights the division between CM and DM. Two stars lie significantly off the zero-age main sequence (ZAMS): the O-star HD~37742 (upper left in Fig.~\ref{fig:survey_targets}a) and the Herbig Be star HD~200775.

\section{Observations} \label{Observations}

\subsection{GMRT} \label{GMRT}

In this section, we describe the setup and reduction of the GMRT survey carried out during GMRT observation cycles 27 and 28 (Table \ref{tab:gmrtflux}). Each star was observed in two frequency bands: the 610 MHz band (effective range 592.4-623.0 MHz) and the 1390 MHz band (effective range 1372.4-1403 MHz). 1390 MHz (L band) observations were carried out between 20 October 2014 and 27 February 2015 under the GMRT project 27\textunderscore 046 (PI: P. Chandra). 610 MHz observations were taken between 24 July 2015 and 17 February 2016 under project 28\textunderscore 075 (PI: P. Chandra). Each target was observed only once in each band.

During cycle 27, all GMRT observations were taken in the 1390 MHz band, with an on-source time typically 1 to 4 hours per source. 610 MHz band observations during cycle 28 were observed with a shorter observing time ($<2$ h). The 610 MHz band was observed in polar mode, which enabled us to image `RR' and `LL' polarization separately. However, the 1390-MHz band is not capable of producing Stokes V images.  Both bands were divided into 256 channels. For each observation, a standard phase calibrator was used which was selected from the VLA calibrator list\footnote{\url{https://science.nrao.edu/facilities/vla/observing/callist}}, and located within 15$^{\circ}$ of the respective target. Each scan of the target was followed by the corresponding phase calibrators for $\sim$5 minutes. For gain and delay calibration, three standard calibrators, i.e., 3C147 (0542+498), 3C48 (J0137+331), and 3C286 (J1331+305) were used according to source visibility. These were typically observed at the beginning and towards the end for 10--15 minutes. Observations with very poor data quality were excluded from further analysis.

All data were analyzed using version 5.7 of the Common Astronomy Software Applications (CASA\footnote{\url{https://casa.nrao.edu/}}) package \citep{McMullin2007}. We used the calibration process described by \cite{Biswas2023}. The calibrated data for the target were averaged over two frequency channels. The approximate usable bandwidth is $\sim$30 MHz for both bands. Imaging of the calibrated target fields was performed by the CASA task `{\it tclean}' along with the `{\it mtmfs}' deconvolver (Multiscale Multi-frequency with W-projection, \citealt{Rau2011}). For each data set, several rounds of phase and amplitude self-calibration were done to improve the image quality. Finally, for the detected targets, a search for polarization was performed by producing Stokes V images from 610 MHz band data. A Stokes I sample image of two example targets is shown in Fig. \ref{fig:gmrt_img}. The final flux densities were measured using the `\textit{imfit}' task of CASA.

For the 16 supplementary sample targets, flux densities were obtained from the literature. Several of the `supplementary' targets were re-observed with the upgraded GMRT, and gives nearly identical level of base radio emission. In uGMRT, the Band 4 (550--900 MHz) and Band 5 (1050--1450 MHz) are the frequency bands approximately equivalent to the 610 MHz and 1390 MHz bands of GMRT, respectively. However, because of the broader bandwidths, the uGMRT observations are much more sensitive. We anticipate that a higher sensitivity of the uGMRT may result in additional detection bias. An example of this case is the detection of HD 55522, which was detected with uGMRT by \cite{Keszthelyi2024}, but was not detected in GMRT observations. However, since we are considering mainly GMRT observations, this bias is minimized.

\begin{figure*}
\centering
\gridline{\fig{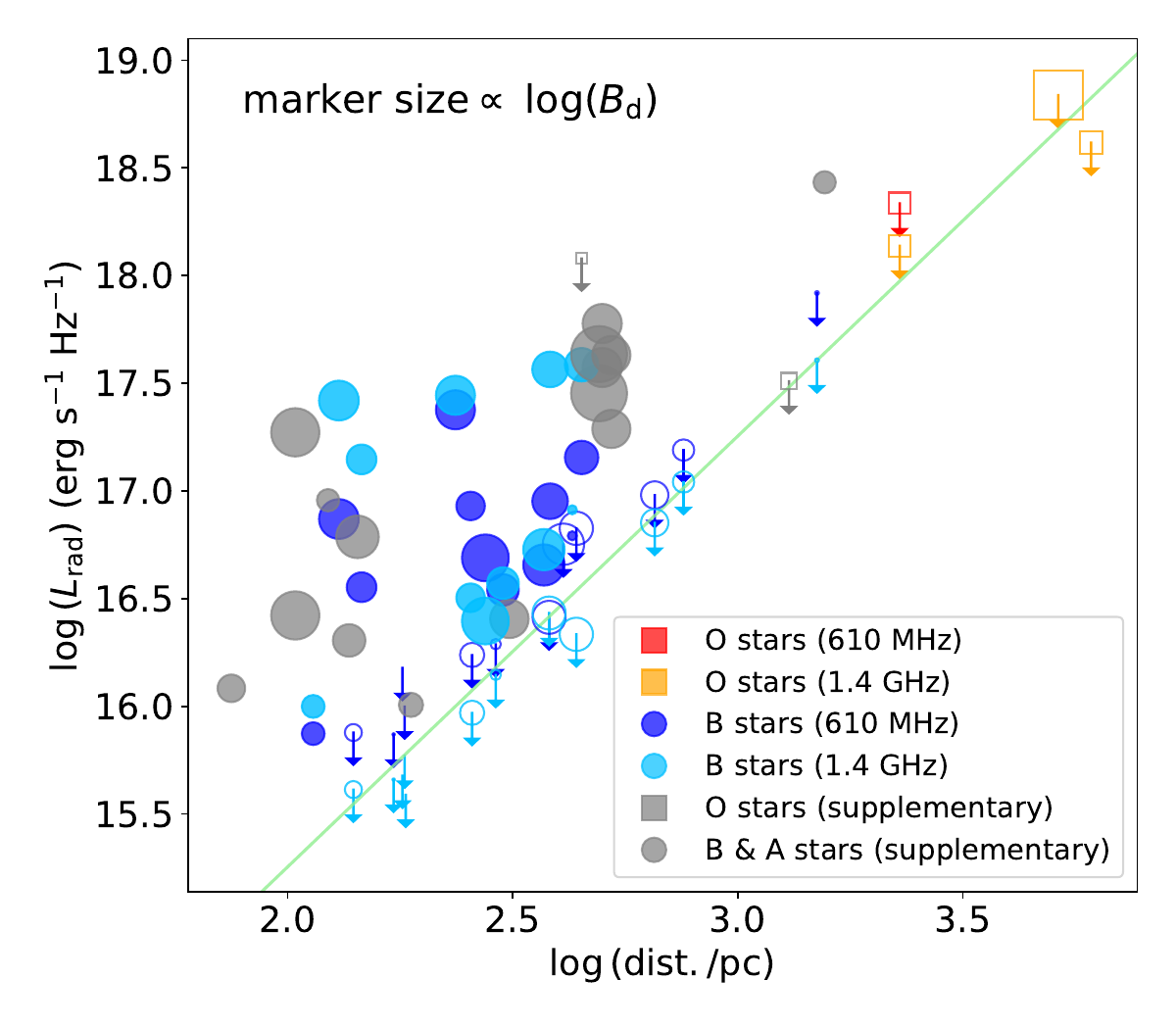}{0.495\linewidth}{(a)} \label{fig:det1}
          \fig{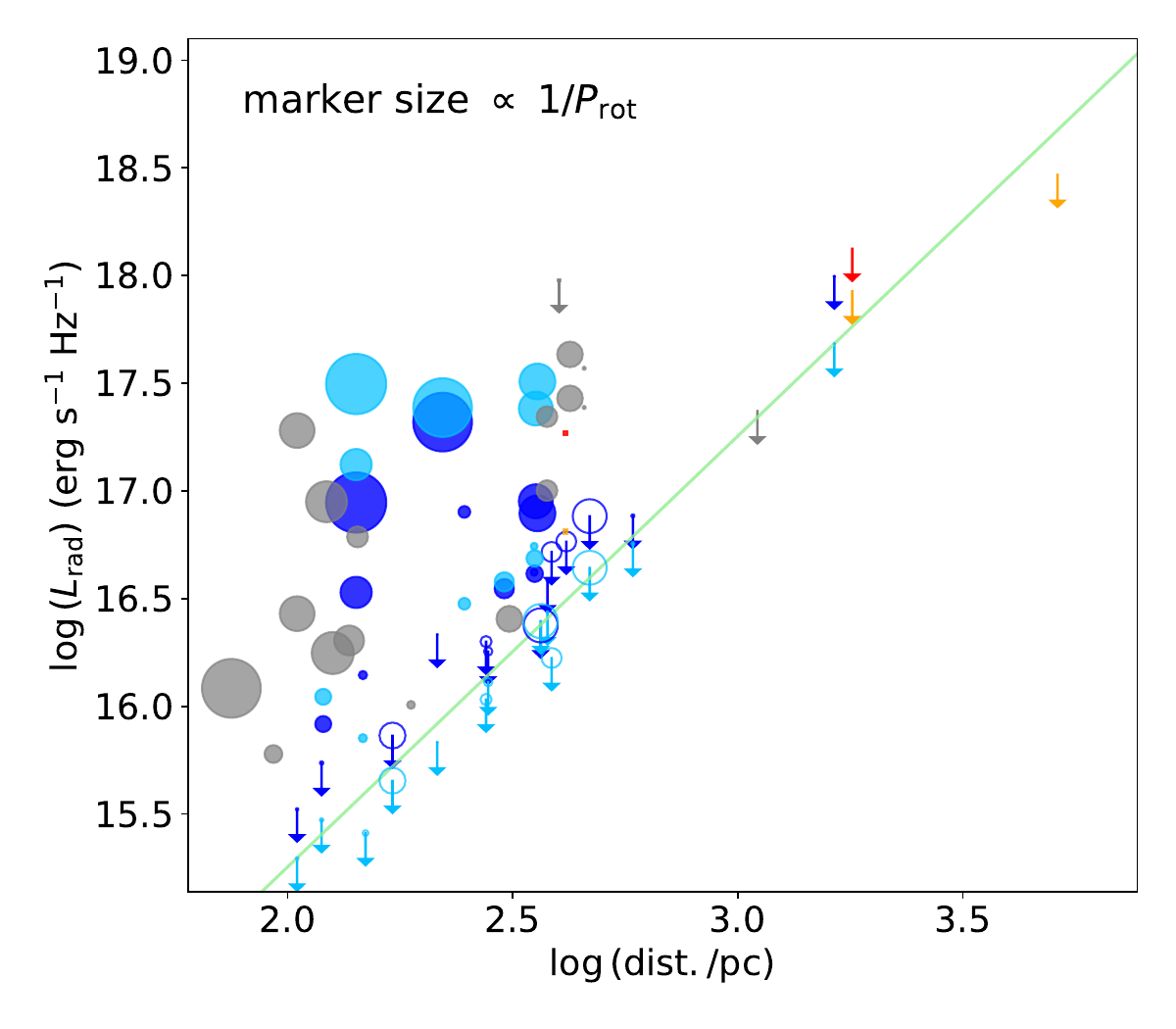}{0.495\linewidth}{(b)} \label{fig:det2}}
    \caption{Radio luminosity vs distance of the stars: (a) marker size proportional to dipolar magnetic field strength ($B_{\rm d}$), and (b) marker size proportional to rotational frequency ($f_{\rm rot}$). The filled squares and circles denote detections of O and B stars respectively, while empty markers denote non-detections. The deep blue and light blue circles represent data from 610 MHz, and 1390 MHz observations of B stars from this project, respectively. Similarly, red and orange squares represent O stars observed at 610 MHz, and 1390 MHz, respectively. The gray squares and circles represent additional O and B (or A) stars observed with GMRT/uGMRT, which are taken from \cite{Chandra2015}, and \cite{Das2022c}. The solid green lines in both figures indicate the $3-\sigma$ detection limit assuming an image noise level of $50 \ \mu$Jy.}
    \label{fig:det_bias}
\end{figure*}

\subsection{TESS} \label{TESS}

Some of the survey targets from this study do not have complete rotational ephemeris. For that reason, we investigated the TESS (Transiting Exoplanet Survey Satellite, \citealt{Ricker2014}) light curves for selected targets. TESS observes a single sector for 27 days in the wavelength range 600--1050 nm. Processed 2-minute cadence light curves were obtained from the MAST archive\footnote{MAST: \url{https://archive.stsci.edu/}}. We investigated the Lomb-Scargle periodogram for the targets of interest and determined the ephemeris. We show the new period estimates for such stars in Table \ref{tab:properties}. Further details on TESS light curves and period estimations are given in Appendix \ref{App:TESS}.

\section{Results} \label{Results}

\subsection{Detection Rate and Bias}

The flux densities and rotational phases of all observations are listed in Table~\ref{tab:gmrtflux}. Of the total of 50 individual observations obtained under projects 27\_048 and 28\_075 (core sample), 3 were rejected due to poor data quality, leaving 47 usable observations.  
Of these, 14 observations correspond to 7 targets were first reported by \cite{Shultz2022}; we re-analyze these data here for consistency. It should be noted that these 16 observations were finally not used by \cite{Shultz2022} during the determination of radio scaling laws.

Across the 28 stars in the core sample, we detect radio emission from 11 objects and obtain non-detections for 17, corresponding to a detection rate of $\sim$40\% for the projects 27\_048 and 28\_075 alone.  We summarize the result as following:

\begin{itemize}
    \item \emph{Detections from projects 27\_048 and 28\_075:} 11 detections; 3 of previously reported in \cite{Shultz2022}.
    \item \emph{Unique sub-GHz detections:} Of 11 detections, 5 stars, i.e. HD~37742, HD~200775, HD~142184, HD~189775, and HD~61556, are detected for the first time at frequencies below 1\,GHz.
    \item \emph{Non-detections from projects 27\_048 and 28\_075:} 17 non-detections, with the sole exception of HD 36526, 16 are unique, i.e. targets with the first ever low-frequency radio observations. Of the 17 non-detections from this project, 4 non-detections were already reported by \cite{Shultz2022}.
    \item \emph{Supplementary detections:} The supplementary sample adds 16 detected stars with compatible archival GMRT/uGMRT measurements from the literature, treated in the same framework for statistical analysis. 
\end{itemize}

In the combined data set of 44 stars (core + supplementary), detections remain strongly associated with certain stellar and magnetospheric properties, as described in the following sections. Figure~\ref{fig:hist_plot} shows the fraction of detections as a function of key stellar parameters.  We did not detect stars with $T_{\rm eff} > 25$\,kK, except for the O-type star HD~37742.  Detections occur predominantly among stars with strong dipolar magnetic fields and/or short rotation periods, with HD~37742 again being the lone exception. A similar pattern is seen in the $R_{\rm A}/R_{\rm K}$ diagram (Fig.~\ref{fig:survey_targets}b), where detected stars overwhelmingly have $R_{\rm A}/R_{\rm K} > 1$ (centrifugal magnetospheres; \citealt{Petit2013}), again excepting HD~37742. Further discussion of this classification is provided in Sect.~\ref{sec:emission}.

\begin{figure*}
     \centering
     \begin{subfigure}[b]{0.45\textwidth}
         \centering
         \includegraphics[width=\textwidth]{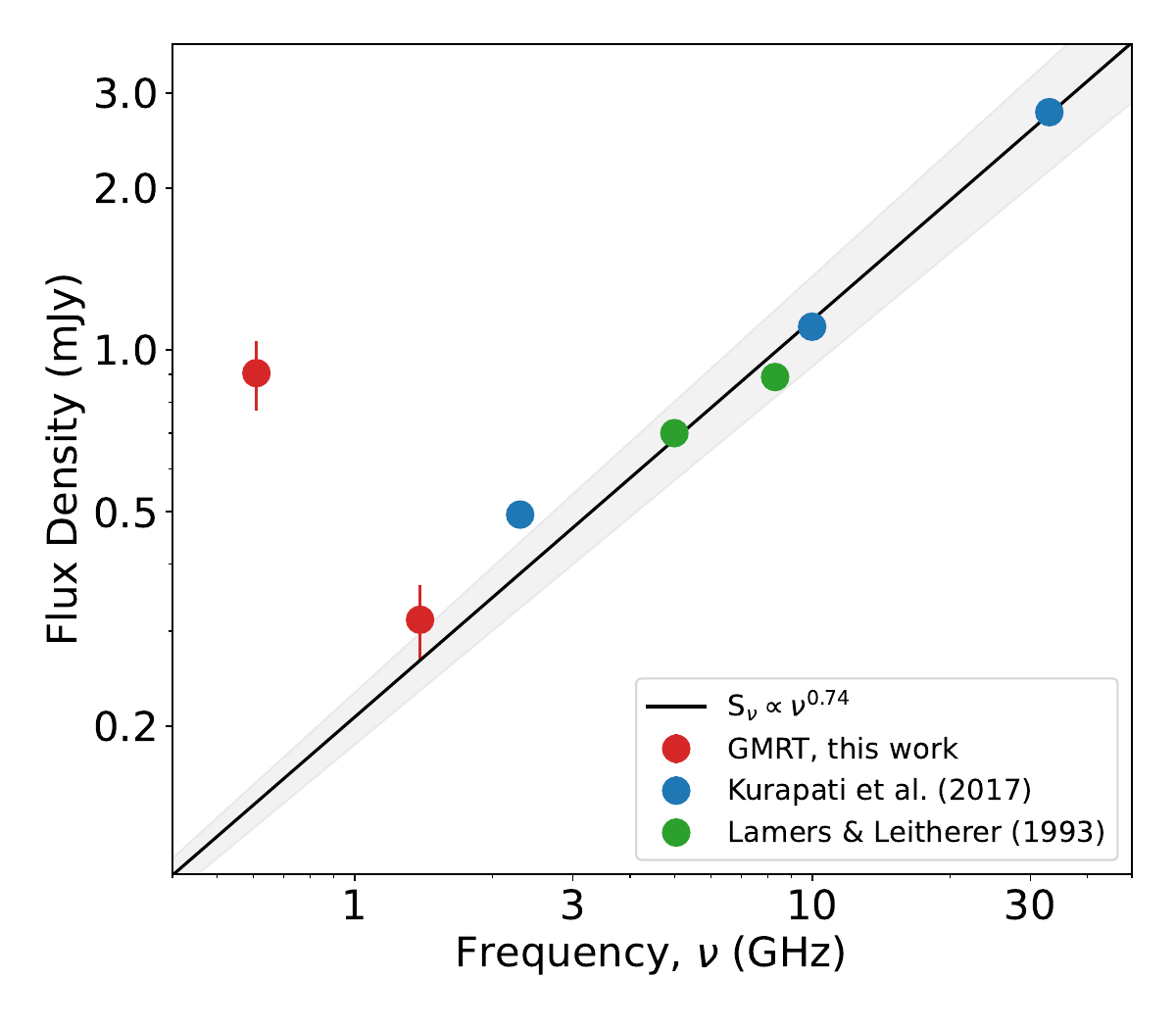}
         \caption{HD 37742}
         \label{fig:37742}
     \end{subfigure}
     \begin{subfigure}[b]{0.45\textwidth}
         \centering
         \includegraphics[width=\textwidth]{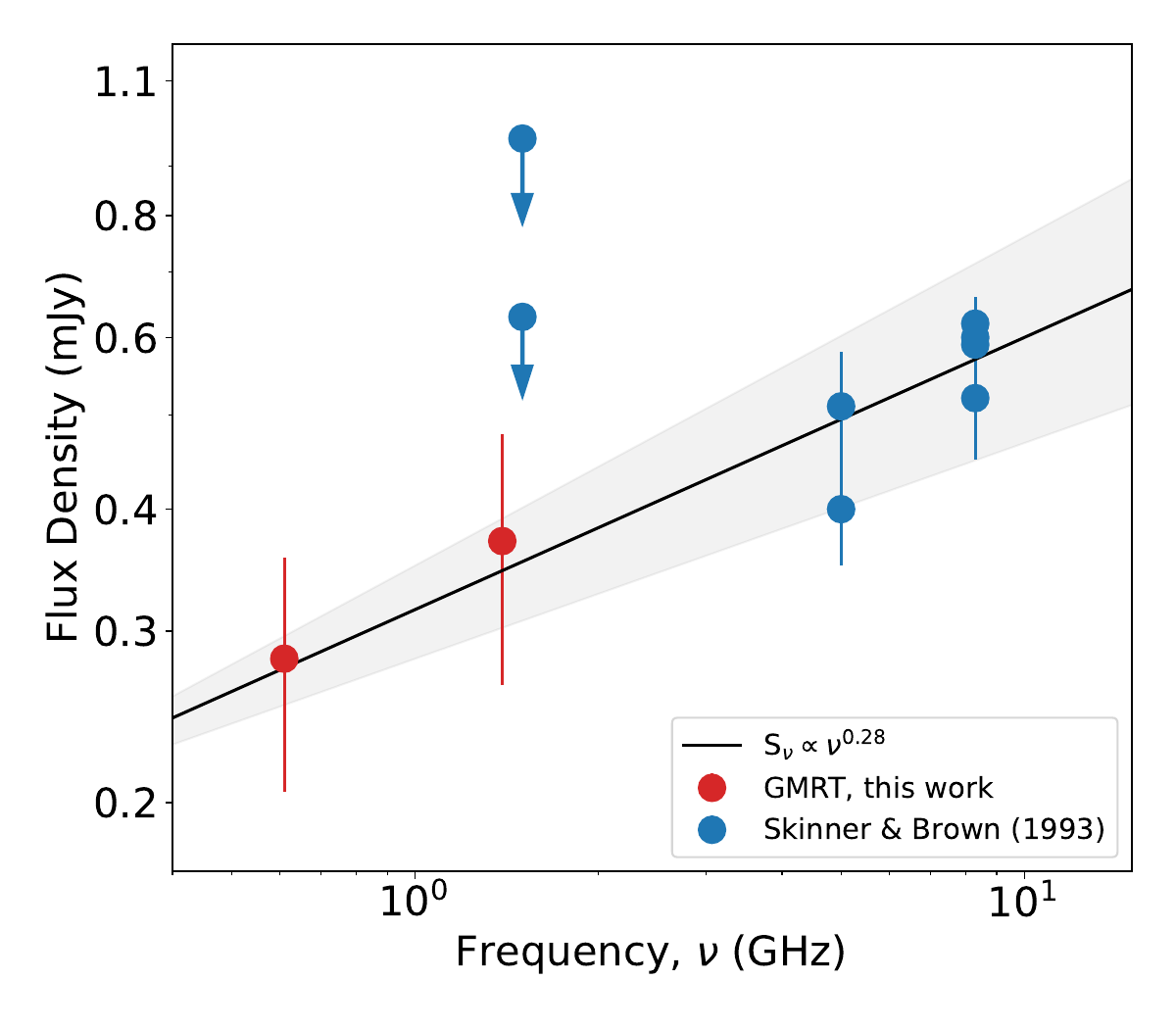}
         \caption{HD 200775}
         \label{fig:200775}
     \end{subfigure}
     \begin{subfigure}[b]{0.45\textwidth}
         \centering
         \includegraphics[width=\textwidth]{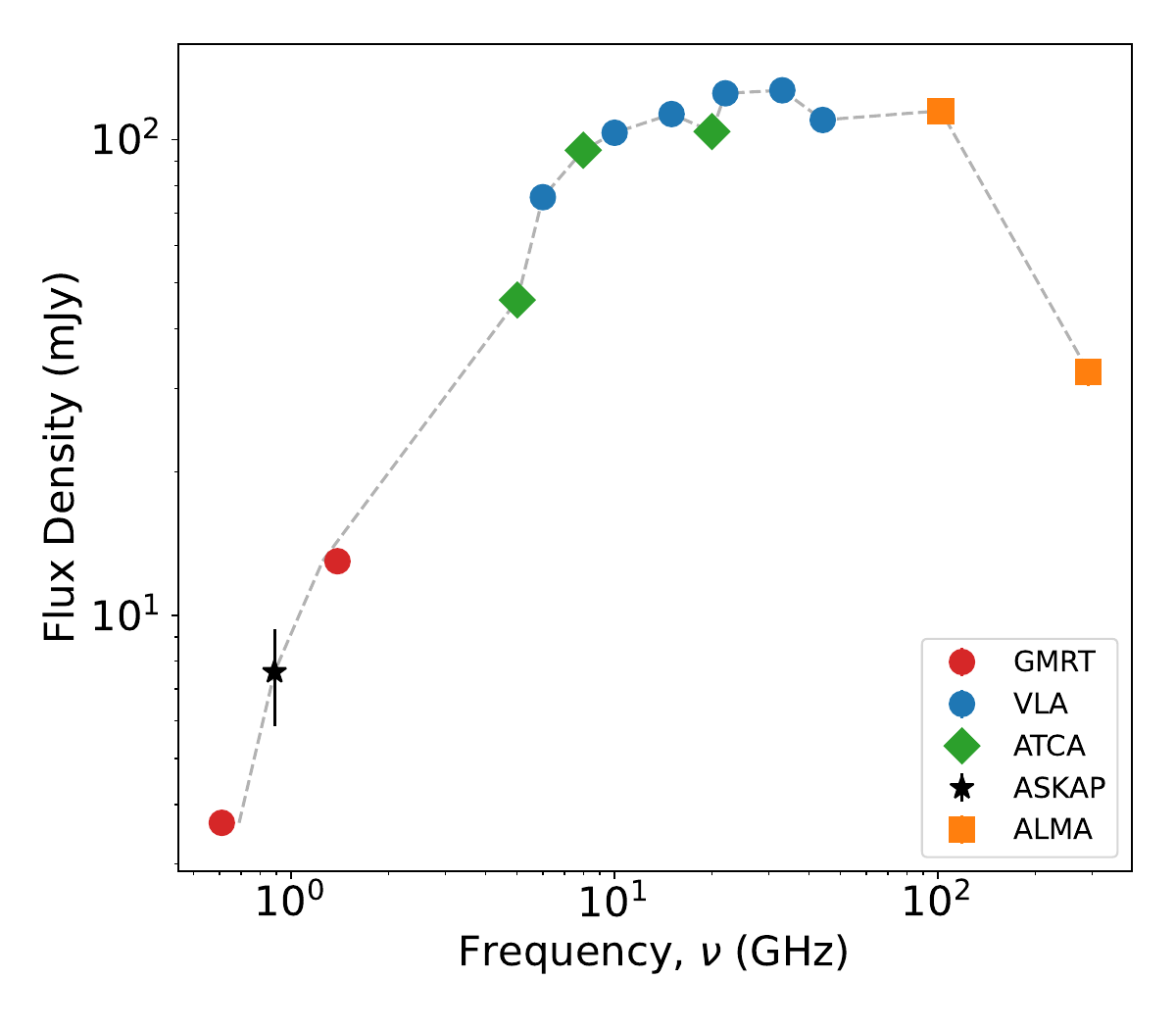}
         \caption{HD 142184}
         \label{fig:spectra}
     \end{subfigure}
     \begin{subfigure}[b]{0.45\textwidth}
         \centering
         \includegraphics[width=\textwidth]{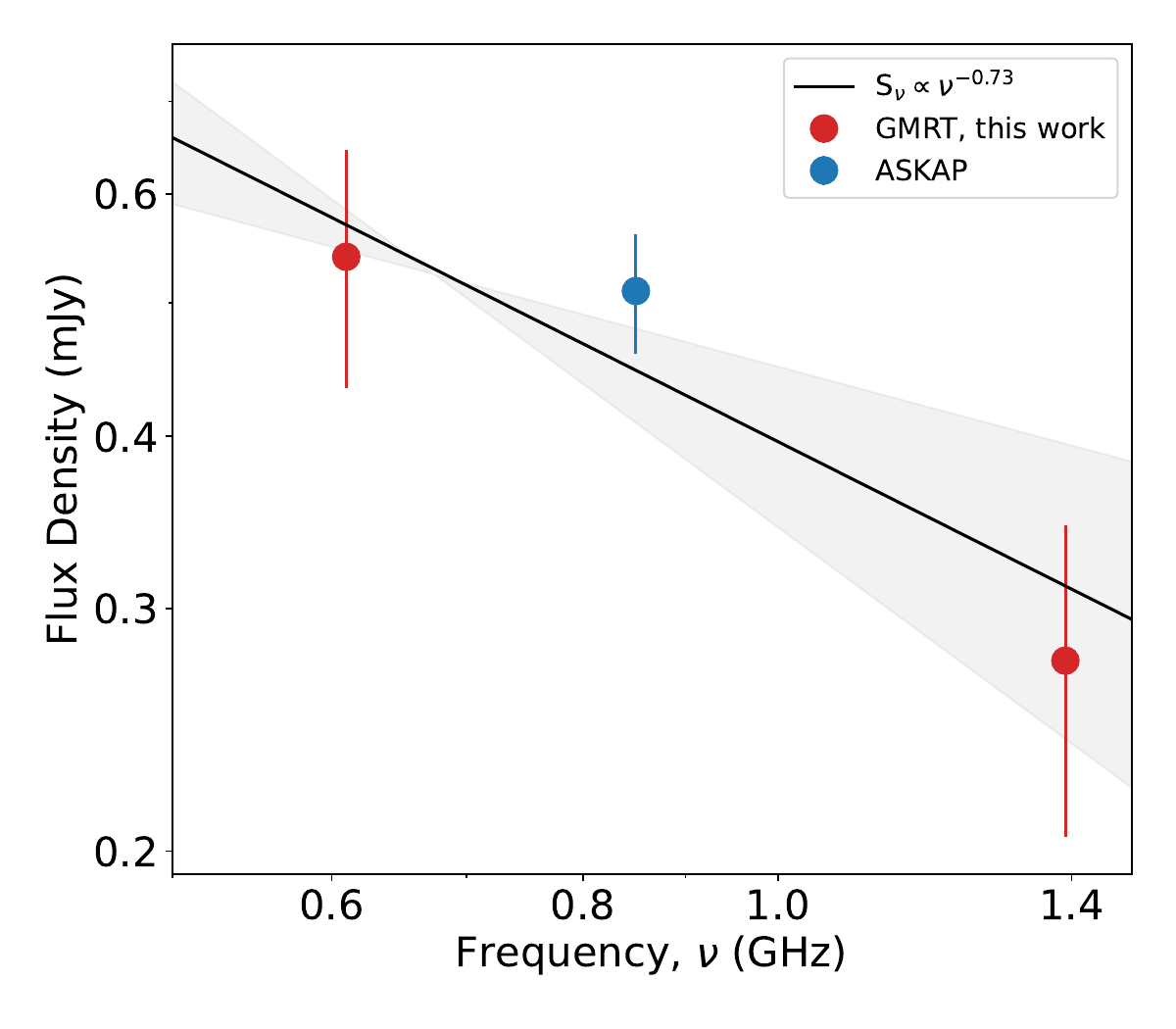}
         \caption{HD 175362}
         \label{fig:175362}
     \end{subfigure}
        \caption{(a) Radio spectra of HD 37742. The 13 cm, 3 cm, and 1 cm observations from \cite{Kurapati2017} are shown in blue, while the 6 cm, and 3.6 cm observations from \cite{Lamers1993} are shown in green.  The fit is performed by excluding the 610 MHz observation. The fitted spectral index is $0.74 \pm 0.04$. (b) Radio spectra of HD 200775. The 3.6 cm, 6 cm, and 20 cm observations taken from \cite{Skinner1993} are shown in blue, while the GMRT flux densities are shown in red. The fit is performed by excluding the non-detections at 20 cm. The fitted spectral index is $0.28 \pm 0.06$. (c) Radio spectra of HD 142184 combining data from different telescopes. GMRT observations from this work is represented by red circle, and the sub-GHz archival ASKAP observations is shown in black pentagon. See text for details on other data points. (d) Similar to the top figures, but for HD 175362. The spectral index is negative in this case. \label{fig:det_spectra}}
\end{figure*}

Figure~\ref{fig:det_bias} illustrates the observed radio luminosity as a function of distance. The solid green line marks the $3\sigma$ detection threshold assuming a representative image rms of $50\ \mu$Jy. Radio luminosities span roughly three orders of magnitude. Rotational modulation due to an oblique dipole is ignored here, as its effect is small compared to the luminosity range observed. In Fig.~\ref{fig:det_bias}a, the marker size is proportional to $\log B_{\rm d}$; in Fig.~\ref{fig:det_bias}b, the marker size is proportional to the rotational frequency. Both panels show that detection is generally favored for stars with stronger magnetic fields or shorter periods, unless they lie near the distance-limited sensitivity threshold of the survey. In contrast, long-period stars cluster near this threshold and are rarely detected.

\subsection{Detected Targets I: First Detection at Sub-GHz Frequencies}

In this subsection, we discuss the targets for which the observations from this study (i.e. GMRT cycle 27 and 28) serve as the first sub-GHz detections.

\subsubsection{HD 37742}

HD 37742 ($\zeta$ Orionis A) is an O9.5I supergiant located in the constellation Orion. It is a binary star with a B0III companion ($\zeta$ Ori B), having an orbital period of 2687 days and eccentricity $e \sim 0.34$ \citep{Hummel2013}. \cite{Bouret2008} observed a weak magnetic field of $60 \pm 10$ G in the primary component, which was later confirmed by \cite{Blazere2015}. Reanalysis of spectropolarimetric data by \cite{Blazere2015} revealed that the dipolar field strength for the star is $\sim$140 G, which was the weakest magnetic field detected on a massive star at that time. No magnetic field was detected on its companion, with an upper limit of 300 G. \cite{Bouret2008} determined the effective temperature to be $T_{\rm eff} = 29500 \pm 1000$ K, with mass $M_* = 40 M_{\odot}$, $\log g = 3.25 \pm 0.10$, and radius  $R_* = 25 R_{\odot}$. \cite{Cohen2006} found that X-ray emission from this star is consistent with the non-magnetic embedded wind shock (EWS) scenario. As the companion star is two orders of magnitude fainter than the primary star, and also in a wide orbit (separation $\sim 400 \ R_*$, \citealt{Rivinius2011}), it is unlikely to observe a colliding wind shock (CWS) in this system \citep{Cohen2014}. 

This system was observed by \cite{Lamers1993} at 6 cm and 3.6 cm. It was also observed in the 3 GHz, 10 GHz, and 33 GHz bands with the Very Large Array (VLA), by \cite{Kurapati2017}. \cite{Kurapati2017} determined a positive in-band spectral index for this target and concluded that the emission was free-free thermal emission. They also found the mass-loss rate obtained by assuming thermal emission to be consistent with the theoretical mass-loss rate. 

We detected this star at both the 610 MHz and 1390 MHz bands, making this the first detection at sub-GHz frequencies of a magnetic O-type star. The radio spectrum of this star is shown in Fig. \ref{fig:det_spectra}a. The 1.4 GHz flux is consistent with the higher-frequency observations, giving a well-constrained power-law index of $0.74 \pm 0.04$, suggesting thermal emission up to 1.4 GHz. However, we notice a significant increase in flux at 610 MHz. This results in a spectral index of $-0.9 \pm 0.2$ between 610 MHz and 1390 MHz. If we consider the flux densities from 1 to 30 GHz, the fitted spectral index is 0.74, typical for a thermal free-free emission. This drastic change in the spectral index at low frequencies is an indication of the presence of nonthermal emission. We do not observe any polarization from this source. Because of the low signal-to-noise ratio, we were unable to obtain a reliable estimate for the circular polarization fraction, and thus the exact emission mechanism could not be estimated with the existing observations. Long term monitoring of this star is necessary to properly disentangle different emission components. In case the emission is dominated by binary interaction at sub-GHz frequencies, the non-thermal emission is expected to vary on the orbital timescale.

\subsubsection{HD 200775}

HD 200775 is a Herbig Be star that illuminates the nebula NGC 7023 \citep{Manoj2006}. This target is a binary system, with two components of similar temperature ($T_{\rm eff} \sim 18.6$ kK), luminosity, radius, and mass ($M_1/M_2 \sim 1.2$; $M_1 = 10.7 M_{\odot}$) in a long period orbit \citep{Alecian2008}. \cite{Bisyarina2015} calculated an orbital period of $1361.3 \pm 2.2$ days (nearly 3.7 years). \cite{Alecian2008} observed a magnetic field from the narrow-line primary component, with a dipolar field strength of $1000 \pm 150$ kG. The authors also determined a rotation period of $\approx 4.328$ days from the variation of the longitudinal magnetic field from this star. We recovered a nearly identical period from the TESS data (see Appendix \ref{TESS}). We observed that the variation of the light curve is non-sinusoidal. \cite{Naze2014} observed over-luminous X-ray emission from this star with $\log (L_X/L_{\rm bol}) > -6$. As the orbit of the system is fairly large, binary interaction between the magnetic and non-magnetic stars will be minimal, and thus we treat this target as a single magnetic star.

\cite{Skinner1990, Skinner1993} observed this star with the VLA and found a positive spectral index. This star was detected in both GMRT bands. The radio spectrum for this target is shown in Fig. \ref{fig:det_spectra}b. We found a consistent spectral index up to the 610 MHz band (spectral index, $\alpha = 0.28 \pm 0.06$). The small positive spectral index hints at the presence of nonthermal emission, which is further discussed in Sec. \ref{sec:emission}.

\subsubsection{HD 142184}

HD 142184 (HR 5907) is a main-sequence B2.5V star with mass 5.5~$M_{\odot}$ and temperature $T_{\rm eff} \sim 17.5$~kK \citep{Grunhut2012_hr5907}. \cite{Grunhut2012_hr5907} discovered a strong magnetic field with a dipolar field strength $>10$ kG. From the modeling of the longitudinal field variation, \cite{Grunhut2012_hr5907} derived a dipolar field strength of $\sim 15.7$ kG, while from the modeling of the Stokes V signatures, they inferred a dipolar field strength of $\sim 10.4$ kG. The origin of this discrepancy is attributed to contributions from higher-order multipolar components. HD 142184 shows significant spectropolarimetric and photometric variability with a period of $\sim 0.508$ days \citep{Grunhut2012_hr5907}, making it one of the most rapidly rotating magnetic hot stars known to date. \cite{Naze2014} observed hard, over-luminous X-ray emission from this star compared to other B-type magnetic stars.

This star is well studied at higher radio frequencies, where it is found to be detectable at centimeter to millimeter bands. \cite{Leto2018} conducted a wide-band study of this star with VLA and the Atacama Large Millimeter Array (ALMA). It was also observed with the Australian Telescope Compact Array (ATCA) by \cite{Murphy2010} as part of the AT20G survey. The radio spectrum for this star is shown in Fig. \ref{fig:spectra}. From VLA observations, \cite{Leto2018} observed several peculiar trends having high brightness temperature that cannot be explained with gyrosynchrotron emission from a simple dipolar field, and suspected the effect of multipolar magnetic field components on the radio emission. We also detected the star at 0.89 GHz in archival ASKAP (Australian SKA Pathfinder) observations showing a remarkable $+$22\% circular polarization \citep{Pritchard2021}.

HD 142184 was detected in both bands. The flux densities are listed in Table \ref{tab:gmrtflux}, showing the brightest radio flux among the survey candidates. The target also shows significant circular polarization in the 610 MHz band with a Stokes V flux density of 0.685 $\pm$ 0.096 mJy. The circular polarization fraction for this observation is thus 18.7\%, similar to the observed fraction with ASKAP. The lower limit of the brightness temperature considering the area of emission to be the same radius as that of the star is found to be $3.43 \times 10^{11}$ and $3.72 \times 10^{11}$ K, respectively, for the 610 MHz and 1390 MHz bands. This target was recently observed with uGMRT and reported in \cite{Biswas2025}, where similar flux densities and circular polarization were found at all rotation phases, showing the presence of non-thermal auroral emission.

\begin{figure*}
    \centering
    \includegraphics[width=\linewidth]{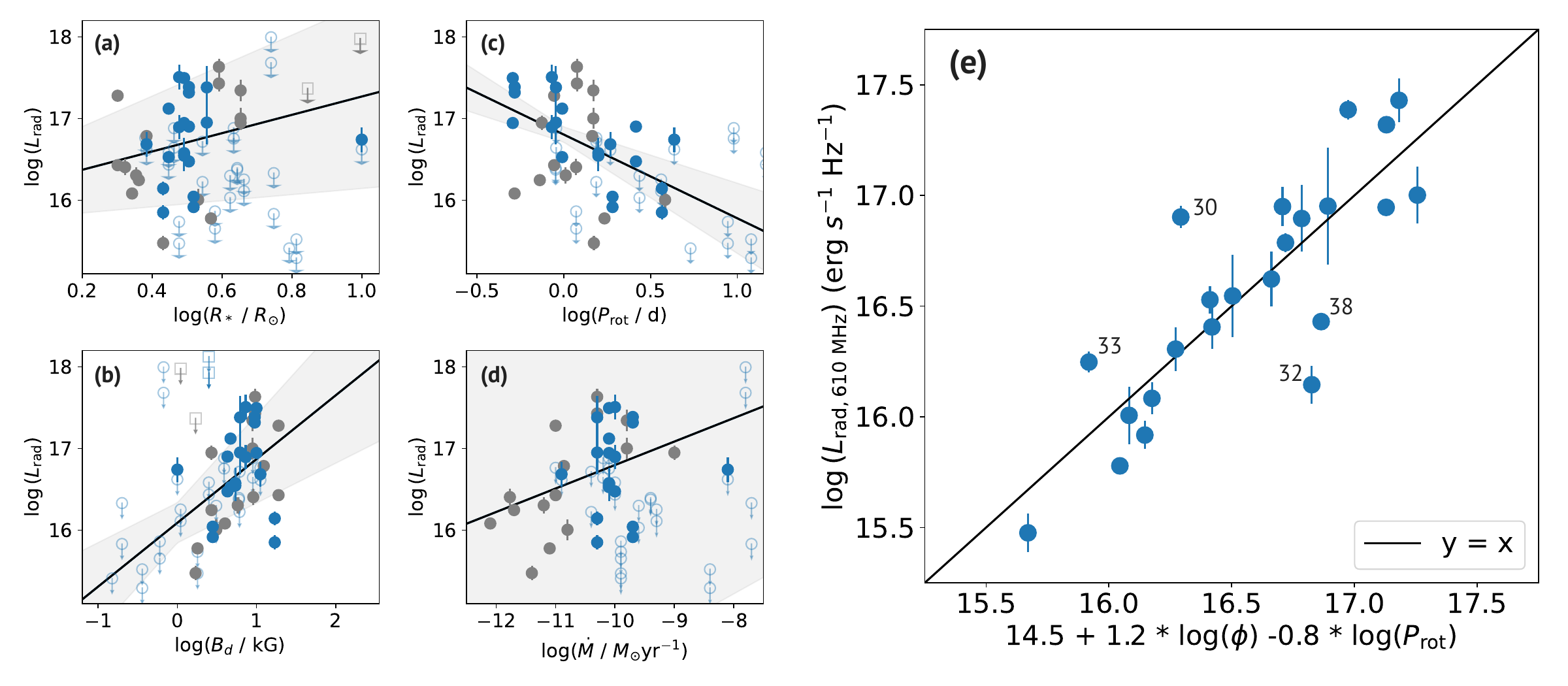}
    \caption{(a-d) Variation of radio luminosity with selected stellar parameters along with their best fits. The blue filled circles represent detected targets from this study, while the gray filled circles represent detected targets from other GMRT studies. The open circles/squares represent the non-detected B and O-type stars, respectively. The black lines represent the best-fit, and the shaded regions represent the uncertainty in the fits. HD 37742 was excluded from regression analysis, and thus not shown in these plots. (e) Two-variable regression of magnetic flux ($\Phi$) and rotation period versus radio luminosity at 610 MHz. Some of the potential outliers are labeled according to their corresponding IDs.  \label{fig:regress} }
\end{figure*}

\subsubsection{HD 189775}

HD 189775 is a magnetic B5III star ($B_p \sim 4.3$~kG). \cite{Lyubimkov2002} determined the photometric temperature to be $T_{\rm eff} = 16.2 \pm 0.6$ K, while \cite{Shultz2019b} determined a temperature of $T_{\rm eff} = 17.5 \pm 0.6$ K from spectroscopic measurements. \cite{Shultz2018b} reported magnetic field measurements for this target, and noticed asymmetries in the longitudinal magnetic field curve, which are indications of departure from a purely dipolar surface magnetic field geometry. No X-ray observations have been reported in the literature, and we also did not find a definite detection from the archival XMM-Newton observation of this target.

We detected this star at both frequencies, with a higher flux density at lower frequency bands. The 610 MHz flux density is $\sim 1.1$ mJy/beam, while the 1390 MHz flux density is $\sim 0.41$ mJy/beam. This immediately implies a negative spectral index $\alpha = -1.2 \pm 0.3$. The negative spectral index hints toward the presence of nonthermal emission. However, because the observations were performed at different rotational phases, the actual spectral index may be different. The observing phases for both observations (0.36 and 0.80) are close to the magnetic null of this star (see \citealt{Shultz2018b}), making coherent emission via ECME highly probable \citep{Das2022}.

This star is the only target in this sample for which these GMRT observations act as the first evidence of radio emission. Also, no evidence of magnetosphere was previously found in this star from optical or UV measurements, making these radio observations the first tentative evidence of a magnetosphere in this star.

\subsubsection{HD 175362}

HD 175362 is a B-type magnetic chemically peculiar (He-weak) star \citep{Renson2009}. The star is a moderately fast rotator with a rotation period of $\sim 3.67$ days and a dipolar magnetic field strength of $\sim 17$ kG \citep{Shultz2018b, Shultz2019}. This star was observed by \cite{Linsky1992} in the 1.5 GHz, 5 GHz, and 8.3 GHz bands with the VLA. This star was not detected at 1.5 GHz or 8.3 GHz. However, it was detected at 5 GHz in two distinct epochs, with an average flux density of $0.32 \pm 0.05$ mJy. \citealt{Polisensky2023} detected a possible radio flare from this star from the VLA Commensal Sky Survey, at the $\sim 350$ MHz band.

We detected this target in both GMRT bands. Between the 5 GHz band of VLA and the 1390 MHz band of GMRT, we get a flat spectral index ($\alpha = 0.1 \pm 0.5$). However, the inter-band spectral index from the GMRT bands shows a negative turnover, with a spectral index $\alpha = -0.8 \pm 0.4$. The negative spectral index at low frequencies suggests a possible presence of coherent emission.

\subsection{Detected Targets II: Stars with existing Sub-GHz observations}

We detected six additional targets from this survey (HD 182180, HD 35298, HD 142990, HD 35502, HD 176582, and HD 61556). Although these observations were taken during 2014-2015, new observations of these targets have been reported in the literature.   Coherent emission was detected from all of these stars except HD 35502 \citep{Das2022, Das2022c}. Although the uGMRT observations of HD 35502 were consistent with the existence of a pulse, the coherent nature was not conclusive \citep{Das2022}. Among these 5 stars with detected ECME emission, only HD 176582 was observed to have a negative inter-band spectral index. For HD 61556, we report a smaller value for the basal flux, compared to the uGMRT observations by \cite{Das2022} which may arise due to the slight difference in central frequencies of these two observations. It is worth mentioning that HD 36526, which is not detected in our observations, was later detected through uGMRT observations by \cite{Das2022}, and is the only non-unique non-detection from this study. However, for this star, the basal flux density is consistent with the upper limit of the flux density from our study \citep{Das2022}.

\subsection{Relation to Stellar Parameters}

In order to quantitatively identify the dependence of radio luminosity on various stellar parameters, we performed a regression analysis using one, two, and three-variate regression. We investigated the correlation with the following parameters: effective temperature $(T_{\rm eff} \ / \ {\rm kK})$, stellar radius $(R_* \ / \ R_{\odot}$), stellar mass $(M_* \ / \ M_{\odot}$), rotation period $(P_{\rm rot} \ / \ {\rm d}$), dipolar field strength $(B_{\rm d} \ / \ {\rm kG}$), magnetic field strength at the Kepler radius $(B_{\rm K} \ / \ {\rm kG}$), and mass-loss rate $(\dot{M} \ / \ M_{\odot} {\rm yr}^{-1}$). We also investigated the magnetic flux $(\Phi \ / \ {\rm kG} \ R_{\odot}^2)$ parameter following \cite{Leto2021}. For all analyses, HD 37742 (ID: 6) was excluded, as the radio emission behavior of this star is found to be significantly different from other stars (Sect. \ref{sec:emission}). For the final scaling relationship, two more stars: HD 189775 (ID: 30) and HD 215441 (ID: 37) were excluded as outliers where we suspect additional emission mechanisms.

\begin{deluxetable}{llllll}
\tablecaption{Regression parameters: The columns represent the variables, the correlation coefficient r; the reduced $\chi^2$; the AIC for the regression; and the best-fit slope.   \label{tab:regression}}
\tablehead{
\colhead{Parameter} & \colhead{$r$} & \colhead{$\chi^2_{\rm red}$} & \colhead{AIC} & \colhead{$P_{\rm F-stat}$} & \colhead{Slope} 
}
\startdata
\multicolumn{6}{c}{1-parameter} \\
$\log (T_{\rm eff} \ / \ {\rm kK})$ & 0.76 & 130 & 16.5 & $10^{-4}$ & $4.1\pm 0.8$ \\
$\log (R_* \ / \ R_{\odot}$) & 0.38 & 188 & 31.5 & 0.086 & $1.9 \pm 1.1$ \\
$\log (M_* \ / \ M_{\odot}$) & 0.67 & 90 & 22.5 & 0.001 & $2.4 \pm 0.6$ \\
$\log (P_{\rm rot} \ / \ {\rm d}$) & -0.38 & 111 & 31.7 & 0.09 & $-0.8 \pm 0.4$ \\
$\log (B_{\rm d} \ / \ {\rm kG}$) & 0.54 & 118 & 27.6 & 0.01 & $0.9 \pm 0.3$ \\
$\log (B_{\rm K} \ / \ {\rm kG}$) & 0.54 & 75 & 15.1 & 0.02 & $0.3 \pm 0.2$ \\
$\log  (\dot{M} \ / \ M_{\odot} {\rm yr}^{-1}$) & 0.57 & 94 & 26.7 & 0.007 & $0.4 \pm 0.1$ \\
$\log (\Phi \ / \ {\rm kG} \ R_{\odot}^2)$ &  0.78  & 65 & 16.8 & $1 \times 10^{-5}$ & $1.2 \pm 0.2$  \\
$\log (\Phi/P_{\rm rot})$ &  0.89  & \textbf{28} & \textbf{1.6} & $ 7 \times 10^{-8}$ & $1.1 \pm 0.1$  \\
\multicolumn{6}{c}{2-parameter} \\
$\log (\Phi \ / \ {\rm kG} \ R_{\odot}^2)$ & 0.84 & 44 & 12.3 & $5 \times 10^{-6}$ & $1.2 \pm 0.2$ \\
$\log (P_{\rm rot} \ / \ {\rm d}$) & & & & & $-0.8 \pm 0.3$ \\
\multicolumn{6}{c}{3-parameter} \\
$\log (B_{\rm d} \ / \ {\rm kG}$) & 0.84 & 45 & 14.7 & $4 \times 10^{-5}$ & $1.3 \pm 0.2$ \\
$\log (R_* \ / \ R_{\odot}$) & & & & & $3.0 \pm 0.7$ \\
$\log (P_{\rm rot} \ / \ {\rm d}$) & & & & & $-0.5 \pm 0.2$ \\
\enddata
\end{deluxetable}

We use the $\chi^2$ minimization technique to obtain the best fit. For each variable (or set of variables), the Pearson’s correlation coefficient $r$ was calculated to check the level of correlation. $r$ values of $-1$ and $+1$ represent the extreme correlation cases, i.e., ideal anticorrelation and ideal correlation, respectively. The value $r=0$  represents the absence of correlation. In each case, we obtained reduced $\chi^2$ values and AIC (Akaike information criterion) to estimate the quality of the linear regression fit. These values provide a relative comparison of the quality of a given model based on the fit and the degrees of freedom (or number of variables). We also estimate the probability of the null hypothesis from F-statistics ($P_{\rm F-stat}$).  A simultaneous lower value of reduced $\chi^2$, AIC, and $P_{\rm F-stat}$ will indicate a better fit despite the additional model parameters.

\begin{figure}
    \centering
    \includegraphics[width=\linewidth]{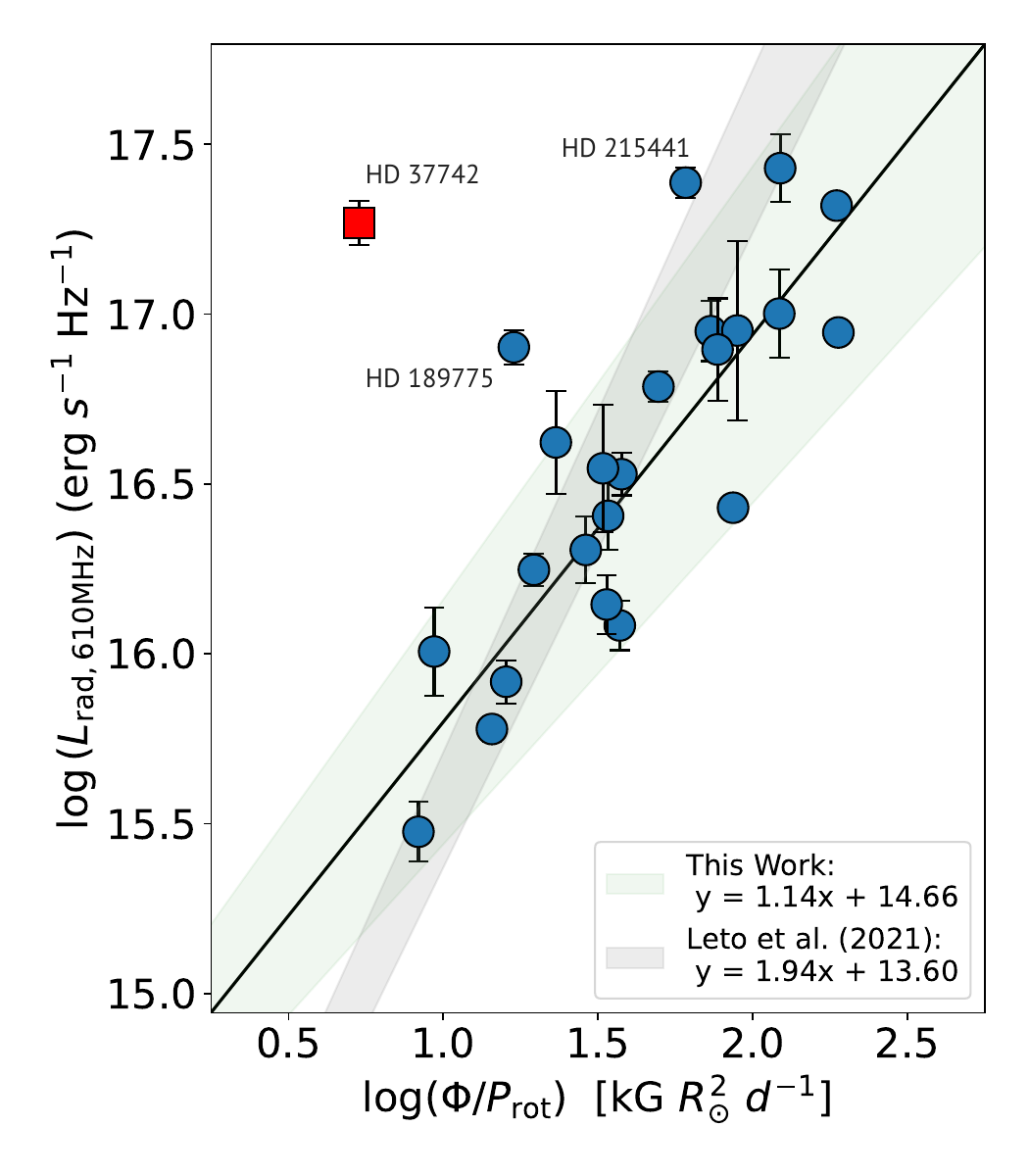}
    \caption{Final scaling relationship with radio luminosity values from 610 MHz bands, or uGMRT band 4. The three annotated stars were excluded from the fit. The green band represent the uncertainty in the fit. The gray band represents the extent of scaling relationship obtained by \cite{Leto2021} from the higher frequency survey. }
    \label{fig:scaling}
\end{figure}

\begin{figure*}
\centering
\gridline{\fig{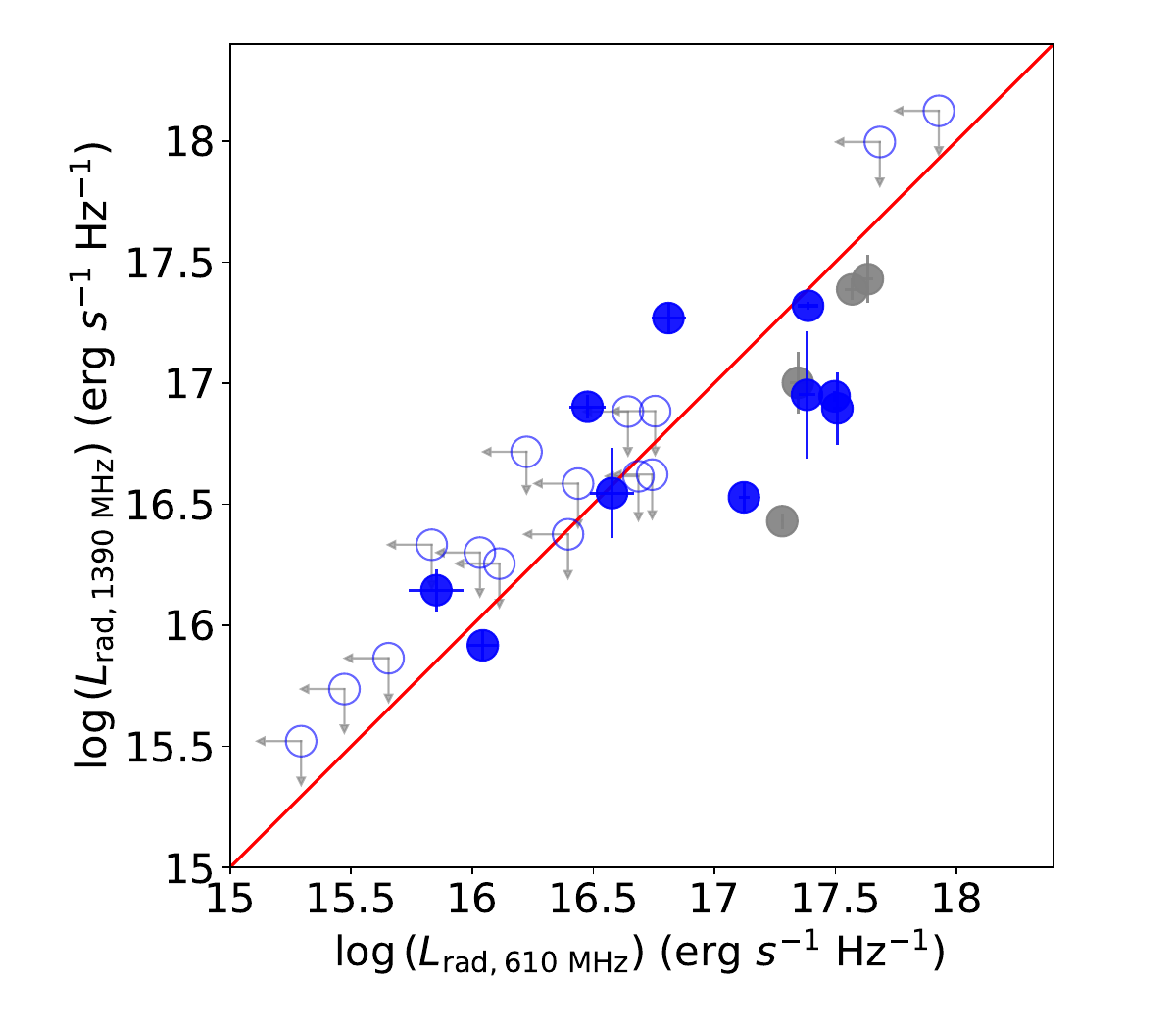}{0.49\linewidth}{(a) } \label{fig:spectra_comp}
          \fig{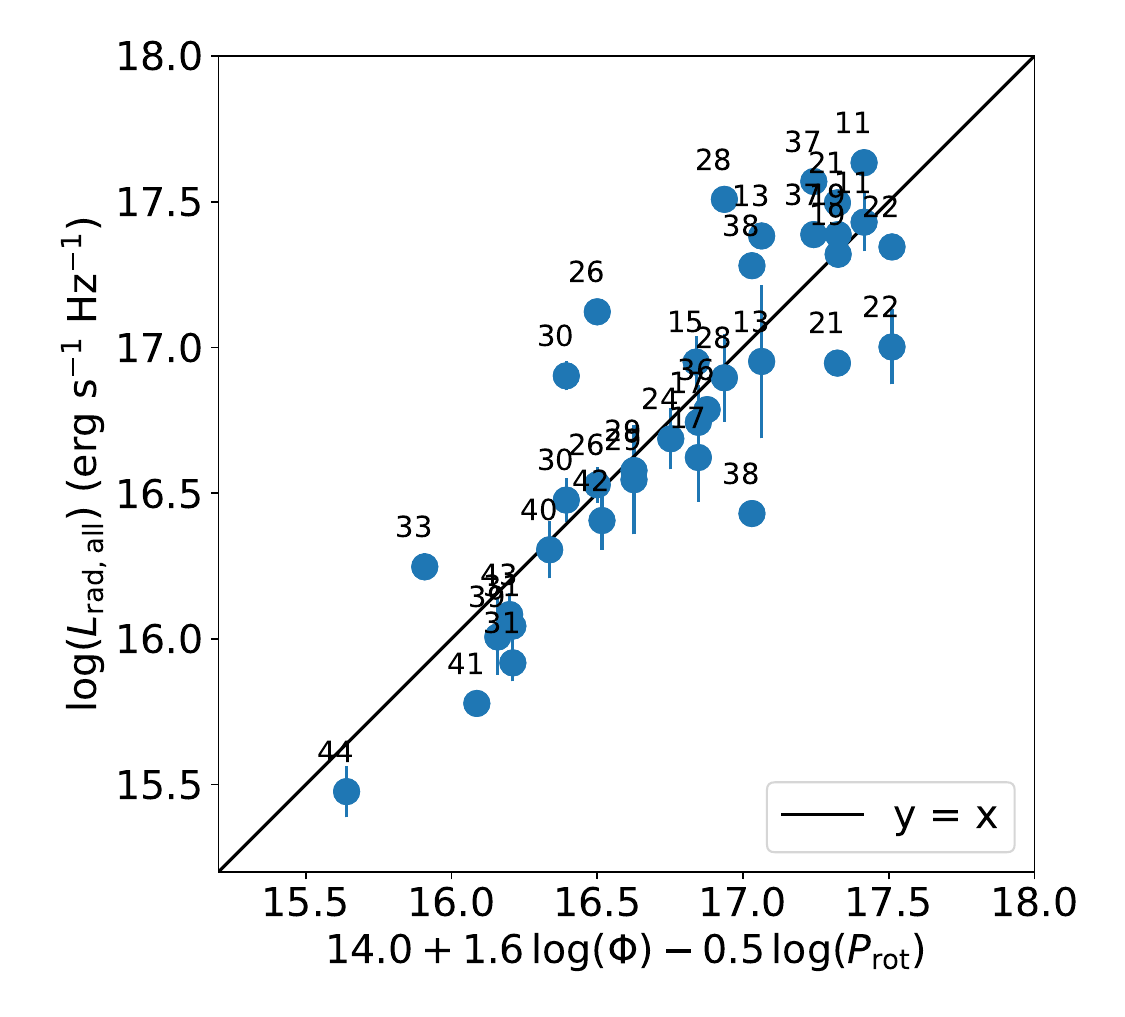}{0.485\linewidth}{(b) } \label{fig:scaling_all}}
    \caption{Inclusion of 1390 MHz bands: (a) Comparison of radio luminosities obtained from the 610 MHz and 1390 MHz bands. The red line represents the $x=y$ line. Filled symbols represent detections (at both bands), while empty circles represent non-detections (again, at both bands). Blue markers represent the core sample and the gray markers represent supplementary samples.  (b) Two-variable regression of magnetic flux ($\Phi$) and rotation period versus radio luminosity. }
    \label{fig:b45_compare}
\end{figure*} 

The results of the regression analysis are shown in Table \ref{tab:regression}. For four selected variables ($R_*$, $B_{\rm d}$, $P_{\rm rot}$, and $\dot{M}$), we show the variation of radio luminosity and the best fit to the detected targets in Fig. \ref{fig:regress}(a-d). Although the radio luminosity varies strongly with the dipolar magnetic field strength and rotation period, it does not vary significantly with the wind properties. Magnetic flux ($\Phi = B_{\rm d} R_*^2$) shows a strong correlation with luminosity. When combined with the rotation period, the two-variable regression gives the best fit, with the lowest $\chi^2$ and AIC, given that all detected stars except HD 37742 are used. The two-variable regression can be represented as $L_{\rm rad} \propto \Phi^{1.2 \pm 0.2} P_{\rm rot}^{-0.8 \pm 0.3}$. The two-variable regression of magnetic flux and rotation period versus radio luminosity is shown in Fig. \ref{fig:regress}e. The reason for the removal of HD 37742 from all regression analyses is explained in further detail in the following section.

It should be noted that for the final scaling relationship and the multiple-variable regressions, we included only the radio-loud targets observed at the 610 MHz bands of GMRT, or Band 4 of uGMRT (550--900 MHz), to maintain consistency. Apart from HD 37742, we excluded two more outliers: HD 189775 (ID:30) and HD 215441 (ID:37). The resulting single-variable regression is given by:

\begin{equation}
    L_{\rm rad} = 10^{14.7 \pm 0.2} \times (\Phi / P_{\rm rot})^{1.1 \pm 0.1} .\label{eq:scaling1}
\end{equation}

This expression results in the lowest $\chi^2$ (=28), AIC (=1.6), and $P_{\rm F-stat}$ (=$7 \times 10^{-8}$). The final scaling relationship plot is shown in Fig. \ref{fig:scaling}. The inclusion of 1390 MHz data in the regression analysis is discussed in the following section.

\subsection{Inclusion of 1390 MHz band in Regression Analysis} \label{app:b5}

We searched for any possible trend in the spectral indices that may have affected the comparison between the 610 MHz and 1390 MHz data. However, we have observations at both bands for only a small fraction of targets. Among these, we did not find any trend in the spectral indices (see Fig. \ref{fig:b45_compare}a). This result, to some extent, justifies the use of data from the two bands interchangeably while predicting radio luminosity.  The non-detections have in general higher upper limits at 1390 MHz, which is an effect of shorter observing times. However, it should be kept in mind that the spectral indices obtained from these data are only a rough estimate, as the observations at different bands correspond to different rotational phases of the stars.  

Inclusion of the 1390 MHz data in the two-parameter regression analysis increases reduced $\chi^2$ ($=328$), AIC ($= 14.2$), while also reducing $P_{\rm F-stat}$ (=$5 \times 10^{-11}$). The final luminosity is then given by: $L_{\rm rad} \propto \Phi^{1.6 \pm 0.2} P_{\rm rot}^{-0.5 \pm 0.2}$. The resulting regression is shown in Fig. \ref{fig:b45_compare}b.

\section{Discussion} \label{Discussion}

\subsection{Thermal vs. nonthermal Emission \label{sec:emission}}

First, we investigate the contribution of free-free emission. An estimate of the thermal flux density can be calculated by assuming a spherical wind structure using the relation \citep{Bieging1989}: 

\begin{equation}
    S_{\nu, \rm thermal} = 23.2 \  (\gamma \  g_{\rm ff} \nu )^{2/3} D^{-2} \left( \frac{\dot{M} \ Z}{v_{\infty}} \right)^{4/3} \ {\rm Jy},
\end{equation}

\noindent
where $\dot{M}$ is the mass-loss rate in $M_{\odot} \ {\rm yr}^{-1}$, $\nu$ is the frequency in Hz, $\mu$ is the mean ionic weight, $\gamma$ is the mean number of electrons per ion, $Z$ is the mean ionic charge, $v_{\infty}$ is the terminal velocity in ${\rm km s}^{-1}$, $D$ is the distance to the star in kpc and $g_{\rm ff}$ is the free–free Gaunt factor, given by:

\begin{equation}
    g_{\rm ff} = - 1.66 + 1.27 \log (T_{\rm wind}^{3/2}/ (Z/\nu)),
\end{equation}

\begin{figure*}
\centering
\gridline{\fig{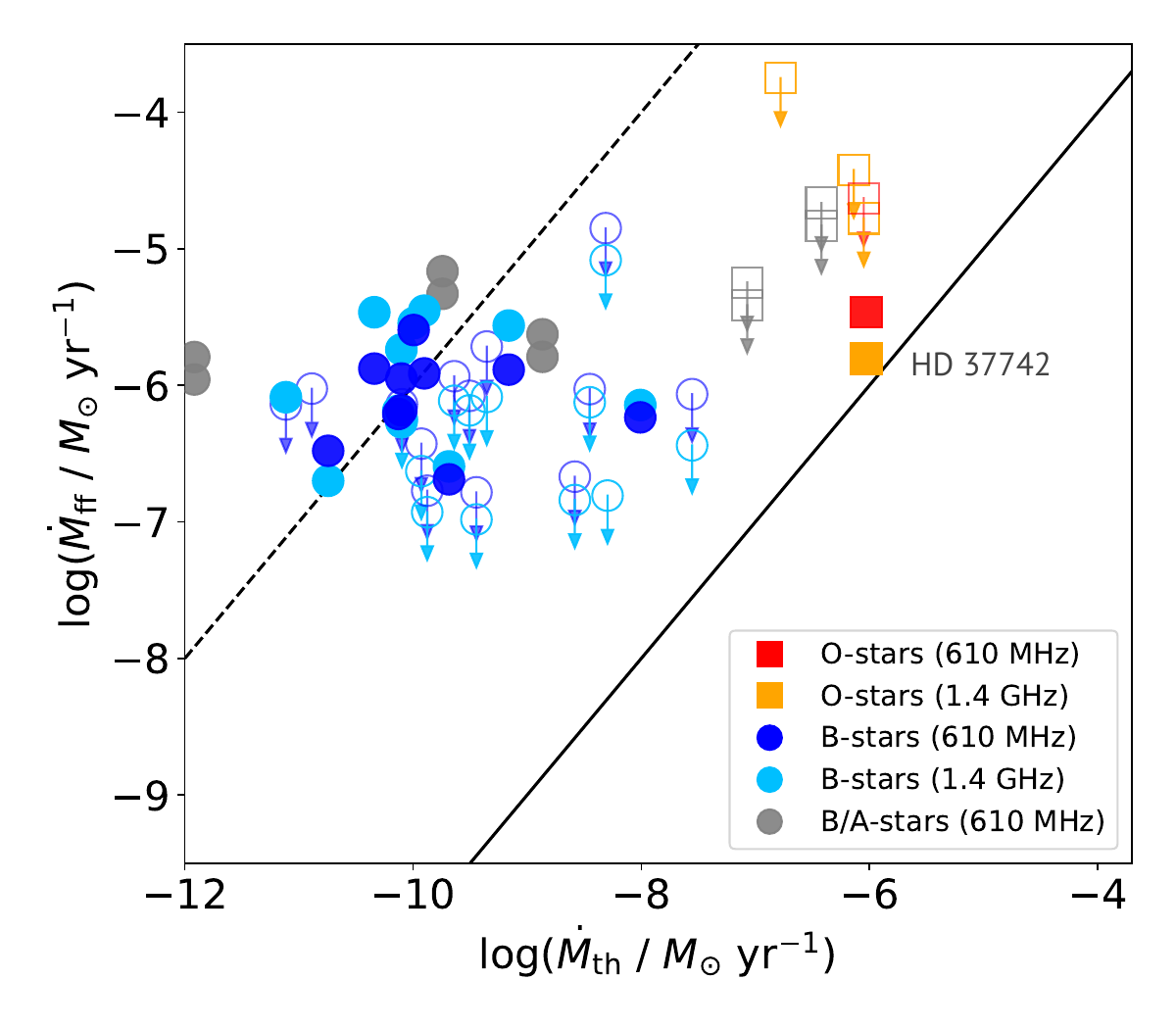}{0.5\linewidth}{(a)} \label{fig:M_dot}
          \fig{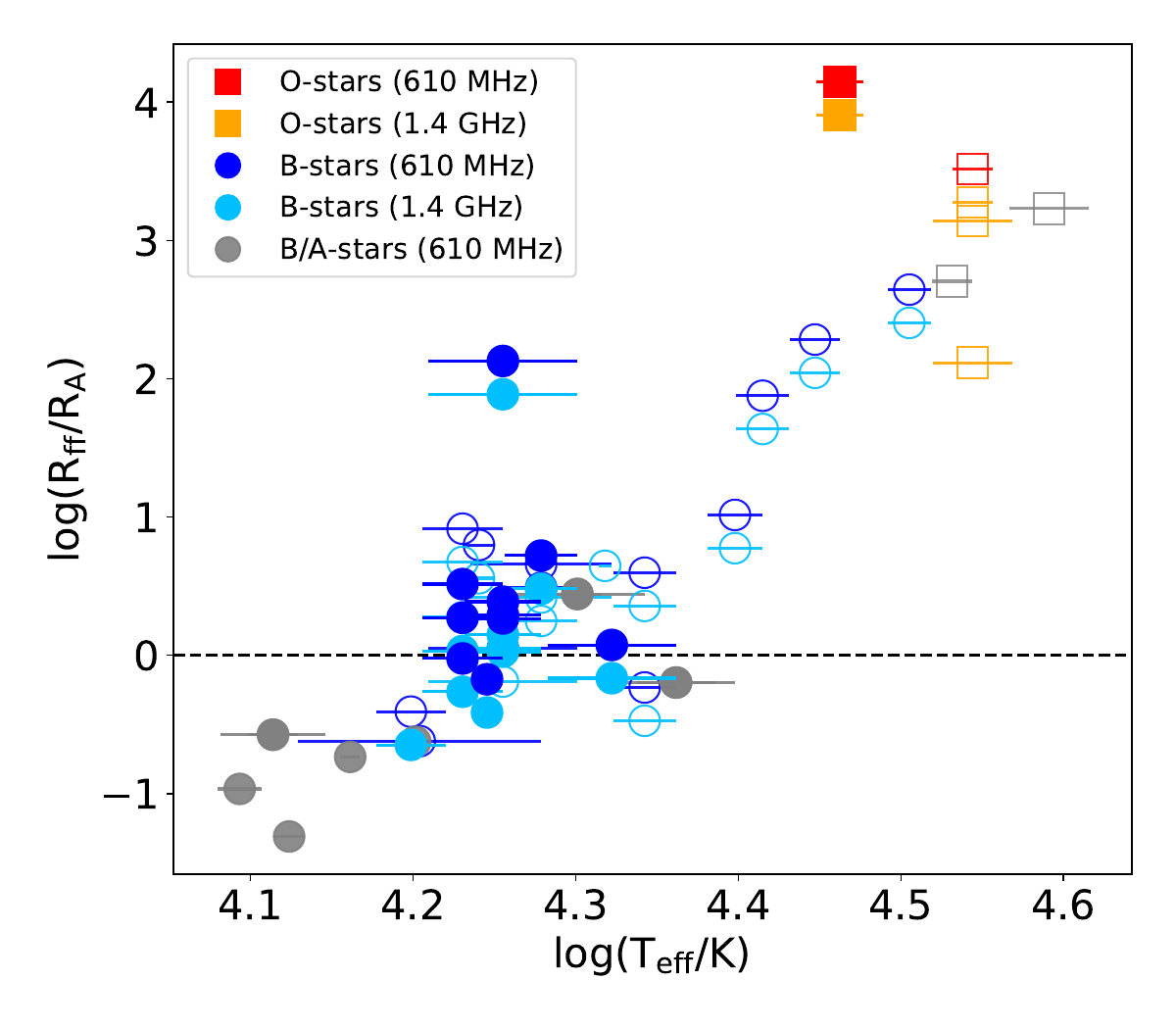}{0.5\linewidth}{(b)} \label{fig:Rff}}
    \caption{Free-free characteristics of the sample: (a) Comparison of theoretically predicted mass-loss rates ($\dot{M_{\rm th}}$) and mass-loss rate assuming a completely free-free thermal emission ($\dot{M_{\rm ff}}$). The black solid line represent $\dot{M_{\rm th}} = \dot{M_{\rm ff}}$, while the dashed line represent $\dot{M_{\rm ff}} = 10^4 \dot{M_{\rm th}}$.  (b) Variation of the ratio of free-free radius $R_{\rm ff}$ and Alfv{\'{e}}n radius $R_{\rm A}$ with effective temperature. The black dashed horizontal line represents the $R_{\rm ff} = R_{\rm A}$ boundary. For stars with $R_{\rm ff} > R_{\rm A}$, the effect of free-free absorption is predicted to be significant. For both figures, the markers are same as in Fig. \ref{fig:det_bias}, with uniform marker size.   }
    \label{fig:free_free}
\end{figure*}

\noindent
where $T_{\rm wind}$ is the local temperature (in K) in the radio photosphere, typically taken to be half of the effective temperature.  In this case, we apply an indirect approach by comparing the mass-loss rate obtained from this diagnostic (assuming 100\% thermal emission) with the theoretical mass-loss values. The comparison between the two mass-loss rates is shown in Fig. \ref{fig:free_free}a. For all targets except HD 37742, we get the free-free mass-loss rate to be three to six orders of magnitude higher than the theoretical mass-loss rate values. Note that mass-loss rates can be suppressed further because of magnetic confinement, and thus this factor may increase. This significant disparity proves that thermal emission has little to no contribution in the observed emission for almost all stars. However, for HD 37742, the radio luminosity at 1390 MHz bands is indeed consistent with the free-free emission.  In 610 MHz band, there is again a mismatch between the observed and theoretical mass-loss rate of HD 37742. Combined with the negative spectral index, this treatment gives evidence of nonthermal emission from HD 37742 at sub-GHz frequencies.

For the supplementary sample, a search for ECME has been conducted in previous studies (e.g. \citealt{Das2022, Das2022c}). For these stars, the basal radio flux is used in this study, which is free from ECME contribution. As we have already established, the emission mechanism is nonthermal at these frequencies. Thus, radio emission from these stars comes primarily from a gyrosynchrotron mechanism originating from single star magnetospheres. For the core survey sample, we did not detect circular polarization, except for HD 142184. This result, along with the inadequacy of free-free emission, also establishes the role of the gyrosynchrotron in this subset of samples.

\subsection{Free-Free Absorption and Non-detections}

Free–free absorption (FFA) is a process in which a photon is absorbed by a free electron in the presence of an ion. For massive stars, the free-free optical depth $\tau_{\rm ff}$ is defined as \citep{Torres2011}:

\begin{equation}
    \tau_{\rm ff} = 5 \times 10^3 \  \dot{M}_{-8}^2 V_{\infty}^{-1}  \nu_{\rm GHz}^{-2} T_{\rm wind}^{-3/2} D_{\rm ff}^{-3},
\end{equation}

\noindent
where, in this case,  $\dot{M}_{-8}$ is the mass-loss rate in units of $10^{-8} M_{\odot}/{\rm yr}$ , $\nu_{\rm GHz}$ is the frequency of observation in GHz, and $D_{\rm ff}$ is the distance from the star (in units of $3 \times 10^{12}$ cm). Due to the strong dependence of $\tau_{\rm ff}$ on frequency, the low-frequency regime ($<1$ GHz) is usually heavily affected by free-free absorption. The radius of the free-free radio photosphere ($R_{\rm ff}$) is then defined as the distance from the star where the free-free optical depth is unity (i.e. $R_{\rm ff}=D_{\rm ff, \tau_{\rm ff}= 1}$).

\begin{figure*}
    \centering
    \includegraphics[width=0.95\linewidth]{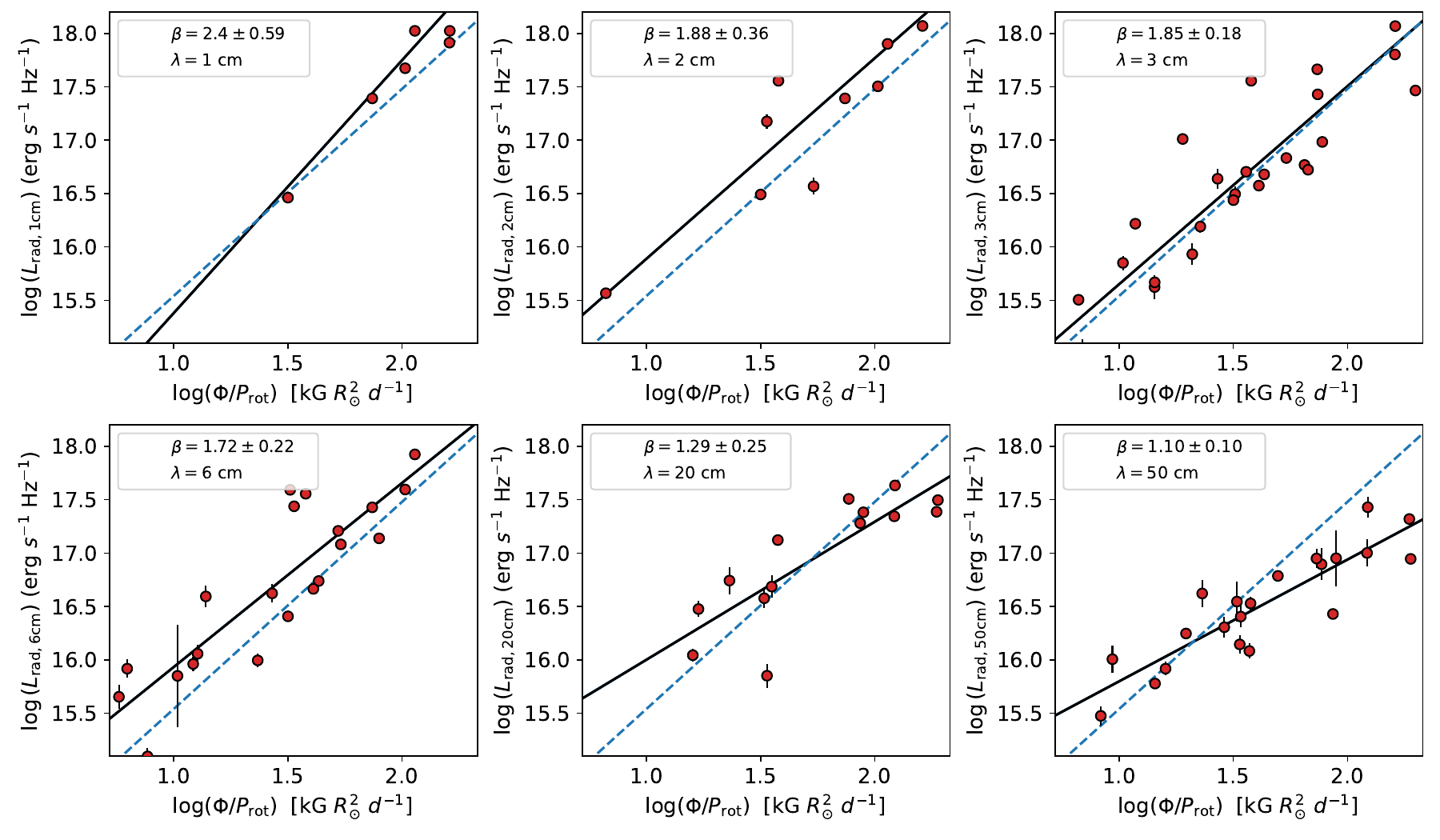}
    \caption{Similar to Fig. \ref{fig:scaling}, but for six different frequency bands separately. Both $\alpha$ and $\beta$ from equation \ref{eq:scaling} are kept as free parameters in all cases. The bounds for x and y axes are kept fixed for comparison. 1 cm, 2 cm, 3 cm, and 6 cm data were taken from \cite{Shultz2022}. The remaining two data sets are from this study. The blue dashed line represent the best-fit line from \cite{Leto2021} and the black solid line is the best fit line from this work.}
    \label{fig:all_freq}
\end{figure*}

We used the theoretical mass-loss estimates for each star to determine the extent of the thermal photosphere, i.e. $R_{\rm ff}$. The variation of the ratio of the free-free radius $R_{\rm ff}$ and Alfv{\'{e}}n radius $R_{\rm A}$ with the effective temperature of the stars is shown in Fig. \ref{fig:free_free}b. Most stars with a high $R_{\rm ff}/R_{\rm A}$ ratio are found to be undetected. $R_{\rm ff}>R_{\rm A}$ signifies that the magnetospheric emission is likely to be hidden within the radio photosphere of the wind; thus most of the nonthermal emission will be absorbed. Even for detected stars, the absorption is expected to significantly decrease the nonthermal flux of the stellar magnetospheres.

\subsection{Scaling Relationship at Sub-GHz}

Previous studies at high frequencies by \cite{Leto2021} and \cite{Shultz2022} have shown that radio luminosity can be scaled with the empirical parameter $\Phi /P_{\rm rot}$ such that:

\begin{equation}
    L_{\rm rad} = 10^{\alpha} \times \left( \frac{\Phi}{P_{\rm rot}} \right)^{\beta}, \label{eq:scaling}
\end{equation}

\noindent
where $\alpha= 13.6 \pm 0.1$, and $\beta = 1.9 \pm 0.1$, as obtained by \cite{Leto2021}. The study of \cite{Shultz2022} with a larger sample revealed that a value of $\beta$ is $1.8 \pm 0.2$, which is consistent with the previous work (i.e. $\sim 2$).   The parameter $\Phi /P_{\rm rot}$ has the physical dimension of the electromotive force. \cite{Shultz2022} showed that in single stars with a CM, the observed dependence can be explained by the CBO mechanism. In this study at lower frequencies, we also found that radio luminosity is most strongly correlated with $\Phi/P_{\rm rot}$, strengthening the role of CBO as an emission mechanism. We did not find any strong correlation of radio luminosity with wind parameters, which also helped us to disfavor alternative theories for gyrosynchrotron emission.

{The scaling law obtained by \cite{Leto2021} was extended up to Jupiter's case using the average luminosity of Jupiter's incoherent radio emission from multi-frequency measurements. However, the applicability of the sub-GHz scaling relation to Jupiter remains uncertain because the magnetospheric conditions and magnetic field strengths differ substantially from those of magnetic massive stars. Using the scaling law obtained from this study and the average flux density of Jupiter in the $\sim610$ MHz band ($4.9 \pm 0.3$ Jy, \citealt{Pater2003}), we found a significant mismatch between the observed ($\log L_{\rm obs} \sim 6.3$ [erg $s^{-1}$ Hz$^{-1}$]) and expected ($\log L_{\rm model} \sim 10.4$ [erg $s^{-1}$ Hz$^{-1}$]) radio luminosities. This discrepancy suggests that the sub-GHz scaling relationship derived for magnetic massive stars may not directly extend to the Jovian regime, likely reflecting the very different magnetospheric conditions and spectral properties of the two classes of objects.

In addition, gyrosynchrotron emission at different observing frequencies probes different regions of the magnetosphere, with higher frequencies generally tracing regions closer to the stellar surface where the magnetic field strength is larger \citep{Ramaty1969}. However, the distinction between “high” and “low” observing frequencies depends on the characteristic magnetic field strength of the object. For Jupiter, whose surface magnetic field is only of the order of a few tens of gauss, 610 MHz likely probes magnetospheric regions relatively close to the planet and may correspond to the high-frequency portion of the spectrum. In contrast, for magnetic massive stars with kG-strength magnetic fields, 610 MHz observations probe comparatively outer magnetospheric regions and may lie near or below the spectral turnover frequency. This difference in spectral and spatial sampling may therefore contribute to the discrepancy between the observed and expected radio luminosities in the Jovian case.}

\subsection{Frequency Dependency of Gyrosynchrotron?}

\begin{figure}
    \centering
    \includegraphics[width=\linewidth]{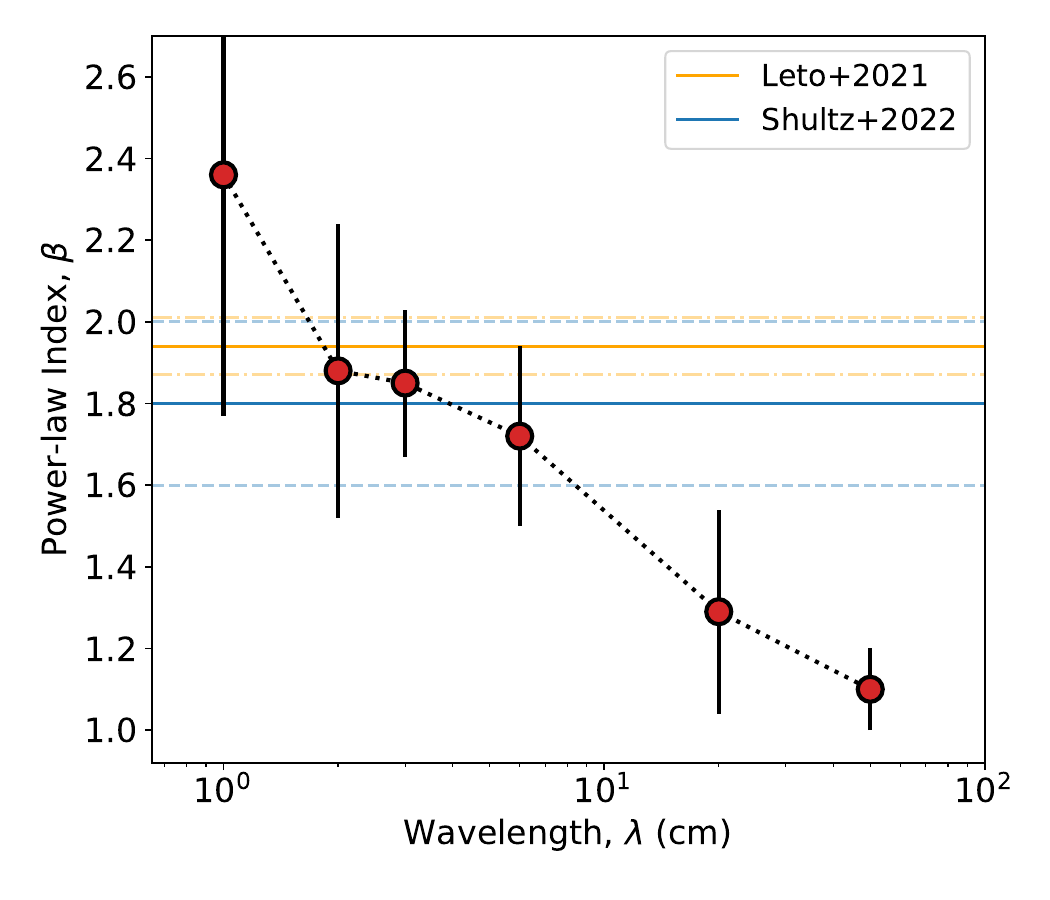}
    \caption{Variation of $\beta$ value from equation \ref{eq:scaling} with wavelength. The blue solid line shows the $\beta$ value obtained by \cite{Shultz2022} from high-frequency study, with blue dashed line indicating the corresponding uncertainty. Similarly, the orange line represents the $\beta$ value obtained by \cite{Leto2021} from their high frequency survey.  }
    \label{fig:beta}
\end{figure}

{ A striking difference from high-frequency behavior is seen in the value of $\beta$. The best-fit value obtained in this work is $\beta = 1.1 \pm 0.1$ (see Table \ref{tab:regression}), significantly lower than the values reported from GHz-frequency studies by \cite{Leto2021} and \cite{Shultz2022}. To investigate this further, we compiled existing radio measurements of magnetic hot stars from \cite{Leto2021}, \cite{Shultz2022}, and this work, and performed regression analysis independently within different observing bands. The resulting best fits are shown in Fig. \ref{fig:all_freq}. We find that the value of $\beta$ decreases systematically toward longer wavelengths (Fig. \ref{fig:beta}), a behavior that was not apparent in previous studies that combined measurements from multiple frequency bands.

The origin of this frequency dependence is likely related to the complex spectral behavior of gyrosynchrotron emission at low radio frequencies. In gyrosynchrotron sources, different observing frequencies probe different regions of the magnetosphere, with higher frequencies generally tracing regions closer to the stellar surface where the magnetic field strength is larger \citep{Ramaty1969}. Consequently, the observed luminosity at a given frequency depends not only on the efficiency of particle acceleration but also on which part of the magnetosphere is being sampled. In addition, several low-frequency attenuation processes, including free-free absorption, synchrotron self-absorption, and Razin suppression, can significantly modify the observed spectra below a few GHz.

Another important effect may be the star-to-star variation of the turnover frequency separating the optically thick low-frequency regime from the flatter GHz-frequency regime. Recent broad-band studies demonstrate that magnetic massive stars do not share a universal radio spectral shape. For example, Fig.~5 of \cite{Das2025} shows substantial diversity in the observed radio spectra, including departures from the commonly assumed flat-spectrum behavior at GHz frequencies. Similarly, recent broad-band modeling by \cite{Leto2026} suggests that stars with larger values of $\Phi/P_{\rm rot}$ may have turnovers shifted toward higher frequencies. In such cases, stars with strong magnetic fields and rapid rotation could appear systematically under-luminous in the sub-GHz regime relative to their luminosities measured at GHz frequencies. This interpretation is qualitatively consistent with the lower value of $\beta$ obtained in the present work.

Therefore, scaling relationships derived at different observing frequencies may not necessarily probe identical magnetospheric conditions across the stellar sample. Instead, the observed frequency dependence of $\beta$ may partially reflect the combined effects of frequency-dependent magnetospheric sampling, varying turnover frequencies, and low-frequency absorption processes. Coordinated broad-band observations and detailed spectral modeling of a larger sample of magnetic massive stars will be essential to disentangle these effects and establish a more complete picture of the origin of nonthermal radio emission in these systems. }

\subsection{Temperature Dependence of Radio Luminosity}

Another difference from the high-frequency behavior is seen in the variation of $L_{\rm rad}$ with the effective temperature. We noticed a moderate correlation between radio luminosity and effective temperature, with an $r$ value of 0.76. \cite{Das2022} also noticed a correlation of ECME flux with temperature at low frequencies, in a sample of main-sequence radio pulse emitters (MRPs). They found that for auroral emission, $L_{\rm ECME} \propto B_{0, \rm max} / (T_{\rm eff} -16.5)^2$. At temperatures lower than the reference value of 16.5 kK, beamed emission is suppressed because of circumstellar density, while at lower temperatures (than 16.5 kK), weak winds are increasingly inefficient for producing strong emission. In this study where the gyrosynchrotron is the primary source of radio luminosity and absorption is a major factor, the possible origin of the temperature dependence may lie in the expression of $\tau_{\rm ff}$. For hotter stars, the terminal velocity and wind temperature increase, and for a specific distance from the star, a hotter star will have lower free-free optical depth. However, this assumption is only speculative and a larger unbiased sample will be necessary to determine a statistically relevant factor.

\subsection{Cases of Exception}

For most of the 44 targets, observed emission can be explained in general as primarily gyrosynchrotron emission that can be explained via the CBO mechanism, while the non-detections can be explained via absorption at lower frequencies. The following are some exceptions/outliers observed in this sample:

\begin{enumerate}
    \item HD 37742 is the only O-star detected in this sample, and to our best knowledge, is the only magnetic O-star detected at sub-GHz frequencies. As discussed in sec. \ref{sec:emission}, the 1390 MHz emission can be explained perfectly with thermal free-free emission. The radio spectrum at frequencies $>1$ GHz show a typical spectra for a free-free emission (Fig. \ref{fig:det_spectra}). However, the sudden change in spectra at 610 MHz, with a high negative spectral index ($-0.9 \pm 0.2$) is a clear indication of nonthermal radio emission. Despite having a very large free-free radius and a weak magnetic field, the detection of nonthermal emission only at such low frequencies is highly unlikely. The scaling relationship clearly shows HD 37742 as an outlier (Fig. \ref{fig:scaling}). Considering the scaling relationship for ECM emission by \cite{Das2022}, it is unlikely to detect coherent emission from this star. The absence of circular polarization also supports this argument. Thus the nonthermal emission might actually come as synchrotron emission from binary interaction. However, the weak wind of the companion  non-magnetic B-star makes this scenario challenging. This star will thus be followed up as a binary candidate in the accompanying paper (Paper-II).
    
    \item For the star HD 200775, we expect a non-negligible contribution from thermal emission along with nonthermal gyrosynchrotron emission. The high value of $R_{\rm ff}/R_{\rm A}$, and moderate mismatch with theoretical mass-loss (Fig. \ref{fig:free_free}), along with a slightly positive spectral index supports this argument. 

    \item Among the non-detections, HD 35912 is the only target for which the observed upper limit is smaller than the expected radio luminosity obtained from equation \ref{eq:scaling1}. This star was recently re-classified as non-magnetic based on sensitive spectropolarimetric observation by \cite{Shultz2018b}. Thus this target can be safely ignored from all analysis, and the non-detection is justified.

    \item For HD 142184, high-frequency radio observations showed a low value of circular polarization fraction and brightness temperature $<10^{11}$. These values are consistent with gyrosynchrotron emission. But both GMRT and ASKAP observations differ from the high-frequency observations regarding this aspect. An increase in circular polarization with decrease in frequency is not usually observed, which also makes this system unique. Such high brightness temperature and circular polarization at low radio frequencies can be indicative of additional emission mechanisms such as ECME. Based on the X-ray properties of the star, \cite{Leto2018} proposed that the star is likely to produce ECME, which could be visible despite the fact that the star does not exhibit magnetic null. A follow-up study with uGMRT by \cite{Biswas2025} revealed that the star is indeed producing ECME at all rotation phases characterized by a high degree of circular polarization. The authors identify the star as a new  Main-sequence Radio Pulse emitter (MRP). 

    \item HD 189775 is the most prominent outlier considering the scaling relationship from this study (and also compared to the scaling law by \cite{Leto2021}. The possible reason behind this discrepancy is unclear, and the observed emission might be contaminated by emission mechanisms other than thermal and gyrosynchrotron emission. Irrespective of the situation, the current observations confirm a nonthermal nature of radio emission which is a first evidence of magnetosphere in this star. These GMRT observations are also the first evidence for the star's radio bright nature.

    \item The B-star HD 215441 (also known as Babcock's star) is found to be slightly over-luminous compared to the scaling relationship. However, this star has large errors in its radius and magnetic field measurements. Furthermore, this star is found to be variable at higher frequencies \citep{Leone1996}. Thus it is possible that the star was observed during highest emission state, and the average emission might follow the obtained scaling law. 

\end{enumerate}

\section{Conclusion} \label{Conclusion}

In this work, we carried out a survey of radio emission from magnetic hot stars at low frequencies ($<1.5$ GHz) with GMRT. This survey is one of the first large systematic surveys carried out at the sub-GHz band. We observed 28 targets and collected information for 16 additional targets observed with GMRT/uGMRT. Out of the 28 core targets, we report five unique sub-GHz detections, and 16 unique non-detections, drastically increasing the sample size of magnetic hot stars observed at low frequencies.  

One of the unique detections, HD 37742 shows a negative spectral index at low frequencies contrast to what was observed at high frequency radio observations. Although X-ray observations of this target exclude the possibility of a colliding wind scenario, the radio observations reveal the existence of nonthermal emission and hint toward the contribution of binarity. Another unique detection, HD 142184 stands out among the population with extremely high brightness temperature and the presence of a high fraction of circular polarization. Although the emission from this single star follows a scaling relationship, the high value of circular polarization and brightness confirm that coherent emission is present in addition to gyrosynchrotron emission.

{The observed radio emission from the stars in the sub-GHz band is predominantly nonthermal in nature. Similar to previous GHz-frequency studies (e.g. \citealt{Leto2021}, \citealt{Shultz2022, Das2025, Leto2026}), we found that the radio luminosity correlates strongly with magnetic field strength and rotation period, qualitatively consistent with expectations from the CBO scenario. However, the scaling relationship obtained at sub-GHz frequencies differs significantly from that derived at higher frequencies. We suggest that this frequency dependence may reflect systematic variations in the gyrosynchrotron spectral shape among stars with different stellar and magnetospheric properties, including possible shifts in the low-frequency turnover of the radio spectrum. In this picture, different observing frequencies may probe different magnetospheric regions, while low-frequency absorption processes such as free-free absorption may further modify the observed emission.}

Although sub-GHz bands are ideal for detailed study of such magnetic massive stars due to the presence of different emission mechanisms (e.g. free-free, ECM, synchrotron, plasma emissions) and different absorption processes (e.g. FFA, SSA), the complexity makes these bands non-ideal for snapshot surveys. A non-detection at one frequency therefore does not necessarily imply the absence of nonthermal radio emission at other frequencies. For example, HD 36526, which was not detected in this survey, is known to exhibit strong ECM emission at similar bands \citep{Das2022c}. {These results highlight the importance of coordinated broad-band radio observations spanning sub-GHz to millimeter wavelengths in order to determine the full gyrosynchrotron spectral shape and locate the turnover and flat-spectrum regions for a large sample of magnetic massive stars. Such measurements will be essential for establishing robust scaling relationships and for developing a comprehensive understanding of magnetospheric radio emission across stars with different magnetic and rotational properties.}

\begin{acknowledgments}
G.A.W. acknowledges support in the form of a Discovery Grant from the Natural Sciences and Engineering Research Council (NSERC) of Canada. The GMRT is run by the National Centre for Radio Astrophysics of the Tata Institute of Fundamental Research. The authors thank GMRT staff for helping with the observations. The National Radio Astronomy Observatory and Green Bank Observatory are facilities of the U.S. National Science Foundation operated under cooperative agreement by Associated Universities, Inc. All data exploited in this paper are available from the corresponding public archives. A.B. acknowledges Dr. Poonam Chandra (PI of the GMRT projects 27\_048 and 28\_075) for their support and permission while using the GMRT data. The relevant TESS data is available at MAST: \dataset[doi: 10.17909/76nz-yt73]{https://doi.org/10.17909/76nz-yt73}.

\end{acknowledgments}

\begin{contribution}

A.B. has reduced and analyzed the GMRT dataset, performed the necessary analysis and primarily wrote the manuscript. G.A.W. supervised the project. B.D. supplied the flux densities for the supplementary sample. All authors shared ideas and contributed while writing the manuscript. 

\end{contribution}

\facilities{GMRT, TESS}

\software{CASA \citep{McMullin2007}, 
          numpy \citep{harris2020array}, 
          Scipy \citep{Virtanen2020}, 
          Astropy \citep{astropy:2013, astropy:2018, astropy:2022}
          }

\bibliography{sample631}{}
\bibliographystyle{aasjournal}


\appendix

\setcounter{figure}{0} 
\makeatletter 
\renewcommand{\thefigure}{A\@arabic\c@figure}
\makeatother

\begin{figure*}
\centering
\gridline{\fig{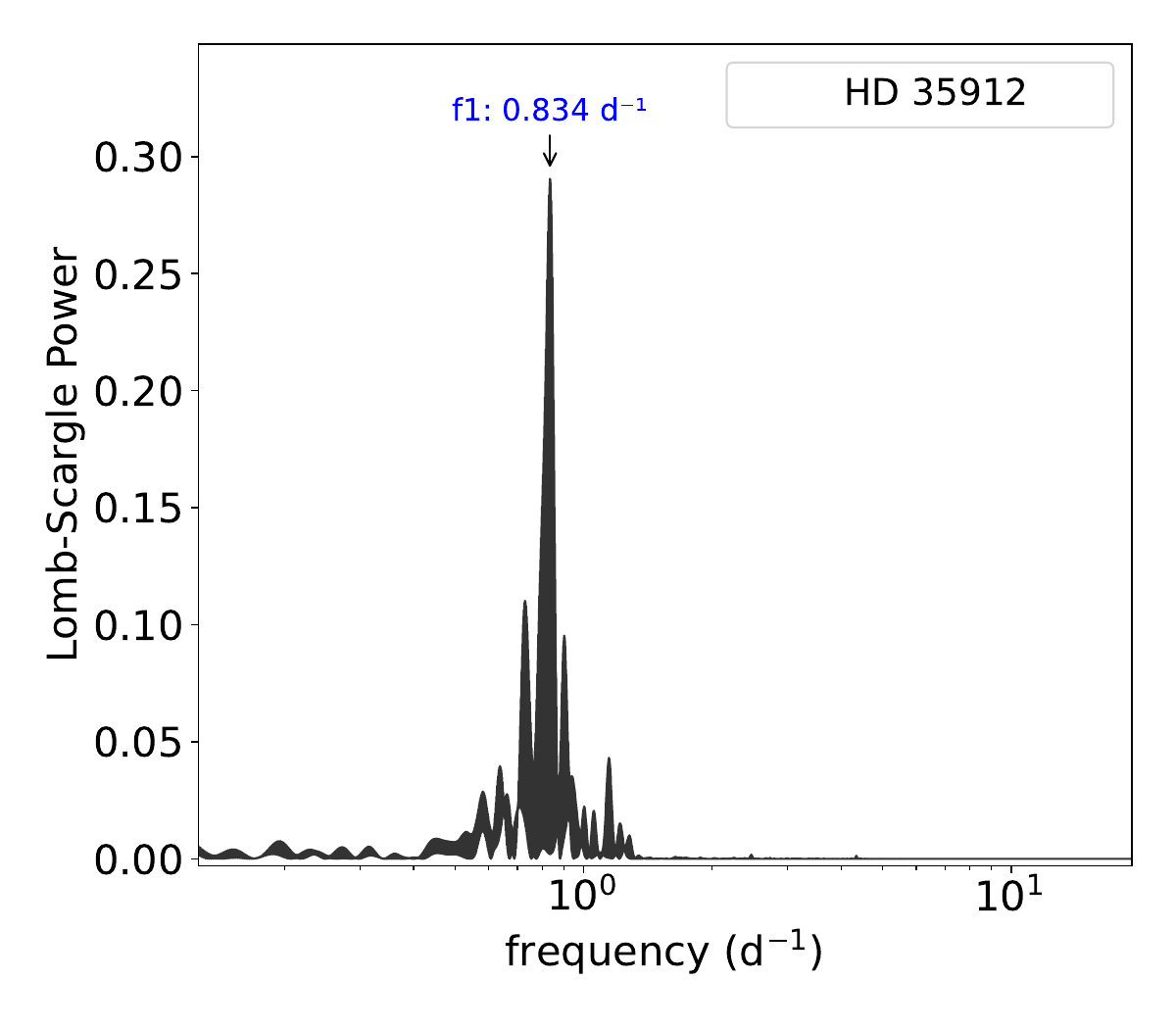}{0.25\linewidth}{(a1) HD 35912: L-S periodogram } 
          \fig{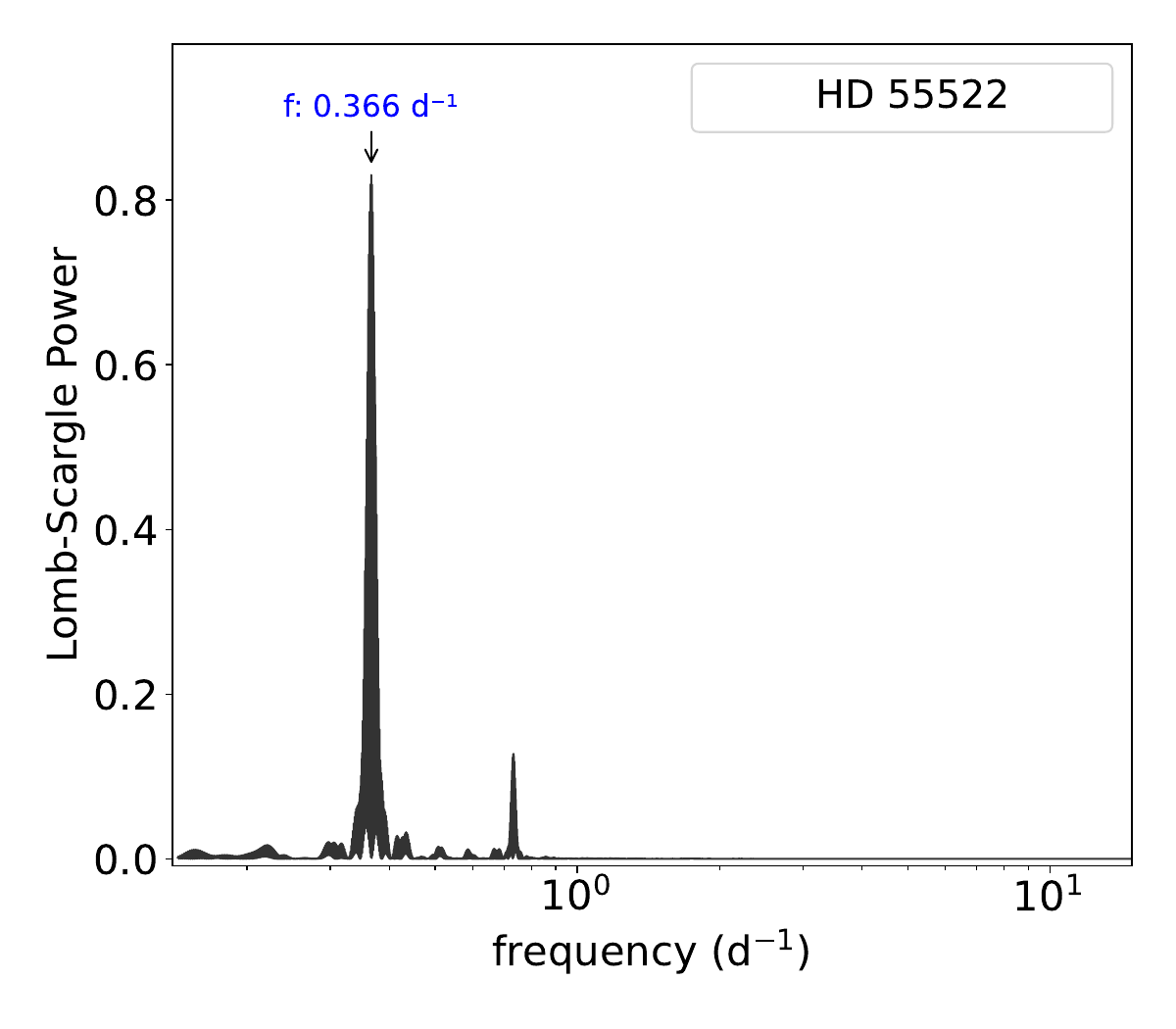}{0.25\linewidth}{(b1) HD 55522: L-S periodogram} 
          \fig{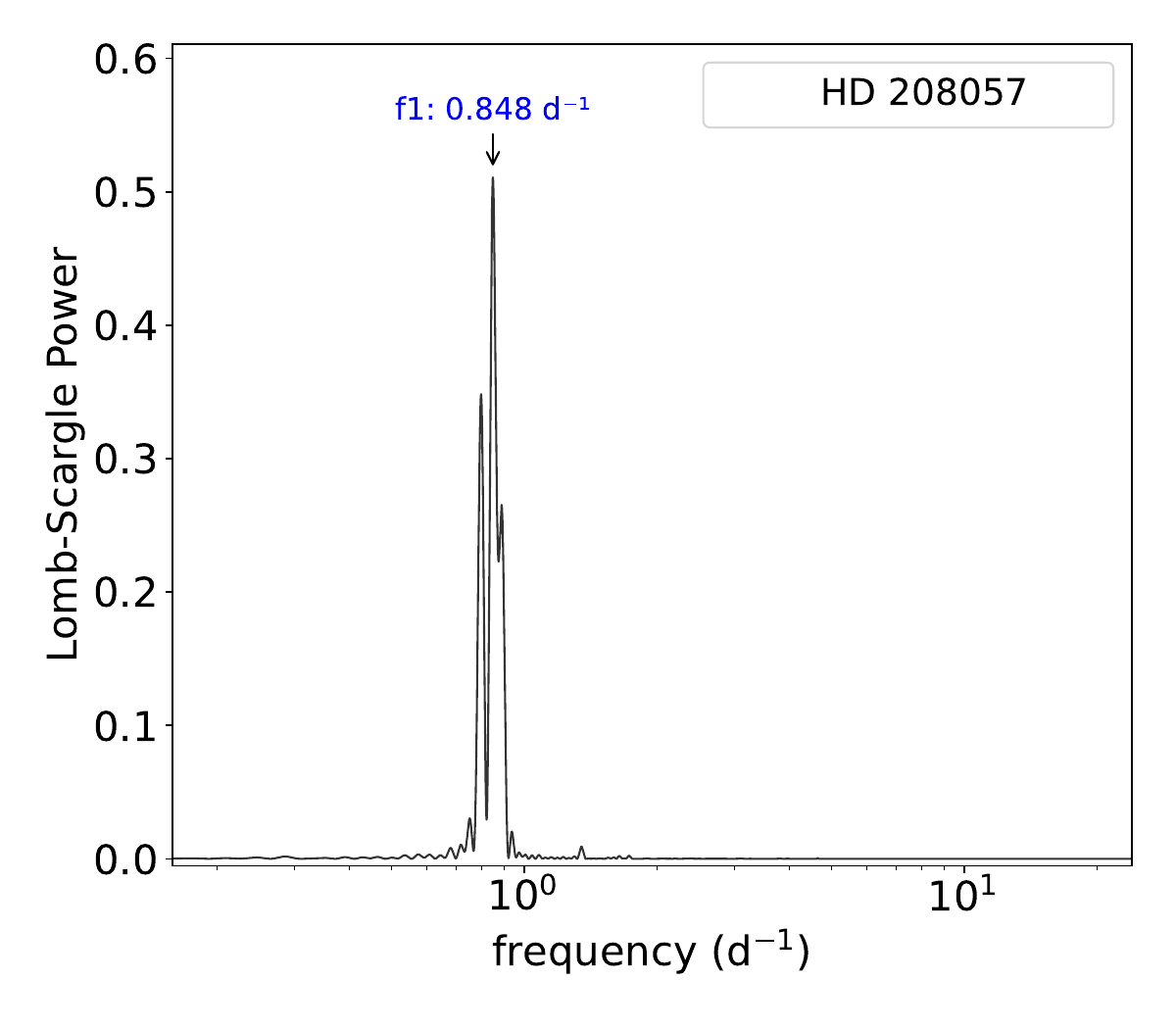}{0.25\linewidth}{(c1) HD 208057: L-S periodogram} 
          \fig{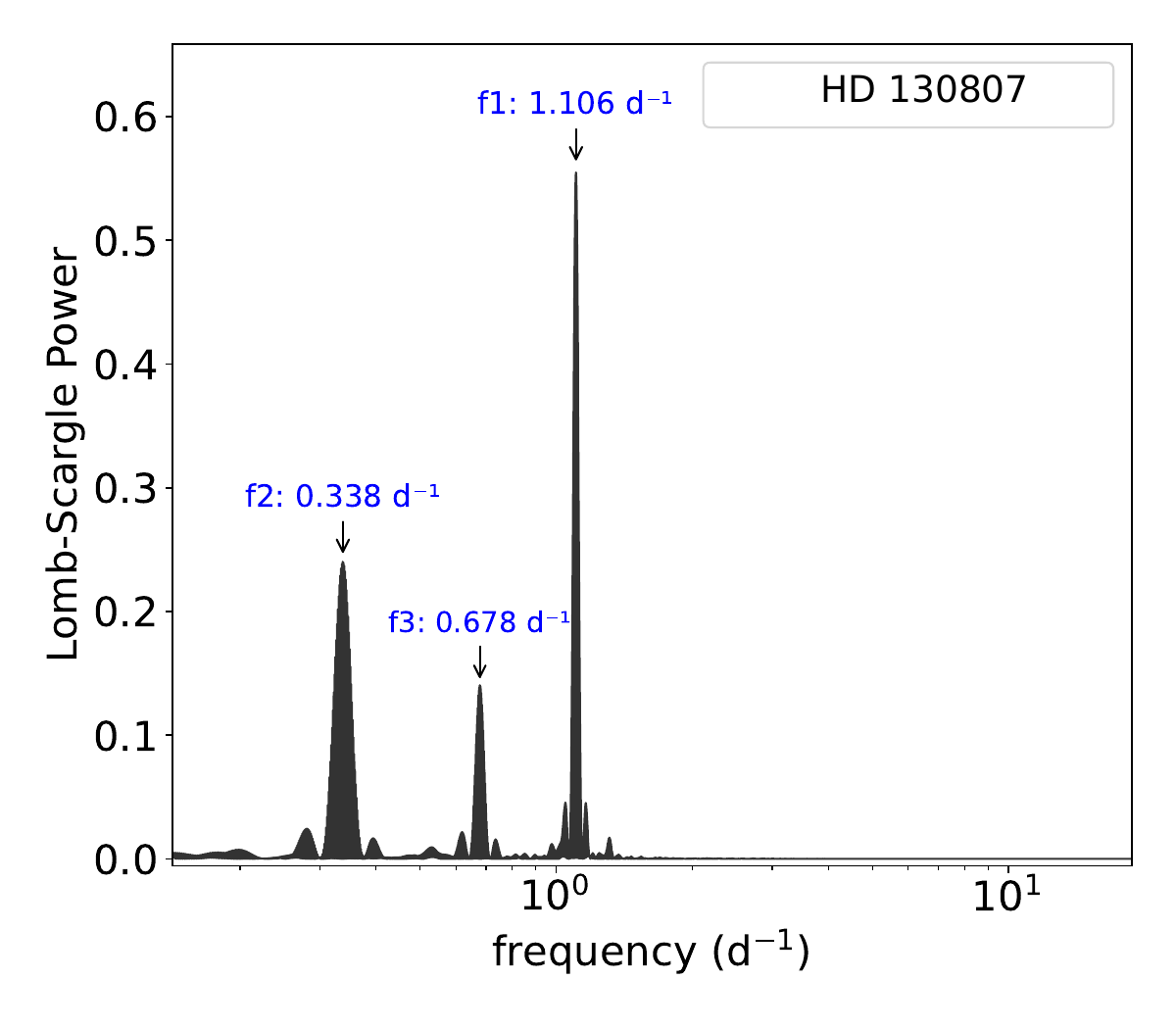}{0.25\linewidth}{(d1) HD 130807: L-S periodogram}}
\gridline{\fig{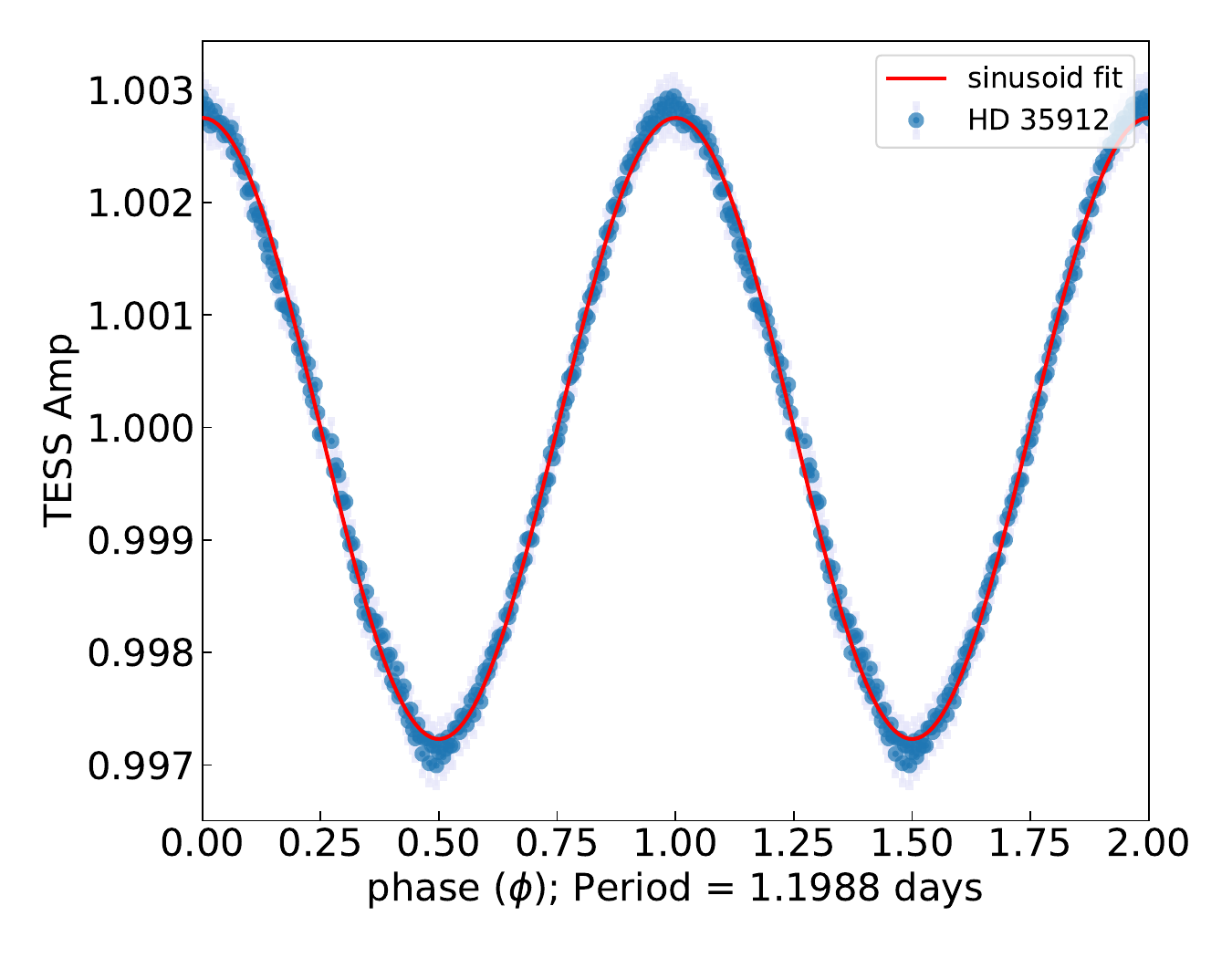}{0.25\linewidth}{(a2) HD 35912: light curve \& fit} 
          \fig{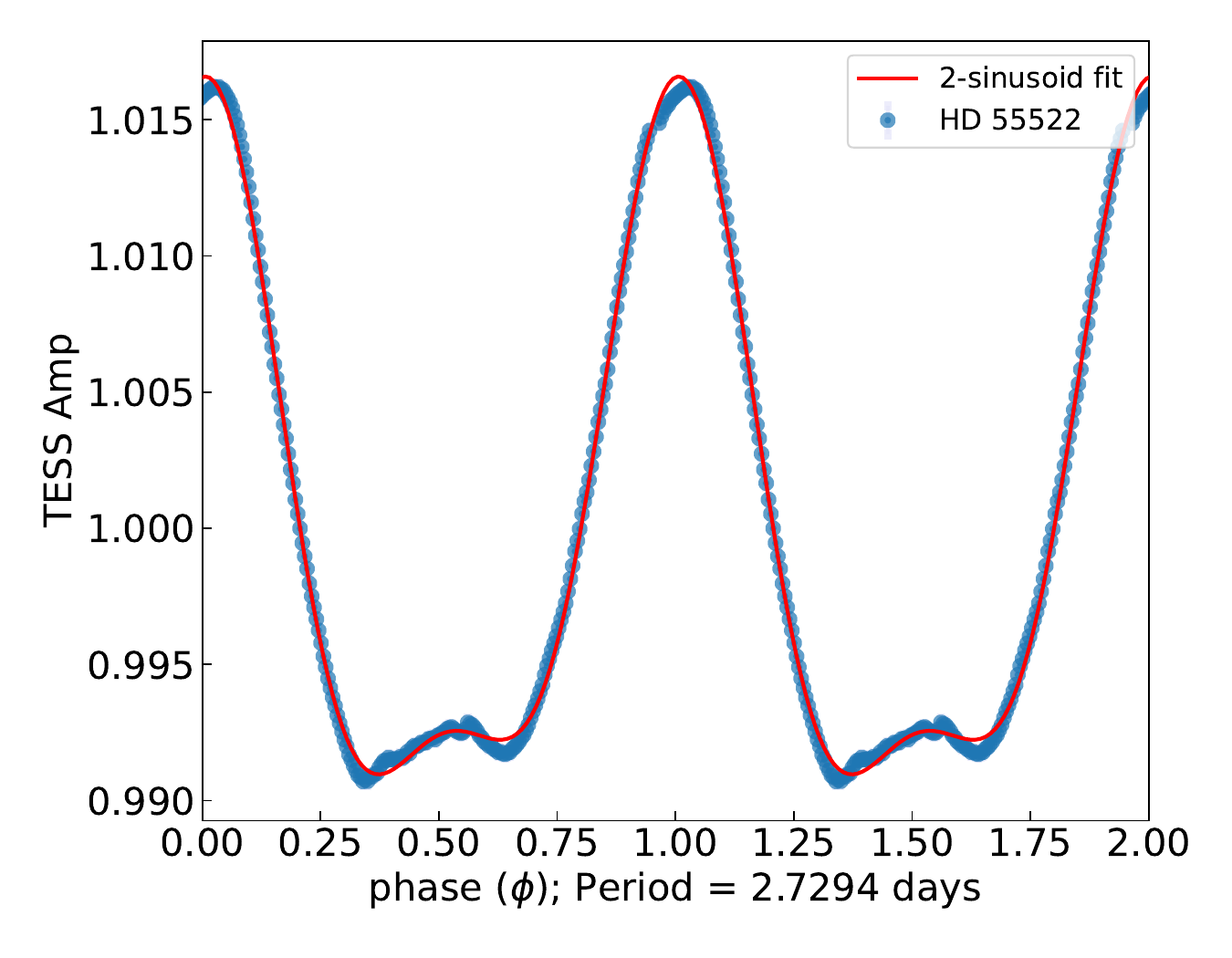}{0.25\linewidth}{(b2) HD 55522: light curve \& fit}
          \fig{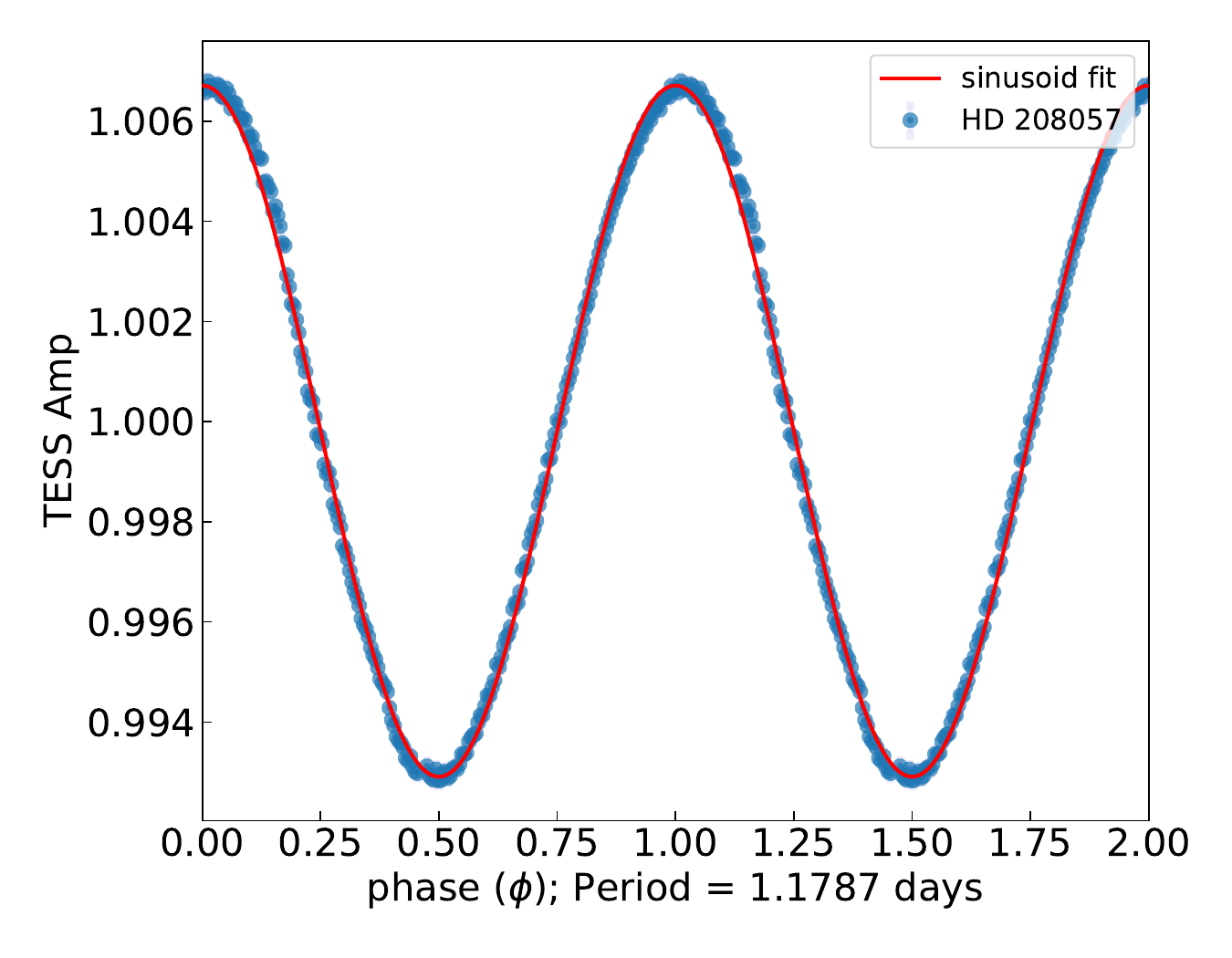}{0.25\linewidth}{(c2) HD 208057: light curve \& fit}
          \fig{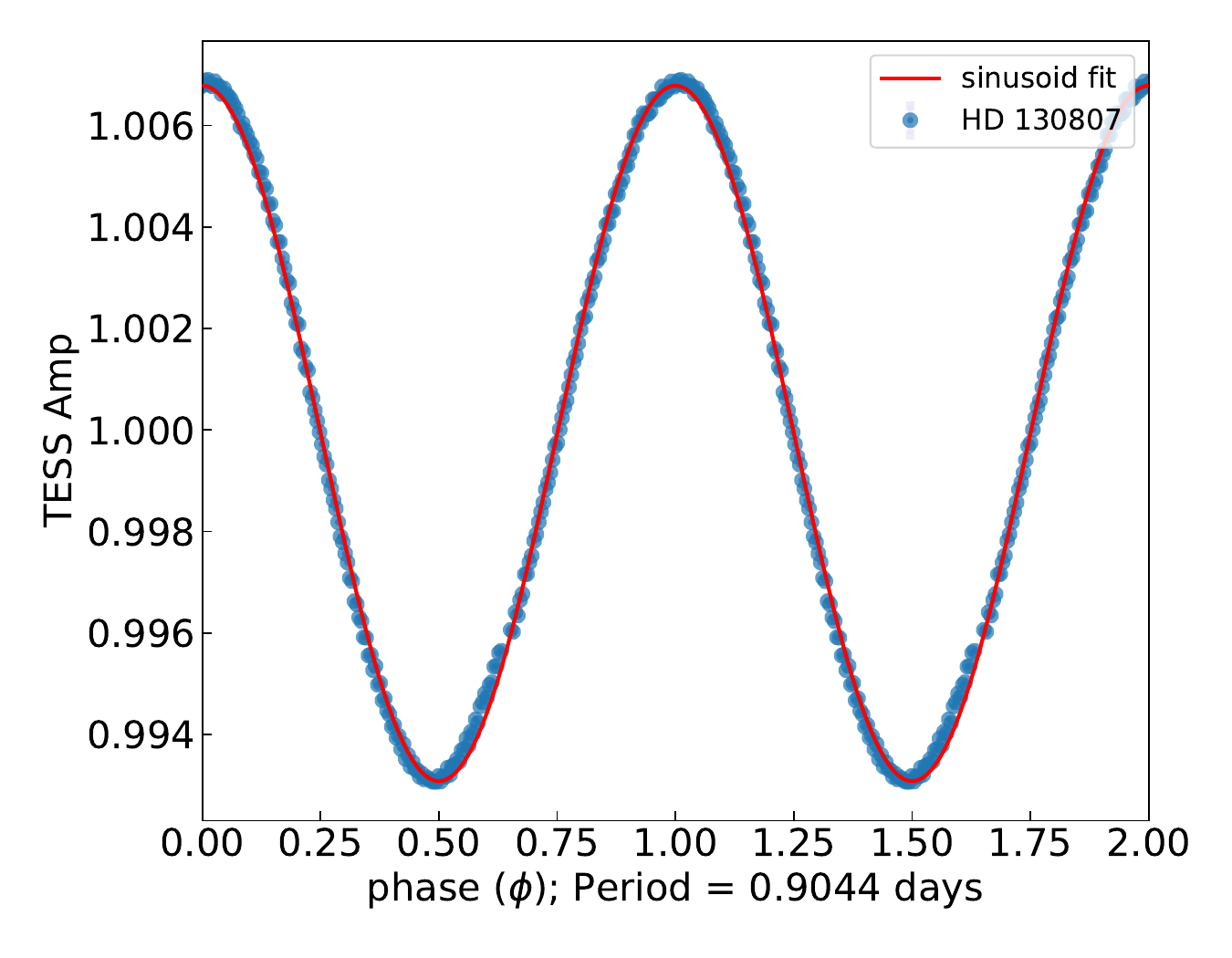}{0.25\linewidth}{(d2) HD 130807: light curve \& fit} }
\gridline{\fig{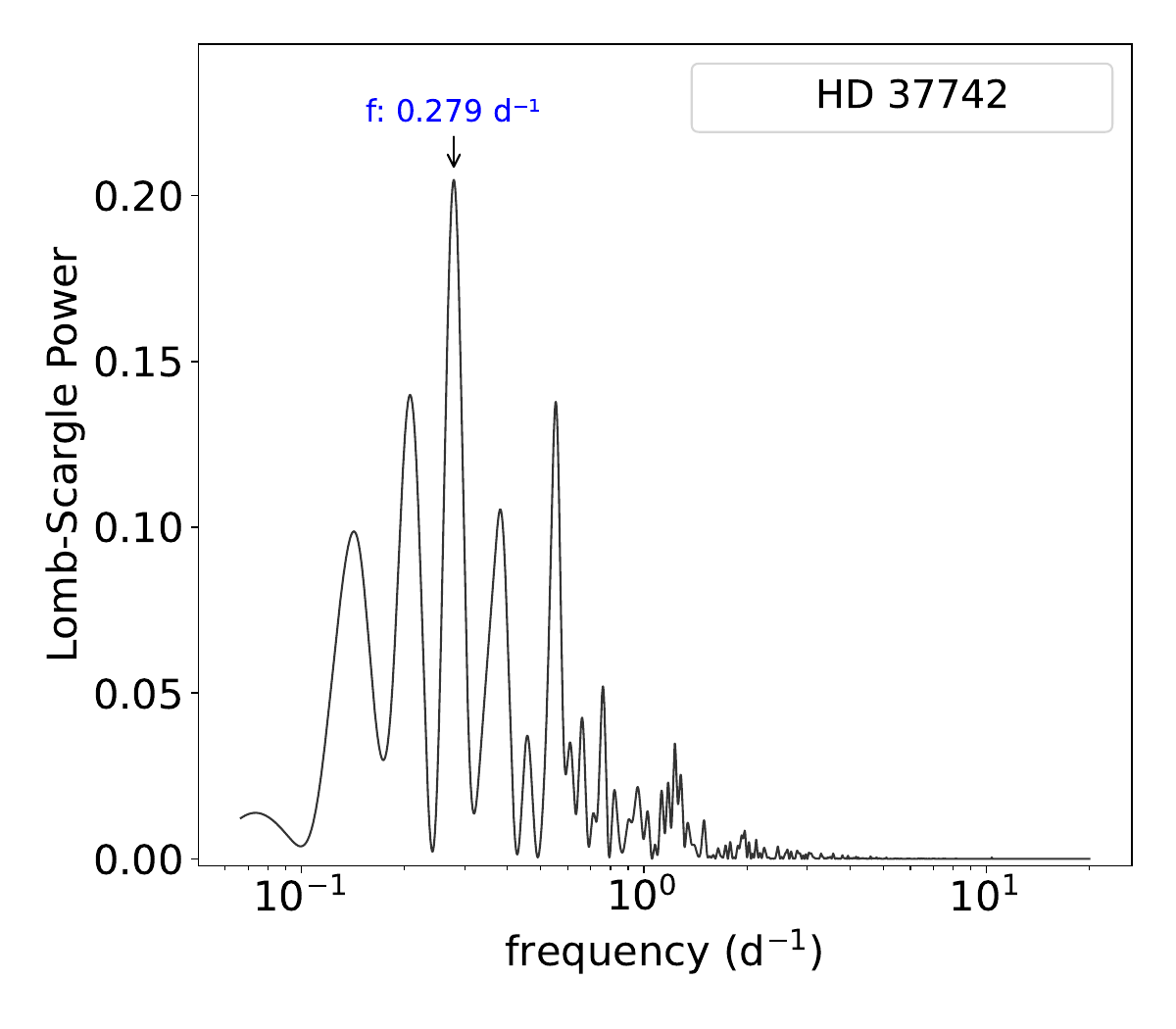}{0.245\textwidth}{(e1) HD 37742: L-S periodogram} \label{fig:per1}
          \fig{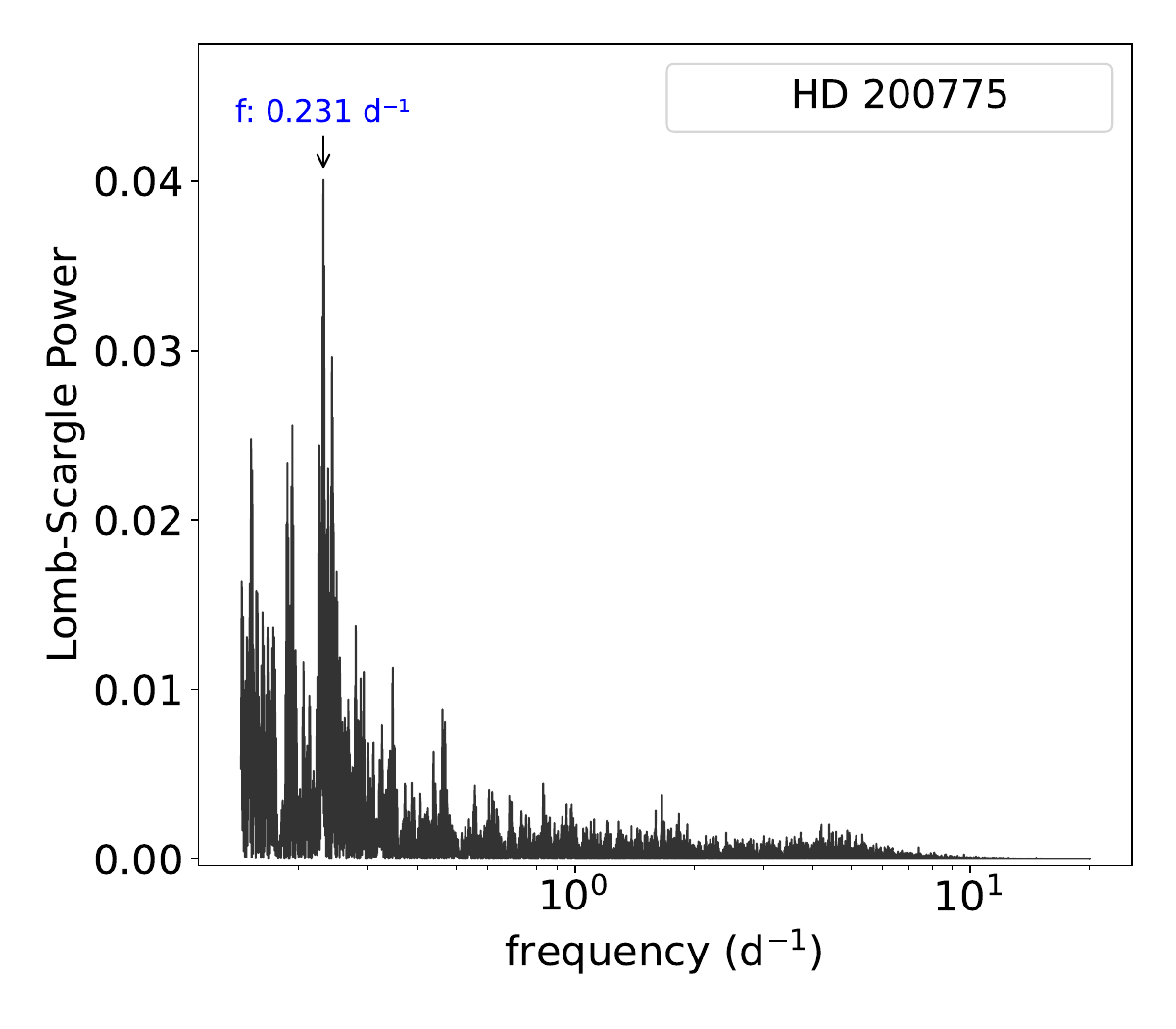}{0.245\textwidth}{f1) HD 200775: L-S periodogram} \label{fig:per2}
          \fig{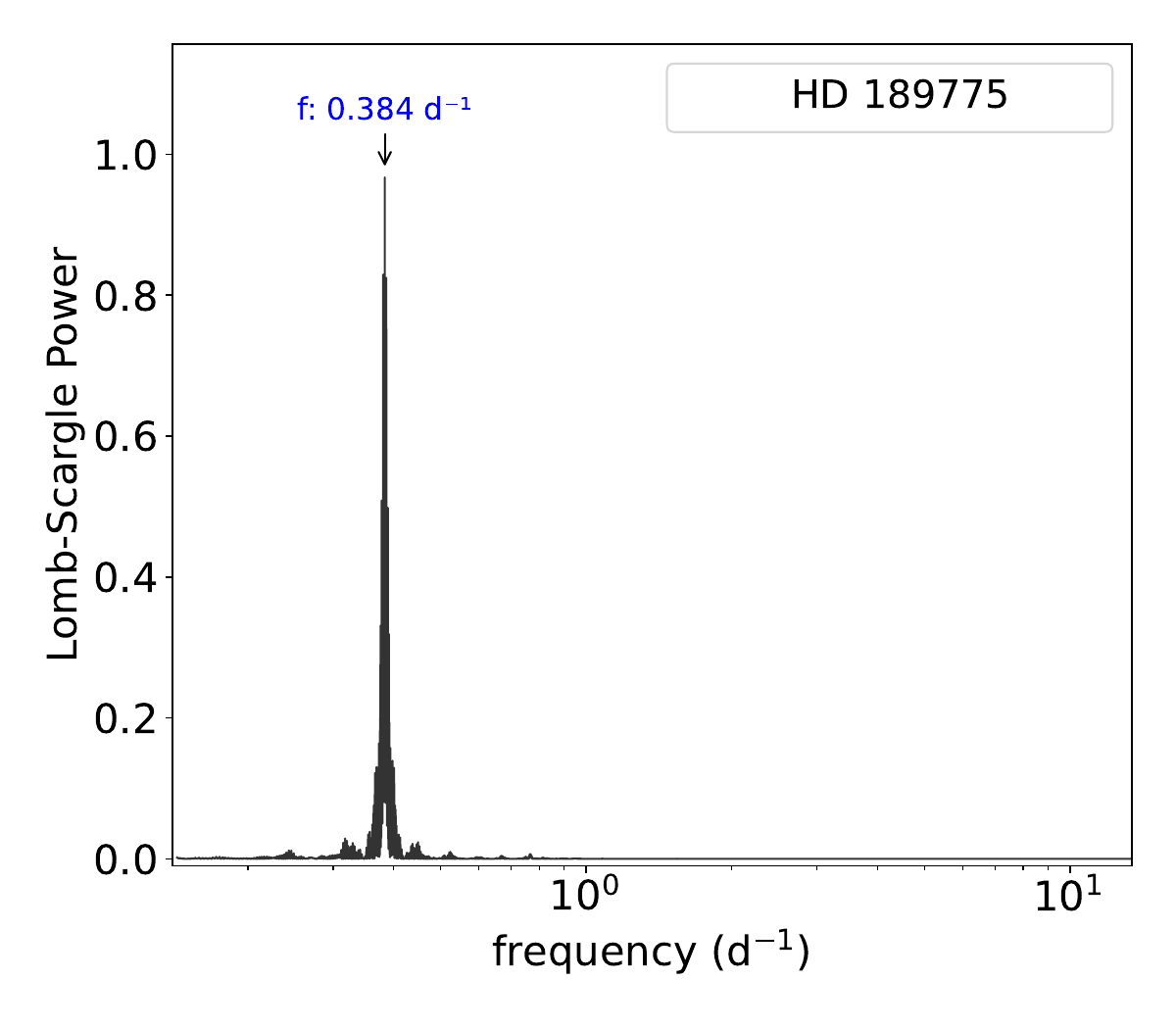}{0.245\textwidth}{(g1) HD 189775: L-S periodogram} \label{fig:per3}
          \fig{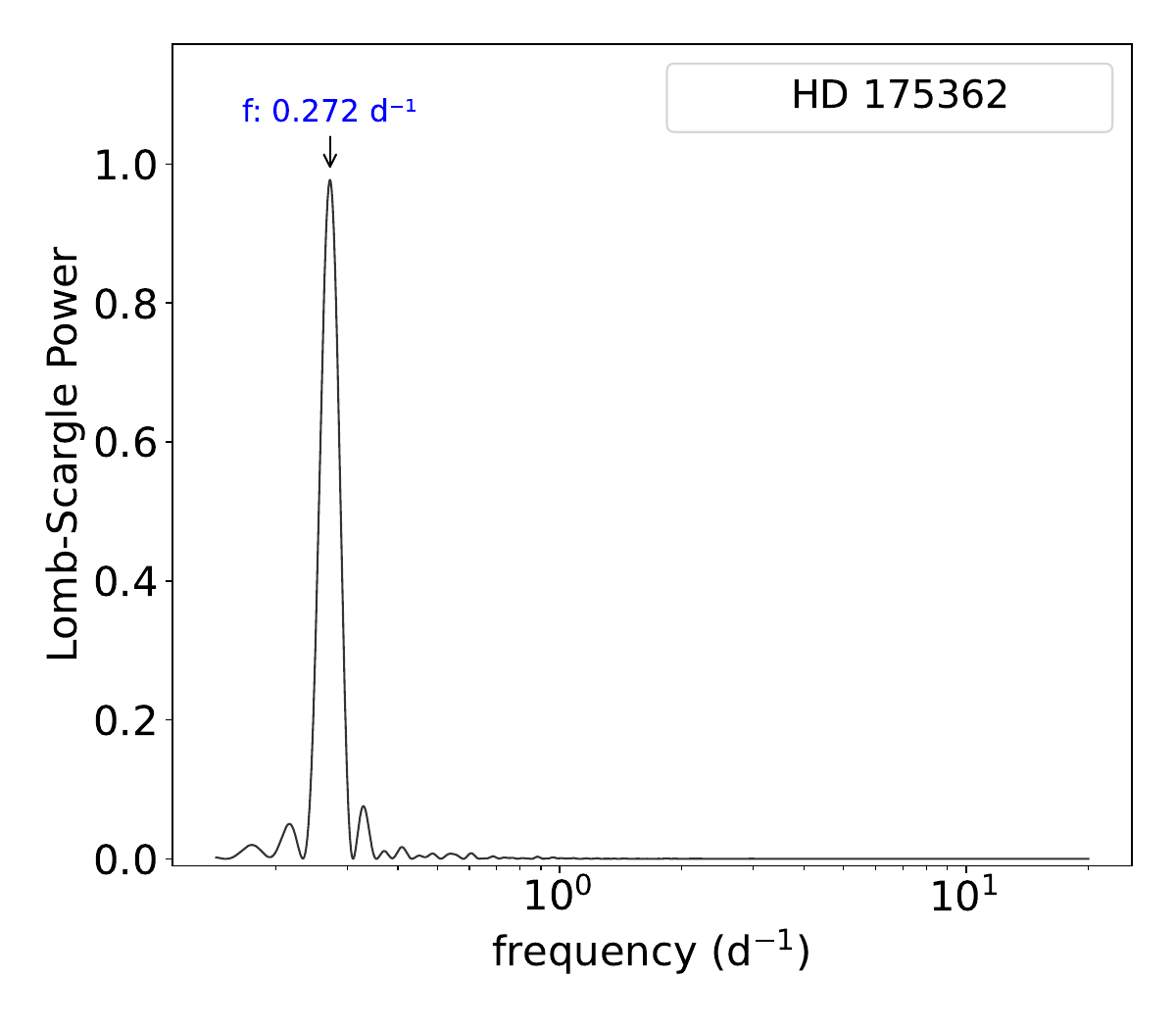}{0.245\textwidth}{(h1) HD 175362: L-S periodogram} \label{fig:per4}}
\gridline{\fig{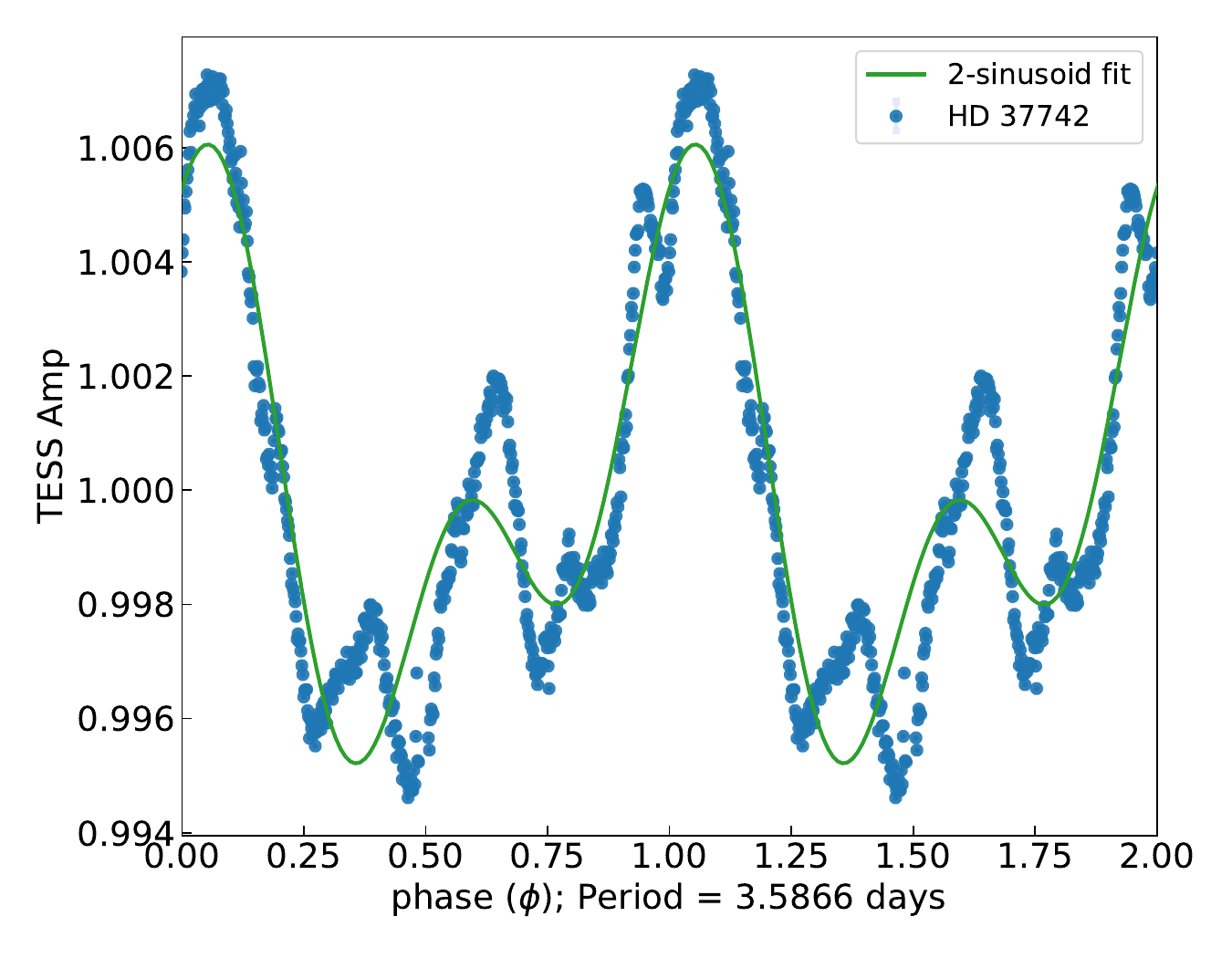}{0.245\textwidth}{(e2) HD 37742: light curve} \label{fig:lc1}
          \fig{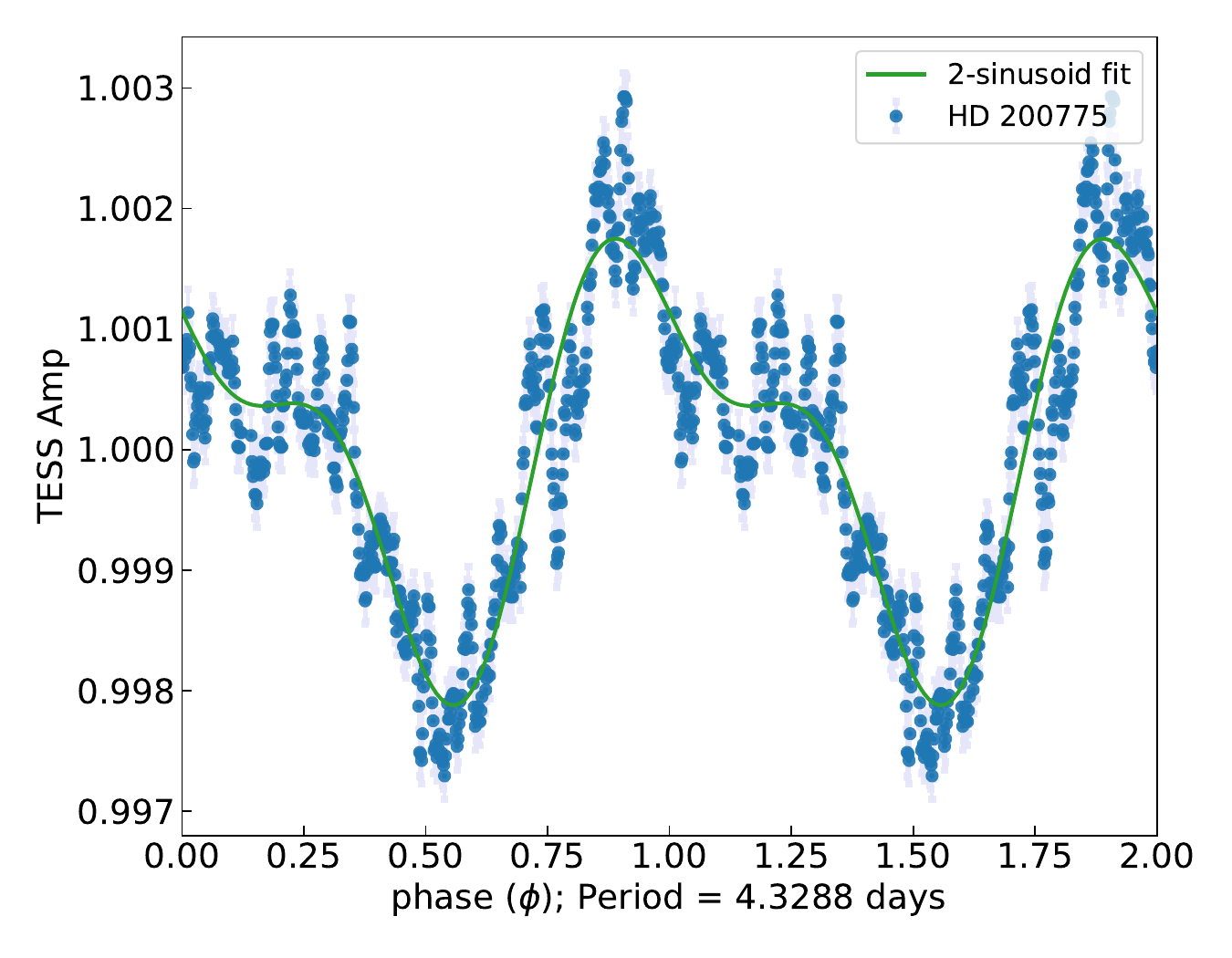}{0.245\textwidth}{(f2) HD 200775: light curve \& fit} \label{fig:lc2}
          \fig{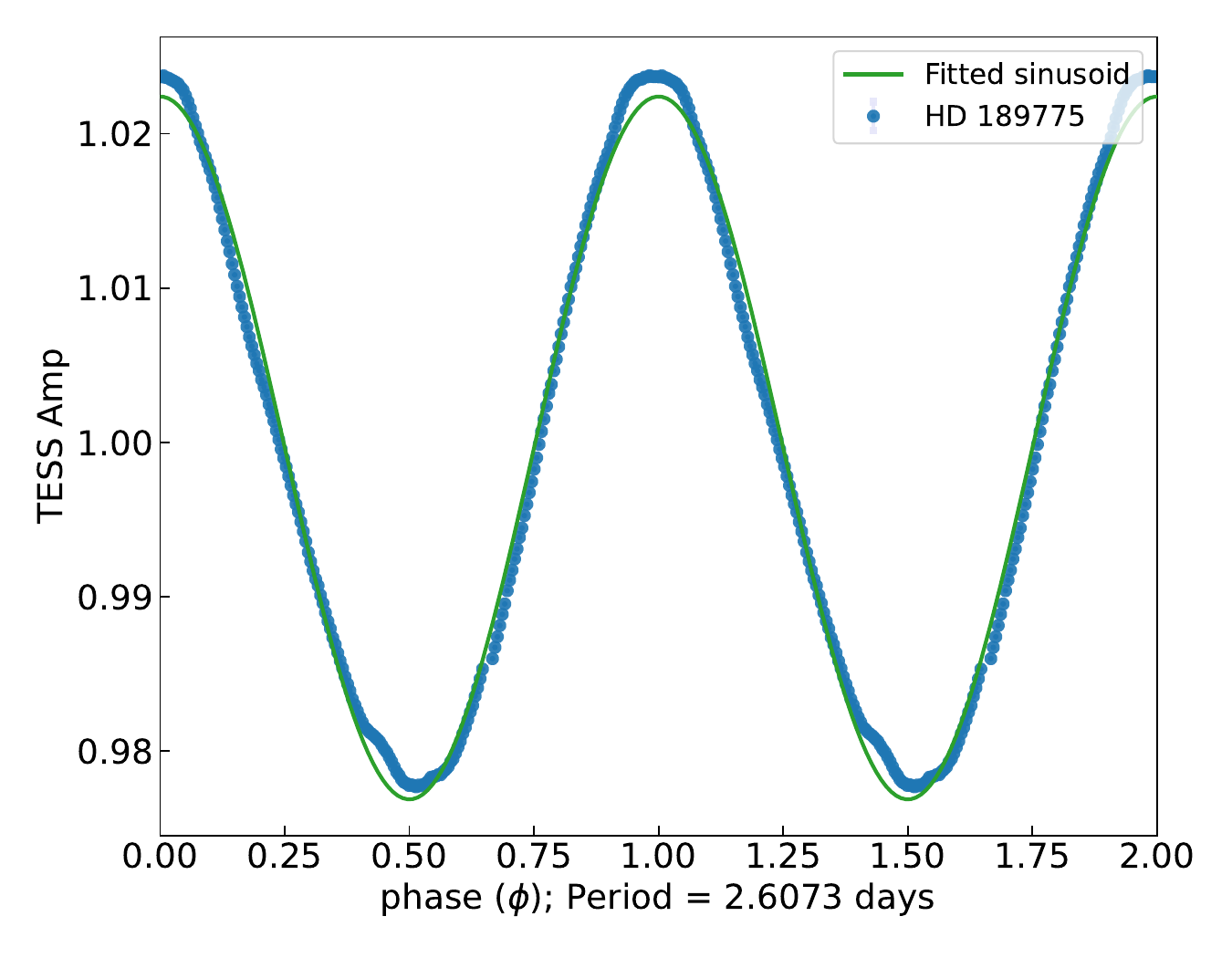}{0.245\textwidth}{(g2) HD 189775: light curve \& fit} \label{fig:lc3}
          \fig{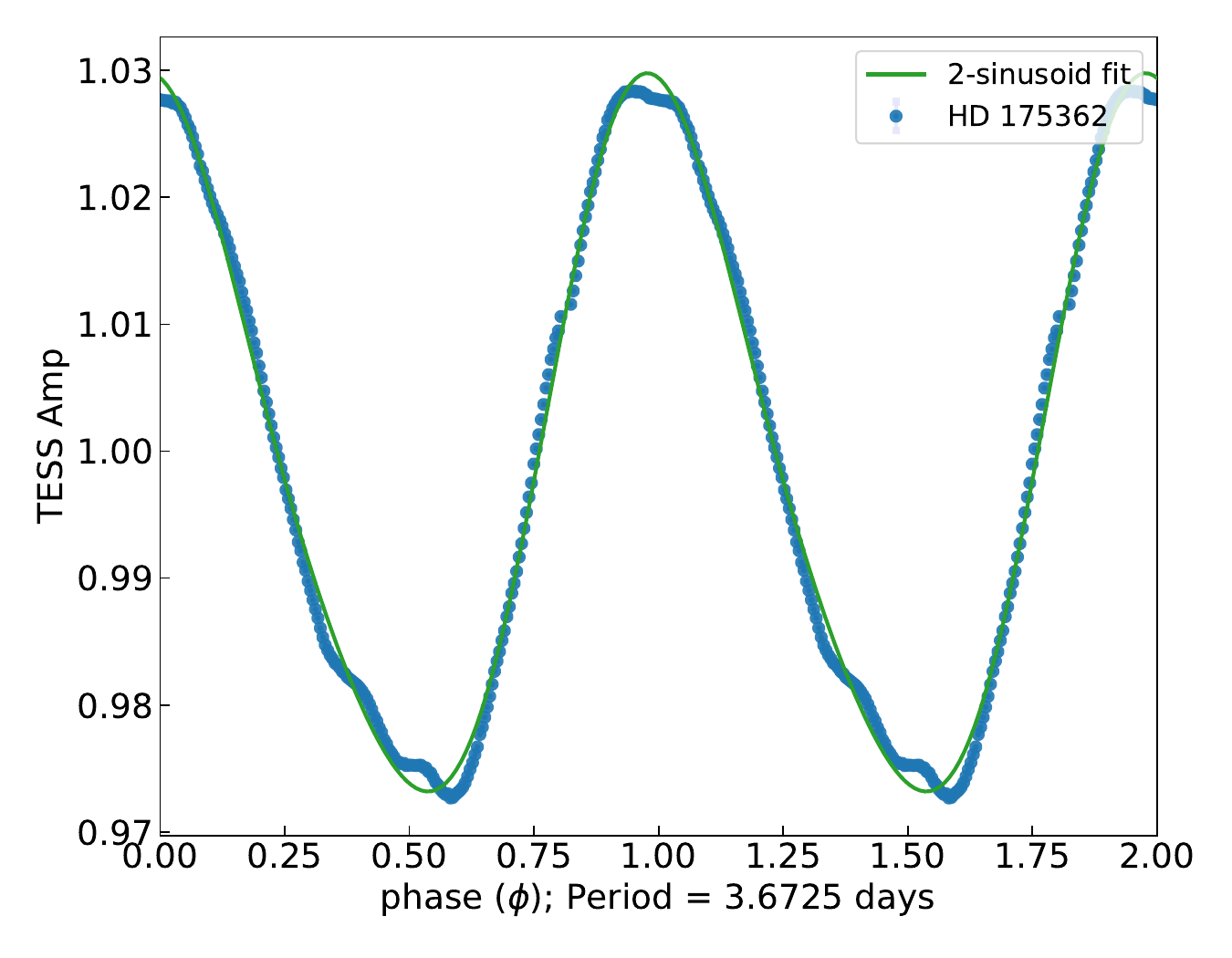}{0.245\textwidth}{(h2) HD 175362: light curve \& fit} \label{fig:lc4}}
\caption{\textbf{(a--d)} Lomb-Scargle periodogram and phased light curve of the targets with unique non-detections for which either the literature does not contain enough information for phase calculation (except HD 130507). The periods are selected from the L-S periodogram. All the phased light curves are binned with $\Delta \phi_{\rm rot}=0.005$. The red curves in the bottom row plots represent the fitted light curves. \textbf{e--h} LS periodogram and phased light curve of the unique detections (except HD 142184). For the first three stars, the first harmonic is also fitted along with their corresponding best frequencies. Only HD 37742 does not give the rotation frequency as the best frequency. \label{fig:tess_per0}}
\end{figure*}

\section{targets with Unique Non-detections} \label{sec:App_non_det}

This section briefly introduces the targets that were not detected in either frequency bands (under projects 27\textunderscore 048 and 28\textunderscore 075), and record high-frequency radio observations found in the literature (if any). All uGMRT flux density upper limits for these targets are mentioned in Table \ref{tab:gmrtflux}.

\subsection{CPD-28 2561}

CPD $-28^o$ 2561 is a chemically peculiar Of?p star \citep{Walborn2010}. Only five galactic Of?p stars are known to date (CPD $-28^o$ 2561, HD 191612, NGC 1624-2, HD 148937, and HD 108), of which three are survey targets from this work. \cite{Hubrig2011} detected a magnetic field in CPD $-28^o$ 2561, which was later confirmed by \cite{Wade2015}. \cite{Wade2015} estimated the dipolar field strength to be $\simeq 2.6$ kG, and a rotation period of 73.41 days. \cite{Kurapati2017} reported a non-detection for this star at 3 and 13 cm, with the VLA.

\subsection{HD 191612}

HD 191612 is a binary system consisting of a magnetic primary star of spectral type O8f?p, and a secondary of spectral type B1V \citep{Howarth2007}. The primary star hosts a dipolar magnetic field of strength of $\sim2.5$ kG \citep{Wade2011}. The rotation period of the primary star is 537.2 days \citep{Wade2011}, which is much shorter than the orbital period of $P_{\rm orb}\sim 1542$ days \citep{Howarth2007}. Similarly to CPD $-28^o$ 2561, this target was also observed in higher frequencies with VLA by \cite{Kurapati2017}, which resulted in non-detection in all frequency bands (2-37 GHz).

\subsection{NGC 1624-2}

NGC 1624-2 is an O7f?cp star \citep{Walborn2010}, possessing the strongest magnetic field among the detected magnetic O-stars \citep{Wade2012}. \cite{Wade2012} detected a dipolar magnetic field of $\sim20$ kG, which is about one order of magnitude higher than the other magnetic O stars (having $B_{\rm p} <$ kG). Similarly to HD 191612, it also has a slow rotation period ($P_{\rm rot} \sim 158$ days, \citealt{Wade2012}) and consequently a DM. This star was a part of the GMRT pilot survey \citep{Chandra2015}, which resulted in a non-detection in the 1.4 GHz band. It was also not detected at higher frequencies (2-37 GHz, \citealt{Kurapati2017}).

\subsection{HD 149438}

HD 149438 ($\tau$ Sco) is a B0.2V star that hosts a magnetic field that significantly deviates from a global dipolar structure \citep{Donati2006}. In addition to the existence of a complex magnetic field, this star stands out from other OB stars due to its hard X-ray emission and peculiar wind diagnostics observed in ultraviolet (UV).  From the temporal variability of the Zeeman signatures, \cite{Donati2006} derived a rotation period of $\simeq41.03$ days for this star. This target was observed by \cite{Kurapati2017} with VLA in the 1-13 cm bands, resulting in a non-detection.

\subsection{HD 66665}

HD 66665 is a $\tau$ Sco analogue star that shows similar UV spectra \cite{Petit2011}. \cite{Petit2011} derived a dipolar magnetic field strength of 0.67 kG for this target. Although it was a $\tau$ Sco analogue, no conclusive evidence of hard X-ray emission was found for this star \citep{Ignace2013}. This star was also not detected in any VLA bands \citep{Kurapati2017}.

\subsection{HD 205021}

HD 205021, more commonly known as $\beta$ Cep, is the prototype of the $\beta$ Cep class of pulsating stars. It has a pulsation period of 4 h and 34 min due to the radial $p$ mode. This star hosts a dipolar magnetic field of strength 0.26 kG \citep{Shultz2019}. Although $\beta$ Cep is in a multiple system \citep{Pigulski1992}, its companions are separated enough to consider this as a single target for this survey. However, this star was not detected with GMRT (this work, \citealt{Shultz2022}) or VLA \citep{Kurapati2017}.

\subsection{HD 163472}

HD 163472 (V2052 Oph) is a He-strong B1V star \citep{Neiner2003} hosting a dipolar magnetic field of strength 1.1 kG \citep{Shultz2019}. This star was also not detected in the VLA bands \citep{Kurapati2017}.

\subsection{HD 184927}

HD 184927 is a slowly rotating ($P_{\rm rot} \sim 9.531$ d), He-strong star with a strong magnetic field \citep{Yakunin2015, Wade1997}. From TESS photometry, we obtained a similar period of 9.5301 days. This star was observed by \cite{Leone1994} with VLA in the 6 cm band, which resulted in a non-detection.

\subsection{HD 37776}

HD 37776 (Landstreet's Star; V901 Ori) is an mCP star with spectral type B2V. It hosts a complex non-dipolar magnetosphere \citep{Kochukhov2011}. This star has a remarkable change in its period in the last few decades \citep{Mikulasek2008, Mikulasek2020}. However, in table \ref{tab:gmrtflux}, we use the ephemeris given by \cite{Adelman1997}.

\subsection{HD 3360}

HD 3360 ($\zeta$ Cas) is a slowly pulsating B-type star with a weak magnetic field \citep{Briquet2016}. This star was also not detected with VLA \citep{Kurapati2017}.

\subsection{HD 35912}

HD 35912 is a less studied B-type star in the Orion star-forming region \citep{Simon2010}. \cite{Bychkov2005} reported a period of 0.89786 days. However, after investigating the TESS data for this target and performing a period analysis, we found no known period. We found 1.1988 days to be the only significant period in the periodogram (Fig. \ref{fig:tess_per0}-a1), and therefore treat it as the rotation period for this target. \cite{Shultz2018b} did not detect any magnetic field from this target using high-resolution spectropolarimetric observations. Thus, the non-detection is well justified, and this star is excluded from any analysis.

\subsection{HD 55522}

HD 55522 is a B2IV/V type star that hosts a magnetic field of dipolar strength of 3.1 kG \citep{Shultz2019, Briquet2007}. \cite{Briquet2007} found a period of $2.729 \pm 0.001$ days. However, a reference HJD was not mentioned in the literature, and thus we performed a period analysis with the TESS data. We found the same rotation period from the periodogram (Fig. \ref{fig:tess_per0}-b1). This target was detected with uGMRT by \cite{Keszthelyi2024} near the magnetic maximum, but resulted in a non-detection near magnetic null. Re-analysis of the GMRT data resulted in a deeper $3-\sigma$ upper limit of flux density of $0.19$ mJy.

\subsection{HD 208057}

HD 208057 (16 Pegasi) is a magnetic B3V star with a magnetic field strength of 0.6 kG \citep{Henrichs2009, Shultz2019}. \cite{Henrichs2009} found a period of 1.441 days from longitudinal magnetic field variation. However, reference HJD was not found in the literature and thus a follow-up period analysis was performed for this star. We found a period of 1.1787 days as the only significant period (Fig. \ref{fig:tess_per0}-c1), and thus adopted it as the rotation period.

\subsection{HD 130807}

HD 130807 is a He-weak magnetic star from the Sco-Cen association \citep{Alecian2011}. \cite{Buysschaert2019} derived a rotation period of 2.95333 days from BRITE Constellation data, which is in agreement with the longitudinal mode variation. In the TESS light curve, we found 0.9044 days as the most significant frequency, which was also observed by \cite{Buysschaert2019}. However, the longitudinal fields do not vary with this period, and thus we selected the original period reported in the literature. However, the origin of this period of $\sim0.9$ days is not known with confidence and could be a g-mode variation.

\subsection{HD 37058}

HD 37058 is a He weak magnetic B-type star \citep{Glagolevskij2007}. \cite{Pedersen1979} reported a period of $14.062\pm 0.075$ days for this target. However, no reference HJD was mentioned in the literature. A reliable period analysis could not be performed with TESS, as this target was observed in only one sector. For this reason, the phase of the radio observations was not calculated. This star was observed by \cite{Linsky1992} with VLA, but it was not detected.

\subsection{HD 125823}

HD 125823 is a He-weak B-type magnetic chemically peculiar star \citep{Borra1983, Renson2009, Shultz2022}. No radio emission from this star was detected from 6 cm VLA observations \citep{Drake1987}, or GMRT observations (this work).

\subsection{HD 36526}

HD 36526 is one of the two targets from this study that is not a unique non-detection. \cite{Das2022c} detected this star with the uGMRT at Band 4 (550--900 MHz), and found evidence of ECME emission near one of the magnetic nulls. The observations of \cite{Das2022c} put a deeper upper limit on the basal flux density ($\sim 0.1$ mJy/beam) for this target.

\section{TESS Period Analysis} \label{App:TESS}

For general Fourier analysis of the TESS light curves, we used the {\textsc{Period04}\footnote{\url{http://period04.net/}}} package \citep{Lenz2005}. For a comprehensive period extraction, we used the Lomb-Scargle-based pre-whitening package {\textsc{pywhiten}\footnote{\url{https://pywhiten.readthedocs.io/}}} \citep{Stacey2022}. Fig. \ref{fig:tess_per0} shows four example periodograms and the corresponding phased light curves. Each phased light curve was fitted with the best period assuming sinusoidal variation. For stars showing a strong first harmonic in the periodogram, a two-component fit was applied. One notable exception to this process is HD 130807, where the strongest frequencies do not represent the rotation frequency, but rather a g-mode frequency \citep{Buysschaert2019}. For the uniquely detected targets, we investigated the TESS light curves, despite having known ephemerides in the literature (Fig. \ref{fig:tess_per0}e--f). We noticed identical frequency values from the TESS light curves and from the literature. For HD 37742, we were unable to find the known rotation period. The best frequency in this case is close to the first harmonic of the known $\sim7$ day period. However, this star was observed in only one sector with TESS, and thus the absence of longer periods in the periodogram is somewhat expected.

\end{document}